\documentclass[fleqn,usenatbib]{mnras}

\usepackage{mathptmx}

\usepackage[T1]{fontenc}

\DeclareRobustCommand{\VAN}[3]{#2}
\let\VANthebibliography\thebibliography
\def\thebibliography{\DeclareRobustCommand{\VAN}[3]{##3}\VANthebibliography}

\usepackage{graphicx}	
\usepackage{amsmath, tabu}	
\usepackage{amssymb}	
\usepackage{adjustbox}
\usepackage{makecell, multirow}
\usepackage{xcolor}
\usepackage{float}
\usepackage{placeins}
\usepackage{tabularx}

\newcommand{\Msun}{\ensuremath{\,{\rm M}_\odot}}                  
\newcommand{\Rsun}{\ensuremath{\,{\rm R}_\odot}}                  
\newcommand{\Lsun}{\ensuremath{\,{\rm L}_\odot}}                  
\newcommand{\Teff}{\ensuremath{T_{\rm eff}}}
\newcommand{\dsct}{$\delta$ Scuti}
\newcommand{\gdor}{$\gamma$ Doradus}
\newcommand{\kms}{\,km\,s$^{-1}$}                                 
\newcommand{\vsini}{\ensuremath{v\sin i}}                         
\newcommand{\Msunnom}{\hbox{$\mathcal{M}^{\rm N}_\odot$}}
\newcommand{\Rsunnom}{\hbox{$\mathcal{R}^{\rm N}_\odot$}}
\newcommand{\Lsunnom}{\hbox{$\mathcal{L}^{\rm N}_\odot$}}
\newcommand{\reff}[1]{#1}

\title[The pulsating eclipsing binary KIC 9851944]{Physical properties of the eclipsing binary KIC 9851944 and analysis of its tidally-perturbed p- and g-mode pulsations}

\author[Z.\ Jennings et al.]{
Z.\ Jennings$^{1}$\thanks{E-mail: z.jennings@keele.ac.uk},
J.\ Southworth$^{1}$ K.\ Pavlovski$^{2}$ T.\ Van\ Reeth$^{3}$
\\
$^{1}$\,Astrophysics Group, Keele University, Staffordshire, ST5 5BG, UK \\
$^{2}$\,Department of Physics, Faculty of Science, University of Zagreb, 10000 Zagreb, Croatia\\
$^{3}$\,Institute of Astronomy, KU Leuven, Celestijnenlaan 200D, B-3001 Leuven, Belgium\\
}

\date{Accepted XXX. Received YYY; in original form ZZZ}

\pubyear{2022}

\begin{document}
\label{firstpage}
\pagerange{\pageref{firstpage}--\pageref{lastpage}}
\maketitle

\begin{abstract}
Stars that are both pulsating and eclipsing offer an important opportunity to better understand many of the physical phenomena that occur in stars, because it is possible to measure the pulsation frequencies of stars for which the masses and radii are known precisely and accurately. KIC 9851944 is a double-lined detached eclipsing binary containing \reff{two F-stars} which show both pressure and gravity mode pulsations. We present an analysis of new high-resolution spectroscopy of the system and high-quality light curves from the \textit{Kepler} and TESS space missions. We determine the masses and effective temperatures of the stars to 0.6\% precision, and their radii to 1.0\% and 1.5\% precision. The secondary component is cooler, but larger and more massive than the primary \reff{so is more evolved}; both lie inside the $\delta$~Scuti and $\gamma$~Doradus instability strips. We measure a total of 133 significant pulsation frequencies in the light curve, including 14 multiplets that each contain between 3 and 19 frequencies. We find evidence for tidal perturbations to some of the \reff{p- and g-modes}, attribute a subset of the frequencies to either the primary or secondary star, and measure a buoyancy radius and near-core rotational frequency \reff{for the primary component}. KIC 9851944 is mildly metal-rich and MIST isochrones from the MESA evolutionary code agree well with the observed properties of the system for an age of 1.25~Gyr.

\end{abstract}




\section{Introduction}
\label{sec:introduction}
Binary and multiple systems make up the vast majority of all observed medium- and high-mass ($\textit{M}>1.15 \Msun$) stellar objects in the galaxy \citep{Duchene_Krauss_2013}. For double-lined spectroscopic binary systems (SB2s), where light emitted from both components is visible in the stellar spectrum, the radial velocity (RV) variations of each component throughout the orbit can be measured. Combining this with the analysis of the light variability due to geometrical effects when the components also eclipse one another, \reff{i.e., eclipsing binaries (EBs)}, \reff{can lead} to estimations of the mass and radius to within $1\%$ \citep{Torres_2010}, making these objects our best source of \reff{precise} fundamental stellar parameters. Since the methods used to derive the properties of the components in \reff{double-lined eclipsing binaries (DLEBs)} do not rely on theoretical stellar models, these objects are critical for testing and verifying stellar evolution theory \citep{Pols_1997, Pourbaix_2000, Lastennet_VallsGabaud_2002, deMink_Pols_Hilditch_2007}. 

Such measurements \reff{for the masses and radii of stars} allow for a precise determination of evolutionary status and age (e.g., \citealt{Higl_Weiss_2017}) \reff{and} the calibration of interior physics. As examples, the mass dependence of convective core overshooting \citep{Claret_Torres_2016, Claret_Torres_2017, Claret_Torres_2018, Claret_Torres_2019}, and the amount of core mixing in massive stars has been probed using \reff{DLEBs} \citep{Pavlovski_2018,Tkachenko+20aa}.

\reff{It has been known since the 1970s that stellar pulsations occur in the components of EBs \citep[e.g.,][for the system AB Cassiopeiae]{Tempesti_1971}.} Oscillation signals measured in the light curves of pulsating stars can be studied using asteroseismology. This is an alternative way by which \reff{accurate} stellar parameters can be derived \citep{Aerts++10book} and is a powerful tool for probing the internal structures of stars \citep{Aerts_2017, Chaplin_Miglio_2013}. The analysis of space-based photometry of pulsating stars has revealed information about important phenomena such as internal rotation, core overshooting and angular momentum transport (e.g., \citealt{Saio_2015, Lovekin_Guzik_2017}). 

It is advantageous to \reff{study pulsating stars in EBs} because analysing the intrinsic variability due to oscillations of the stellar interior and atmosphere as well as the light variability during eclipses leads to independent constraints on theoretical models \citep{Guo_2019, Liakos_2021, Miszuda_2022}. The parameters derived from EBs can also be used to constrain the seismic models \citep[e.g.,][]{Sekaran_2020}, \reff{and this} can be useful, for example, in mode identification for \dsct\ stars \reff{which} is difficult \citep{Streamer_2018, Murphy_2021}. 
The complimentary nature of these two methods of analysis makes pulsating stars in EBs (\reff{particularly DLEBs}) the best objects to use to refine our knowledge of stellar structure and evolution \citep{Guo_2016}.  \citet{Lampens_2021} and \citet{Southworth_2021} have reviewed the impact of space missions on the study of EBs with pulsating components, and on binary stars in general, respectively.

\dsct\ stars are a class of short-period (15\,min to 8\,h; \citealt{Uttyerhoeven_2011, Aerts++10book}), multiperiodic pressure mode (p-mode) pulsators. They are dwarfs or subgiants located at the lower end of the classical instability strip \reff{in the Hertzsprung-Russell (HR) diagram} \citep{Breger_2000, Dupret_2005b, Murphy_2019}. Their spectral types are A2 to F5, corresponding to typical effective temperatures (\Teff s) of 6500\,--\,9500~K \citep{Liakos_2021}. Their masses range from 1.5 to 2.5\Msun\ \citep{Aerts++10book, Yang_2021}, which places them in the transition region between lower-mass stars with radiative cores and thick outer convection zones, and massive stars with convective cores and thin outer convective zones \citep{Yang_2021}. They thus provide an opportunity to study the structure and evolution of stars in this transition region \citep{Bowman_Kurtz_2018}. Low-order, non-radial modes are generally observed for these stars and these modes are driven by the $\kappa$ mechanism acting in the partial ionization zone of He\,{\sc II} \citep{Pamyatnykh_1999,Breger_2000, Antoci_2014, Murphy_2020}. Higher-order non-radial pulsations have been observed in some \dsct\ stars such as $\tau$\,Peg \citep{Kennelly_1998}, where the $\kappa$ mechanism may not be sufficient to explain the modes, and may instead be attributed to the turbulent pressure in the hydrogen convective zone \citep{Antoci_2014, Grassitelli_2015}.

The \gdor\ class of pulsators are \reff{also located at the lower end of the classical instability strip in the HR diagram; they exist} near the red edge of the \dsct\ instability strip \citep{Yang_2021}. Such stars pulsate with typical periods between 8 and 80\,h \citep{Handler_1999} in high order gravity modes (g-modes) \citep{Kaye_1999}, believed to be excited by the interaction of convection and pulsations \citep[e.g.][]{Guzik_2000, Grigahcene_2005, Dupret_2005b}. Typical masses are between 1.3 and 1.8 \Msun\ \citep{Hong_2022, Aerts_2023} with spectral types of F5 to A7. The high radial order of their pulsations puts them in the asymptotic regime, meaning their oscillations are equally spaced in period for non-rotating, non-magnetic, chemically homogeneous stars \citep{Tassoul_1980}. Deviations from a homogeneous spacing emerge due to chemical gradients and rotation \citep{Sekaran_2020}, where the former are influenced by the effects of diffusive mixing; mixing reduces the steepness of chemical gradients. Period spacing diagrams can therefore be used to derive information on chemical composition gradients, internal rotation rates and diffusive mixing processes \citep[e.g.][]{Bouabid_2013,Bedding_2015, Saio_2015, VanReeth_2016, Ouazzani_2017, Li_2019,Miglio_2008, Moravveji_2015, Sekaran_2021}. 

Space missions such as \emph{Kepler} \citep{Koch_2010, Borucki_2010}, CoRoT \citep{Baglin_2006} and TESS \citep{Ricker_2015}, have delivered a large amount of photometric data with precisions \reff{unachievable from the ground}. The unprecedented precision has allowed for the detection of extremely low-amplitude frequencies \citep{Murphy_2013, Bowman_Kurtz_2018} while long sequences of continuous observations \citep{Lehman_2013} means that longer-period pulsations, such as those typical of \gdor\ stars, can be studied. The overlap of the \dsct\ and \gdor\ instability strips supports the existence of \dsct/\gdor\ hybrids \citep{Breger_Beichbuchner_1996,Handler_Shobbrook_2002, Yang_2021}. \reff{These were expected to be rare, based on early calculations by \citet{Dupret_2005b}, but the lower detection thresholds provided by space missions has led to the discovery that such behaviour is indeed common \citep{Uttyerhoeven_2011,Grigahcene_2010,Bradley_2015, Balona_2015, Guo_2019}. Later calculations by \citet{Xiong_2016} conform better with these findings}. 

Hybrids have great potential for asteroseismology \citep{Schmid_Aerts_2016}. The p-modes probe the stellar envelope while the g-modes carry information about the near-core regions \citep{Yang_2021, Grigahcene_2010, Saio_2015, Kurtz_2014}. The behaviour also means that information about two different driving mechanisms can be obtained \citep{Hong_2022}.

KIC 9851944 is an EB showing \dsct/\gdor\ hybrid pulsation signatures. Therefore, the object is an ideal candidate for testing our understanding of stellar structure and evolution given the large amount of constraints that can be obtained due to the advantages associated with hybrid behaviour as well as binarity. KIC 9851944 is included in the \emph{Kepler} Eclipsing Binary Catalogue (KEBC; \citealt{Prsa_2011,Kirk_2016}), as well as a study by \citet{Matson_2017} who presented RVs for 41 \emph{Kepler} EBs. \citet{Gies_2012, Giess_2015} studied the eclipse times for KIC 9851944 and found no evidence of apsidal motion or a third body; the object was \reff{also included in a catalogue of precise eclipse times} of 1279 \emph{Kepler} EBs by \citet{Conroy_2014}. 

\reff{\citet{Guo_2016} combined the analysis of \emph{Kepler} photometry with medium-resolution spectra ($R = \lambda/\Delta \lambda \sim 6000$) to determine the atmospheric and physical properties of KIC 9851944; we list these results in Table \ref{tab: previous results}. Evolutionary modelling based on these properties shows the post-main-sequence secondary to be more evolved than the main-sequence (MS) primary. The authors concluded that both components show \dsct\ type pulsations, which they interpreted as p-modes and p and g mixed modes, and attempted to identify the modes by comparison with theoretically computed frequencies; the range of theoretically predicted unstable modes agreed roughly with observations but the authors conclude that mode identification is still difficult in \dsct\ stars, even with constrained mass, radius and \Teff. }

This work aims to be complementary to the work by \citet{Guo_2016} on KIC 9851944; we additionally include observations by TESS in our photometric analysis and combine this with the analysis of high-resolution ($R=60000$) spectroscopic observations. Section \ref{sec:observations} outlines the photometric and spectroscopic observations. \reff{We determine the orbital ephemeris based on the photometric observations in Section \ref{sec:ephem}}. \reff{In Section \ref{sec:RV-analysis} we analyse RVs derived from the spectroscopic observations and in Section \ref{spectroscopic analysis} we present the spectroscopic analysis}. \reff{The analysis of the photometric light curves is given in section \ref{sec analysis of light curve} and in Section \ref{tab:absdim} we present the physical properties of the system}. An investigation of the pulsations is given in Section \ref{sec:asteroseismology}. The discussion and conclusion are given in Sections \ref{sec:discussion} and \ref{sec:conclusion}, respectively.


\section{Observations}
\label{sec:observations}

\subsection{Photometry}
\begin{figure*}
    \centering
    \includegraphics[width = \textwidth]{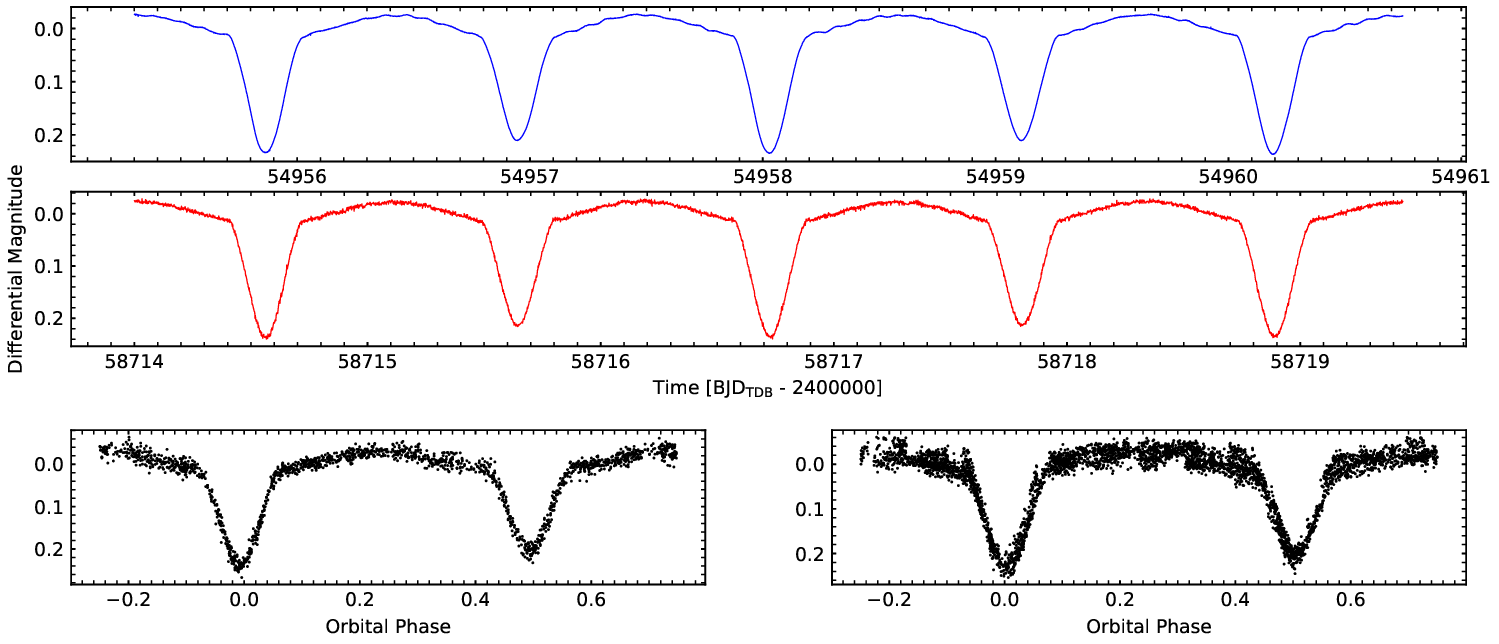}
    \caption{\reff{Cuts of the \emph{Kepler} Quarter 0 (top panel) and TESS Sector 15 (middle panel) processed light curves. Also shown in the bottom panels are the phase-folded light curves observed by WASP between May 2007 and July 2007 (left), and between June 2008 and July 2008 (right).}}
    \label{fig: light-curve plot}
\end{figure*}

The \emph{Kepler} space telescope \citep{Koch_2010} was launched in March 2009 with the primary aim of detecting Earth-like extrasolar planets around solar-like stars. However, the mission has also provided a huge amount of high-quality data on stars and stellar systems \citep{Hong_2022}. The data were collected using two modes of observation \citep{Gilliland_2010}, long cadence (29.5\,min sampling rate) and short cadence (58.5\,s sampling rate). To keep the solar arrays pointed toward the Sun during its Earth-trailing heliocentric orbit, the spacecraft was rotated by $90^{\circ}$ every three months (one quarter). \reff{Primary functions of the \emph{Kepler} photometric analysis module of the \emph{Kepler} science data processing pipeline are responsible for generating the Simple Aperture Photometric (SAP) flux time series of the observed objects, while secondary photometric analysis functions performed by the Pre-Search Data Conditioning (PDC) module support systematic error correction, giving rise to the PDCSAP fluxes \citep[see][for further details on the SAP and PDCSAP measurements]{Kepler_Handbook}.} KIC 9851944 was observed by \emph{Kepler} in six quarters \reff{(0, 12, 13, 14, 16, 17)} in short cadence mode between May 2009 and May 2013. 

The Transiting Exoplanet Satellite Survey (TESS; \citealt{Ricker_2015}) was also designed to discover extrasolar planets and also produced a large amount of high-quality data for many other celestial objects. More than $2\times10^5$ main-sequence dwarfs were observed at 2\,min cadence during the all-sky survey, which was divided into overlapping sectors on the sky. KIC 9851944 was observed by TESS in sectors 14, 15, 41, 54, 55 and 56 between July 2019 and September 2022. \reff{The TESS science pipeline is based on the \emph{Kepler} science pipeline so the architecture and algorithms are similar; the photometric analysis module computes the SAP flux time-series and the PDC module corrects for systematics to compute the PDCSAP fluxes \citep{Jenkins_2016}}. 

The light curves from the \emph{Kepler} quarters and TESS sectors \reff{mentioned above} were downloaded from the Mikulski Archive for Space Telescopes (MAST) \reff{and are used in Section \ref{sec:ephem} to determine the ephemeris of the system, as well as Section \ref{sec analysis of light curve} to obtain the final model of the light curve}. Both SAP and PDCSAP measurements are available; the SAP and PDCSAP fluxes were similar, \reff{which we verified by inspecting the SAP fluxes with the PDCSAP fluxes over-plotted after dividing them by their median flux values to put them on the same scale. Thus,} we used the SAP measurements to avoid possible biases due to the additional processing applied to the PDCSAP data. \reff{A \reff{second}-order polynomial was fitted to fluxes that correspond to positions of quadrature, i.e., the maximum of the ellipsoidal brightening, to estimate systematic trends present in the light curves. Subtracting the difference between this polynomial and the median flux of the light curve yields a model for the local median level of out-of-eclipse flux. This model was then divided out to remove systematic trends.} The residual value of a smoothed version of the light curve subtracted from the observed light curve was calculated and observed fluxes that deviated by more than $1\%$ were rejected. Fluxes were converted to magnitudes and errors were propagated following \cite{Prsa_2018}. \reff{The short cadence \emph{Kepler} light curve from Quarter 0 and the two-minute cadence TESS light curve from Sector 15, after performing these operations, are shown in the top and middle panels of Fig.\ref{fig: light-curve plot}}, respectively.

KIC 9851944 was also observed by the Wide Angle Search for Planets (WASP) telescope \citep{Butters_2010} between May 2007 and July 2010. WASP consisted of two robotic observatories, one in the Nothern Hemisphere at Observatorio del Roque de los Muchachos on La Palma and the other at the South African Astronomical Observatory (SAAO), each with eight 20\,mm telescopes on a single mount. \reff{Observations of KIC 9851944 collected by WASP between May 2007 and July 2007 as well as between June 2008 and July 2008 are shown in the bottom panels of Fig.\ref{fig: light-curve plot} phase folded about the orbital period determined in Section \ref{sec:ephem}}. The data collected by WASP for KIC 9851944 were \reff{only used to constrain the times of primary minima when performing preliminary fits to the \emph{Kepler} and TESS light curves in Section \ref{sec:ephem}, to determine the orbital ephemeris.}

\subsection{Spectroscopy}

A set of 33 spectroscopic observations were obtained for the target using the Hamilton \'echelle spectrograph \citep{Vogt_1987} on the Shane 3\,m telescope at Lick Observatory during two observing runs, one in July 2012 and the other in June 2013. The data were obtained using CCD chip no.\,4, giving a wavelength coverage of 340\,--\,900\,nm over 89 \'echelle orders at a resolving power of $R \approx 60000$. 

The data were reduced using the standard pipeline for the spectrograph \citep[e.g.][]{Fischer+14apjs}. Flat-fields were obtained with a quartz lamp and divided from the spectra. The wavelength calibration was obtained from exposures of a thorium-argon emission lamp taken roughly every hour during the night.

\reff{Details for the normalization of the one-dimensional extracted spectra are given in Sections \ref{sec:RV-analysis} and \ref{spectroscopic analysis} because each of those components of the analysis used different approaches.}

\reff{Table \ref{tab: RVs and S/N for specobs} gives the epochs of each of the 33 spectrosopic observations as well as the signal-to-noise ratio (S/N), estimated as the square root of the counts close to the peak of the best-exposed \'echelle order. Also given are the corresponding RVs of each component that are derived in Section \ref{sec:RV-analysis}.}

\begin{table}\caption{\label{tab: RVs and S/N for specobs} \reff{RVs and S/N corresponding to the spectroscopic observations taken at times given in the BJD column. An asterisk next to the BJD value means that observation was not used to derive the orbital parameters in Section \ref{sec:RV-analysis}}. }
    \centering
    \begin{tabular}{l r@{\,$\pm$\,}l r@{\,$\pm$\,}l l}
    \hline
     BJD & \multicolumn{2}{c}{RV Star A} & \multicolumn{2}{c}{RV Star B} & S/N \\
    \hline
2456133.71296&  $-$122.47&   1.33&  103.45& 1.14& 59 \\
2456133.72724&  $-$119.89&   1.37&  101.95& 1.08& 59 \\
2456133.74151&  $-$118.46&   1.26&  99.54&  1.03& 57 \\
2456133.75578&  $-$114.54&   1.21&  96.79&  0.93& 57 \\
2456133.80101&  $-$104.26&   1.28&  87.87&  1.13& 60 \\
2456133.83920&  $-$94.81&    1.26&  79.68&  1.25& 58 \\
2456133.87945&  $-$84.36&    1.02&  71.67&  1.67& 59 \\
2456133.90977&  $-$77.35&    1.31&  63.03&  1.97& 58 \\
2456133.94924&  $-$64.66&    1.51&  51.37&  2.85& 60 \\
2456133.98702*&  $-$41.04&    5.71&  34.50&  5.52& 57\\
2456134.70808&   117.69&   1.01& $-$121.35& 1.39& 57\\
2456134.72235&   116.88&   0.88& $-$120.01& 1.68& 59\\
2456134.73662&   116.27&   0.92& $-$119.12& 1.65& 58\\
2456134.75090&   114.89&   0.81& $-$118.39& 1.66& 55\\
2456134.78975&   109.50&   0.96& $-$113.29& 1.81& 53\\
2456134.82917&   104.240&   1.02& $-$107.26& 1.72& 50\\
2456134.86653&   97.71&    0.94& $-$99.91&  1.60& 56\\
2456134.91057&   86.88&    0.84& $-$91.01&  1.74& 53\\
2456134.94964&   77.43&    1.15& $-$82.09&  1.50& 54\\
2456134.98739&   65.92&    1.34& $-$72.46&  1.18& 52\\
2456469.93848&   115.04&   1.08& $-$115.25& 1.17& 31\\
2456469.97977&   115.86&   0.98& $-$120.41& 1.65& 55\\
2456470.72325*&  $-$41.06&    2.92&  29.87&  5.18& 41\\
2456470.79028&  $-$69.87&    1.09&  56.92&  2.79& 35\\
2456470.84339&  $-$84.53&    0.94&  69.97&  1.69& 57\\
2456470.90594&  $-$101.34&   1.04&  83.79&  1.11& 56\\
2456470.96852&  $-$116.02&   1.22&  95.28&  1.17& 57\\
2456470.99830&  $-$121.03&   1.23&  99.82&  1.12& 54\\
2456471.71677*&         &       &      &      & 43\\
2456471.77503*&   19.52&    2.84& $-$29.63&  2.40& 30\\
2456471.87251&   58.97&   1.67&  $-$68.30&  1.74& 50\\
2456471.90957&   69.15&   1.34&  $-$79.35&  1.53& 54\\
2456471.93874&   79.31&   1.00&  $-$85.78&  1.15& 54\\
\hline
    \end{tabular}
\end{table}

\section{Orbital ephemeris}
\label{sec:ephem}

A first model of the \reff{\emph{Kepler} and TESS} light curves was obtained using the \textsc{jktebop} code \citep{JKTEBOP}, \reff{which approximates stellar shapes in detached eclipsing binaries based on the simple and efficient NDE biaxial ellipsoidal model, so is computationally fast \citep{Nelson&Davis_1972, Southworth_2004b}}. We fitted for the period $P$, epoch $T_0$, surface brightness ratio $J$, sum of the fractional radii $r_{\rm A}+r_{\rm B}$, ratio of the radii $k = \frac{r_{\rm B}}{r_{\rm A}}$, inclination $i$, the Poincar\'e elements $e\cos\omega$ and $e\sin\omega$, and a light scale factor. We define star~A to be the star eclipsed at the primary (deeper) minimum and star~B to be its companion. 

\reff{We also fitted for the linear coefficients $u$ of the quadratic limb darkening (LD) law while quadratic terms were taken from \cite{Claret&Bloemen_2011} and \cite{Claret_2017} for the \emph{Kepler} and TESS bands, respectively; we used the estimated atmospheric parameters reported in the KIC to choose the appropriate values for the quadratic coefficients. We performed fits to the WASP light curve but the lower quality of this data compared to the \emph{Kepler} and TESS light curves means that including these results in the calculation of the overall preliminary eclipse model would lead to less well determined parameters. Thus, we simply include the epochs of primary minimum 

estimated from the WASP light curves as additional observational constraints on $T_0$ in the preliminary fits to the \emph{Kepler} and TESS light curves.}  

\reff{The adopted values for the light curve parameters from this preliminary analysis were taken as the weighted averages of the results from the fits to the individual \emph{Kepler} quarters and TESS sectors, where the reciprocal of the squared errors from the covriance matrix were used as weights.} 
\begin{figure}
    \centering
    \includegraphics[width = \columnwidth]{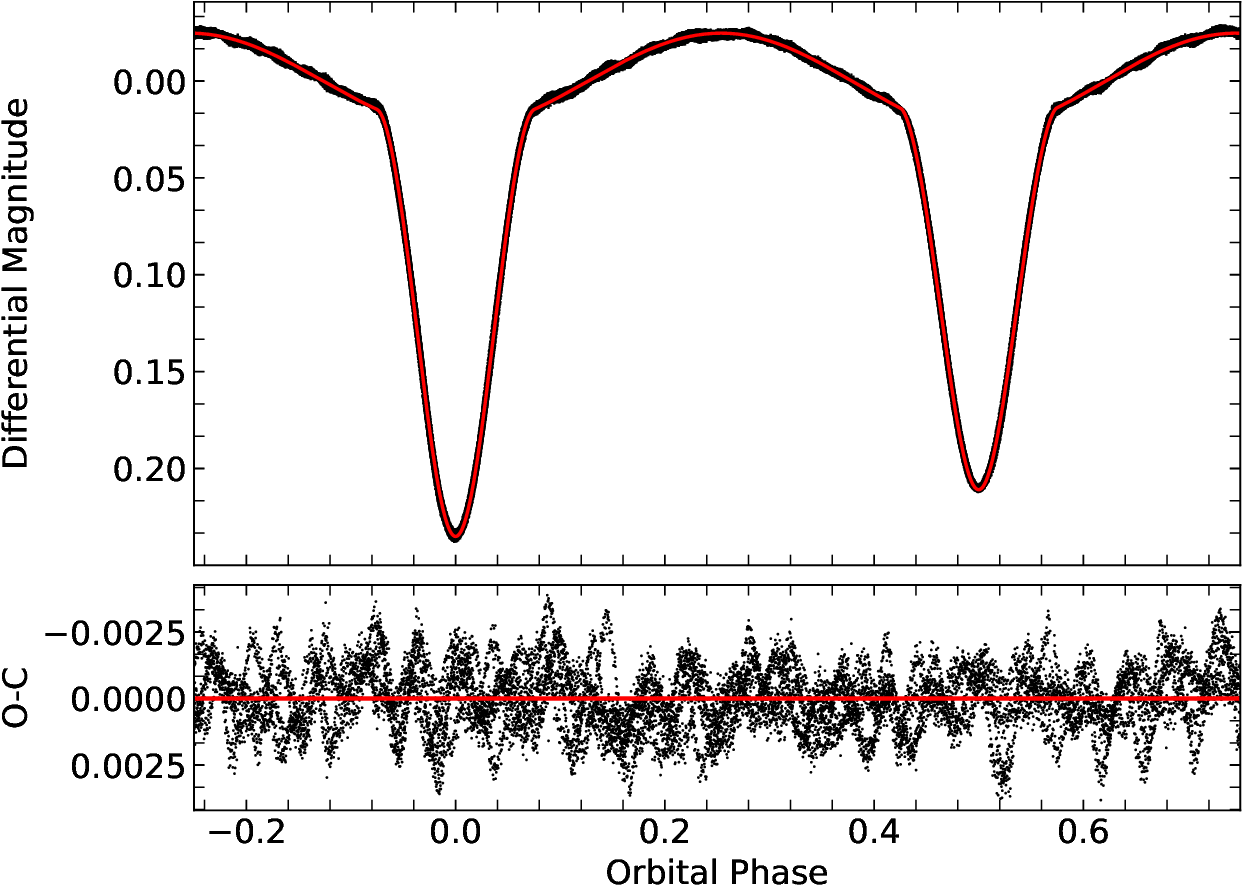}
    \caption{{\sc jktebop} model fit to the phase folded \emph{Kepler} light curve from quarter 0.}
    \label{fig:q0-light-curve-fit}
\end{figure}
\reff{These values are given in Table \ref{tab:pre_lcfit}, and Fig.\,\ref{fig:q0-light-curve-fit} shows the fit to the \emph{Kepler} Quarter 0 light curve; high-frequency variability due to pulsations of the \dsct\ type is clearly visible at all orbital phases in the residuals shown in the lower panel, with amplitudes of around $\pm$0.2\,mmag.}

\reff{The value for $r_{\rm A}+r_{\rm B}$ in Table \ref{tab:pre_lcfit} suggests that the components of KIC\,9851944 are deformed beyond the limits of applicability of the biaxial ellipsoidal approximation \citep{Popper&Etzel_1981} due to their close proximity and thus strong mutual deformations; the ellipsoidal variation ($\sim 0.04$\,mag) observed in Fig.\,\ref{fig: light-curve plot} and Fig.\,\ref{fig:q0-light-curve-fit} is another indication of this. Furthermore,} plotting the results from individual quarters and sectors against each other reveals strong degeneracies between the fitted parameters, specifically $r_{\rm A}+r_{\rm B}$, $k$, $i$, $J$ and the LD coefficients. This is highlighted by the results giving a wide range of ratios of the light contributions of the two stars. \reff{Therefore, we present a detailed analysis of the \emph{Kepler} and TESS light curves using a more sophisticated modelling code, as well as constraints on the light ratio $\ell_{\rm B}/\ell_{\rm A}$ from our spectral analysis, to reliably determine the light curve solution in Section \ref{sec analysis of light curve}.}

\begin{figure}
    \centering
    \includegraphics[width = \columnwidth]{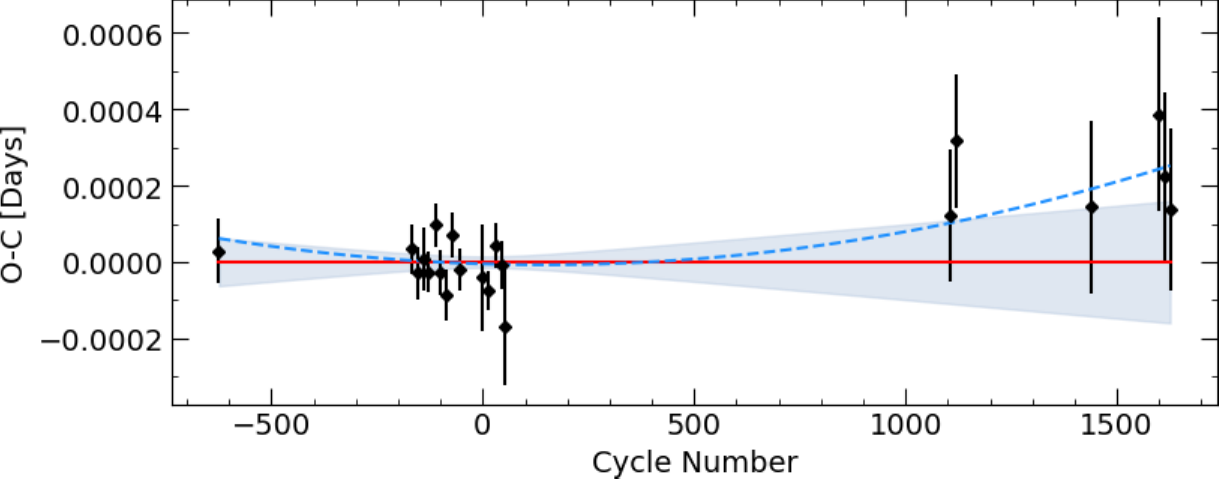}
    \caption{O-C plot from the linear fit (\reff{red line}) to the times of primary minimum (\reff{black dots}), \reff{where the grey shaded region represents 1\,$\sigma$ uncertainty associated to the corresponding calculated values.}. \reff{The blue dashed line represents an attempt to additionally fit for a quadratic term.}}
    \label{fig:ephemeris}
\end{figure}

Among the reliably constrained parameters from this preliminary analysis are the period and epochs of primary minimum from individual quarters and sectors. Therefore, the analysis is useful for establishing the orbital ephemeris. \reff{A linear ephemeris was fit to the resulting values of $T_0$ against orbital cycle using the reciprocal of the squared errors from the covariance matrix as weights; these errors were rescaled during the fitting procedure to yield a reduced chi-squared value of $\chi^2_\nu = 1$. Fig.\,\ref{fig:ephemeris} shows the residuals of the fit and the red line represents the linear ephemeris.} 

\reff{There seems to be a trend in the O-C diagram compared with the linear ephemeris, with data points corresponding to TESS observations ($>$1000 cycles) all appearing above the red line. Thus, we also attempted to fit for a quadratic ephemeris, which is represented by the blue dashed line in Fig.\,\ref{fig:ephemeris}. However, the corresponding quadratic term was a similar size to its errorbar. We therefore decided to stick with the linear} ephemeris which was measured to be:
\\\\
Min I (BJD$_{\rm TDB}$) = 2456308.304072(14) + 2.163901775(98)\,E.
 
\section{Radial velocity analysis}
\label{sec:RV-analysis}

The spectral range between \reff{4400--4800\,\AA}\ is a suitable region to carry out the RV extraction procedure because there are many well-resolved spectral lines and the region does not contain any wide Balmer lines. \reff{Thirteen} \'echelle orders within this range were corrected for cosmic ray spikes by first resampling onto a homogeneous wavelength grid before using a median filter to smooth them, then resampling the orders back to the original grid and subtracting them from the observed spectrum \citep{Blanco_2014}. Pixels where the residual deviated by a threshold percentage, \reff{typically between 5 and 10$\%$}, were masked and their values were corrected for using linear interpolation. The appropriate value for the threshold was investigated for each order individually based on visual inspections of resulting flagged pixels. The blaze signature was removed using the method of \citet{Xu_2019}, where alpha shape fitting \citep[see,][]{Edelsbrunner_1983} combined with local polynomial regression is used to estimate the continuum without dividing out important spectral features. 

Small errors in the continuum estimation are magnified where the blaze function approaches zero. Following \citet{Xu_2019}, a better estimation at the edges can be obtained by taking a weighted average in the overlapping regions and applying weights such that pixels further away from the edge within their own order are favoured. For this to be applied correctly, the orders were re-sampled onto grids with a \reff{common} wavelength step so that the pixel averages were calculated for corresponding wavelength values between neighbouring orders. A useful result from the edge correction procedure is that neighbouring orders share a common edge and can be merged after truncating the orders in the centre of the overlapping regions. The following analysis was carried out after merging the 13 spectral orders \reff{with wavelengths between \reff{4400-4800\,\AA} using this process}. 

\begin{figure}
    \centering
    \includegraphics[width=\columnwidth]{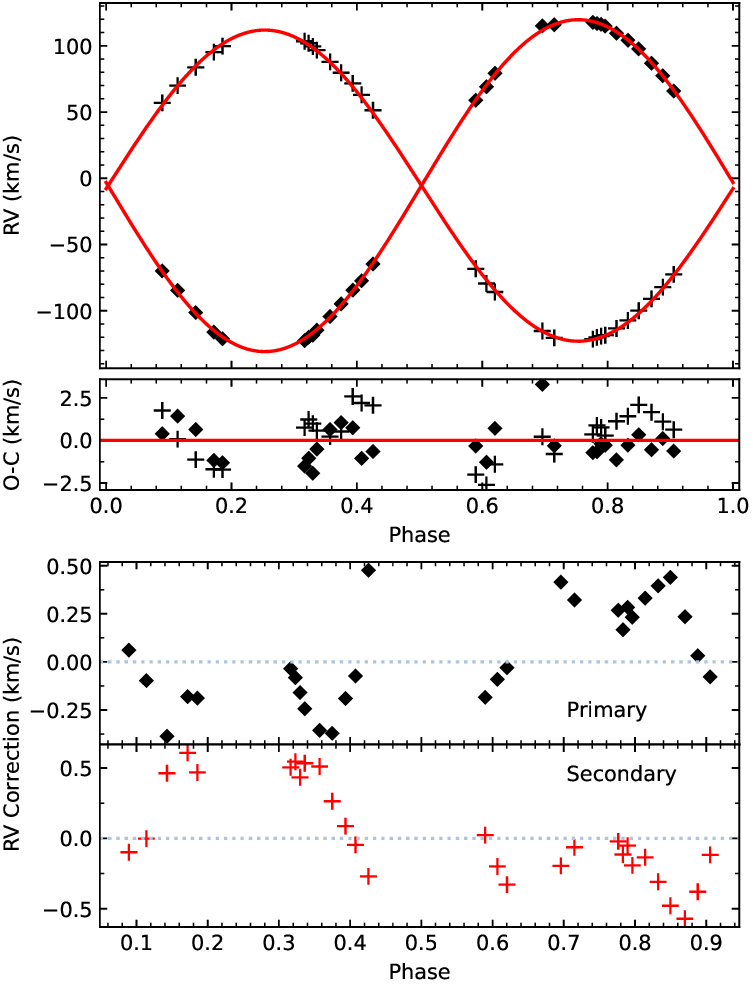}
    \caption{Top: orbital fit to the corrected RVs using the merged approach. Bottom: velocity corrections for the corresponding RVs.}%
    \label{fig:RV}
\end{figure}%
Preliminary estimations for the projected rotational velocity, \vsini, of the components of KIC 9851944 were obtained by cross correlating the observed spectra with a template broadened to a range of \vsini\ values between 0 and 150\kms\ in steps of 10\kms. Observations taken less than 0.125 orbital phases from the eclipses were excluded to minimise issues arising due to lines from each component blending or being eclipsed. Interpolating the peak heights of the correlation functions and taking the maximum yields $\vsini_{\rm A} = 46.3 \pm 0.4$\kms\ and $\vsini_B = 57.2 \pm 0.5$\kms\ for the primary and secondary component, respectively. These values for \vsini\ \reff{were} then used and applied to synthesise primary and secondary templates using the {\sc ispec} code \citep{ispec}. \reff{In the first iteration, the \Teff\ and $\log (g)$ values were taken from the Kepler Input Catalogue (KIC). For the final iteration of the RV extraction, the atmospheric parameters of the templates corresponded to those derived from our atmospheric analysis of the disentangled spectra (see Section \ref{sec:atmospheric parameters}.})

Our own implementation of the \textsc{todcor} two-dimensional cross-correlation algorithm \citep{Zucker&Mazeh_1994} was used to extract the RVs. Systematic errors arise when using the cross-correlation technique to extract RVs because neighbouring peaks and side-lobes in the doubled-peaked CCF disturb each other, and this is related to blending between the spectral lines of the two components \citep{Andersen75aa, Latham_1996}. The result is that the estimated RVs may be shifted relative to the true value, where the size and direction of the shift depends on the phase (see Fig.\ref{fig:RV}), i.e, the relative Doppler shifts between the components \citep{Latham_1996}. 

\reff{This systematic error can be minimised \reff{by correcting} the RVs for this effect \citep[e.g.,][]{Latham_1996,Torres_1997, Torres_2000,Torres&Ribas_2002, Southworth_Clausen_2007}.} A preliminary orbit was obtained by making an initial fit to the RVs extracted using \textsc{todcor}. A synthetic SB2 \reff{model} was then produced with orbital parameters corresponding to the initial fit by Doppler shifting and adding the template spectra weighted by their relative light contributions as estimated by the resulting value for $\ell_{\rm B}/\ell_{\rm A}$ from the initial \textsc{todcor} run. The \textsc{todcor} algorithm was then used to extract the known RVs from the synthetic SB2 model at orbital phases corresponding to the observations. The difference between the calculated RV and the known RV of the model gives an estimation for the magnitude and direction of the systematic error at that orbital phase. The correction was then added to the actual RVs measured from the observed spectra.

This procedure was iterated multiple times using templates with different atmospheric parameters; the top panel of Fig.\,\ref{fig:RV} shows the resulting orbital fit to the corrected velocities \reff{that were derived in our final iteration, where we used templates with atmospheric parameters corresponding to those derived in Section \ref{sec:atmospheric parameters}, and the results are presented in Table \ref{tab:orbital_params}.} The middle panel of Fig.\,\ref{fig:RV} shows the residuals of the fit and the lower panel, split into two for the primary and secondary, gives the corrections that were applied to the extracted RVs. RVs with corrections larger than $2.5$\kms\ were excluded \reff{and are not shown because these RVs correspond to phases of conjunction, where line blending effects are most severe, and negligible information is contained on the velocity semiamplitudes}. The fit was constrained to a circular orbit because attempts to fit for eccentricity yielded values consistent with zero and the study by \citet{Guo_2016} suggests that this system has circularised. \reff{The uncertainties on  the RV measurements were rescaled by a constant factor during the fitting procedure to yield a $\chi^2_r$ of unity. The root mean square (RMS) of the residuals of the fit compared to the RVs for the primary and the secondary are 1.1\,km/s and 1.4\,km/s, respectively.} 

\reff{The differences between our final results for $K_{\rm A}$ and $K_{\rm B}$ presented in Table \ref{tab:orbital_params} and those derived from the initial run (both after applying corrections), which used templates with atmospheric parameters corresponding to those from the KIC, was $\sim0.1\%$. The corresponding difference in $\ell_{\rm B}/\ell_{\rm A}$ was $\sim 9\%$. This shows that the \textsc{todcor} light ratio is more sensitive to the atmospheric parameters of the templates than the derived RVs. We added these differences in quadrature, to the uncertainties derived from the covariance matrix of the fit for the orbital parameters, and to the standard error in the mean value of $\ell_{\rm B}/\ell_{\rm A}$ derived from different observations, in our calculation of the parameter error bars in Table \ref{tab:orbital_params}.}
\begin{table}\caption{\label{tab:orbital_params}Orbital Parameters.}
   \centering
    \begin{tabular}{l r@{\,$\pm$\,}l }
    \hline 
      Parameter&  \multicolumn{2}{c}{Value} \\
     \hline
     $K_{\rm A}$ (km/s)          & 125.282 & 0.269  \\ 
     $K_{\rm B}$ (km/s)          & 117.453 & 0.328  \\
     $\gamma$ (km/s)             & $-$5.52 & 0.16    \\
     $\ell_{\rm B}/\ell_{\rm A}$ & 1.222   & 0.136  \\
                                        
     \hline
\end{tabular} \end{table}

Applying the corrections resulted in an increase in the velocity semiamplitudes $K_{\rm A}$ and $K_{\rm B}$ by $\mathbf{0.2}\%$ and $\mathbf{0.3}\%$, respectively. \reff{This is a small increase}, suggesting that \textsc{todcor} is less sensitive to blending effects than the one-dimensional cross correlation technique. \reff{However, the $0.2\%$ and $0.3\%$ increase in the velocity semiamplitudes translates to a $0.6\%$ and $0.9\%$ increase in the derived masses, which is significant considering that we aim to achieve precisions smaller than these values. This suggests that the corrections are necessary. Furthermore, the corrections are clearly systematic, and depend on the relative Doppler shifts of the components, as shown in the bottom panels of Fig.\ref{fig:RV}.}


\section{Spectral analysis}\label{spectroscopic analysis}

\subsection{Atmospheric parameters}\label{sec:atmospheric parameters}

The method of spectral disentangling (SPD) makes it possible to separate the spectra of individual components from a time-series of binary or multiple star spectra \citep{Simon_Sturm_1994}. SPD has other advantages: the optimal orbital parameters are determined at the same time, and the disentangled spectra have a higher S/N than the original observed spectra. The disentangled spectra contain information on the atmospheric parameters (\Teff, surface gravity, $v \sin i$ and metallicity) of each component. The disentangled spectra are still in the common continuum of the binary system, but can be renormalised if the light ratio between the components is known from other sources, e.g.\ from the analysis of light curves \citep{Hensberge_2000, Pavlovski_2005, Pavlovski_2010}. Alternatively, the disentangled spectra can be fitted with synthetic spectra to determine the light ratio between the components, and this can be useful for partially-eclipsing systems where there is a degeneracy between the radius ratio and light ratio in the system \citep{Tamajo_2011,Tkachenko_2015,Pavlovski_2023}. In well-determined cases, the precision of the light ratio can approach 1\% \citep{Pavlovski_2022}. 

\reff{Since we planned to apply the method of SPD to extract the individual spectra of the components, normalisation of the observed spectra was of critical importance. We used a different approach than that in Section \ref{sec:RV-analysis}, where we extracted RVs. Here, we used the dedicated code described in \citet{Kolbas_2015}. First, the blaze function of \'echelle orders was fitted with a high-order polynomial function. Then, the normalised \'echelle orders were merged. When overlapping regions of successive \'echelle orders are sufficiently long, the very ends were cut off because of their low S/N. \'Echelle orders containing broad Balmer lines, in which it is not possible to define the blaze function with enough precision, were treated in a special way. For these \'echelle orders, the blaze function was interpolated from adjacent orders. This produces more reliable normalisation in spectral regions with broad Balmer lines than the usual pipeline procedures. For recent applications of this approach, please see \citet{Pavlovski_2018, Pavlovski_2023, Lester_2019, Lester_2022, Wang_2020,  Wang_2023}.}

\begin{figure}
    \centering
    \includegraphics[width = \columnwidth]{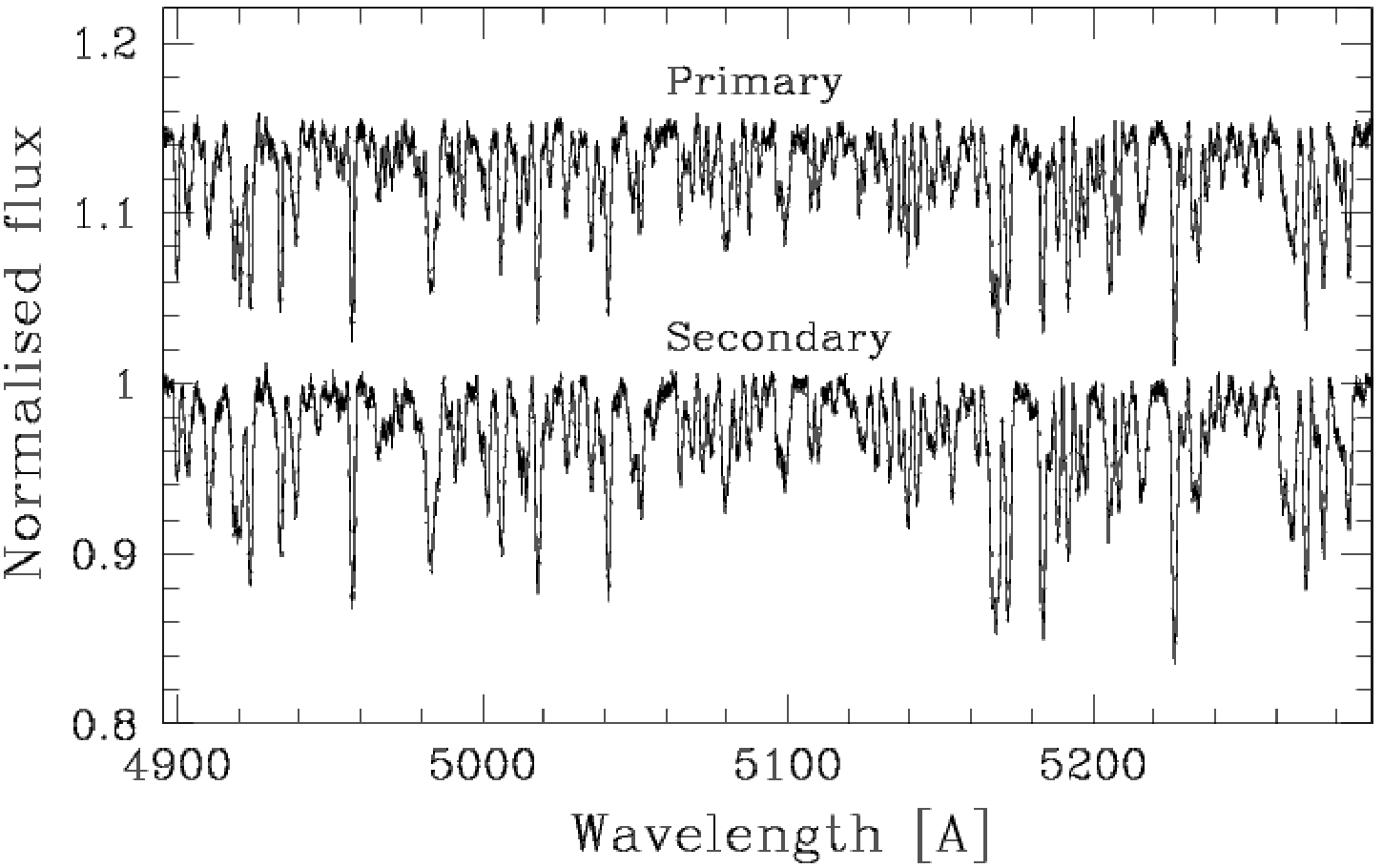}
    \caption{Portion of disentangled spectra in the spectral range 4900--5290~\AA. The similarity between the spectra of both components is obvious.}
    \label{fig:disent}
\end{figure}

SPD was performed in the Fourier domain with the prescription by \citet{Hadrava_1995}. The {\sc FDBinary} code, developed in \citet{Ilijic_2004}, was applied to the time-series of normalised \'echelle spectra of KIC~9851944. {\sc FDBinary} uses the fast Fourier transform approach which allows flexibility in selection of spectral segments for SPD whilst still keeping the original spectral resolution. The orbital parameters, specifically $K_{\rm A}$ and $K_{\rm B}$, were fixed to the solution reported in Table~\ref{tab:orbital_params}, thus SPD was used in pure separation mode \citep{Pavlovski_2010}. At this point the reconstructed individual spectra of the components were still in the common continuum of the binary system. A portion of separated spectra for both components in the spectral range 4900--5290\,\AA\ is shown in Fig.~\ref{fig:disent}.

\begin{figure}
    \centering
    \includegraphics[width = \columnwidth]{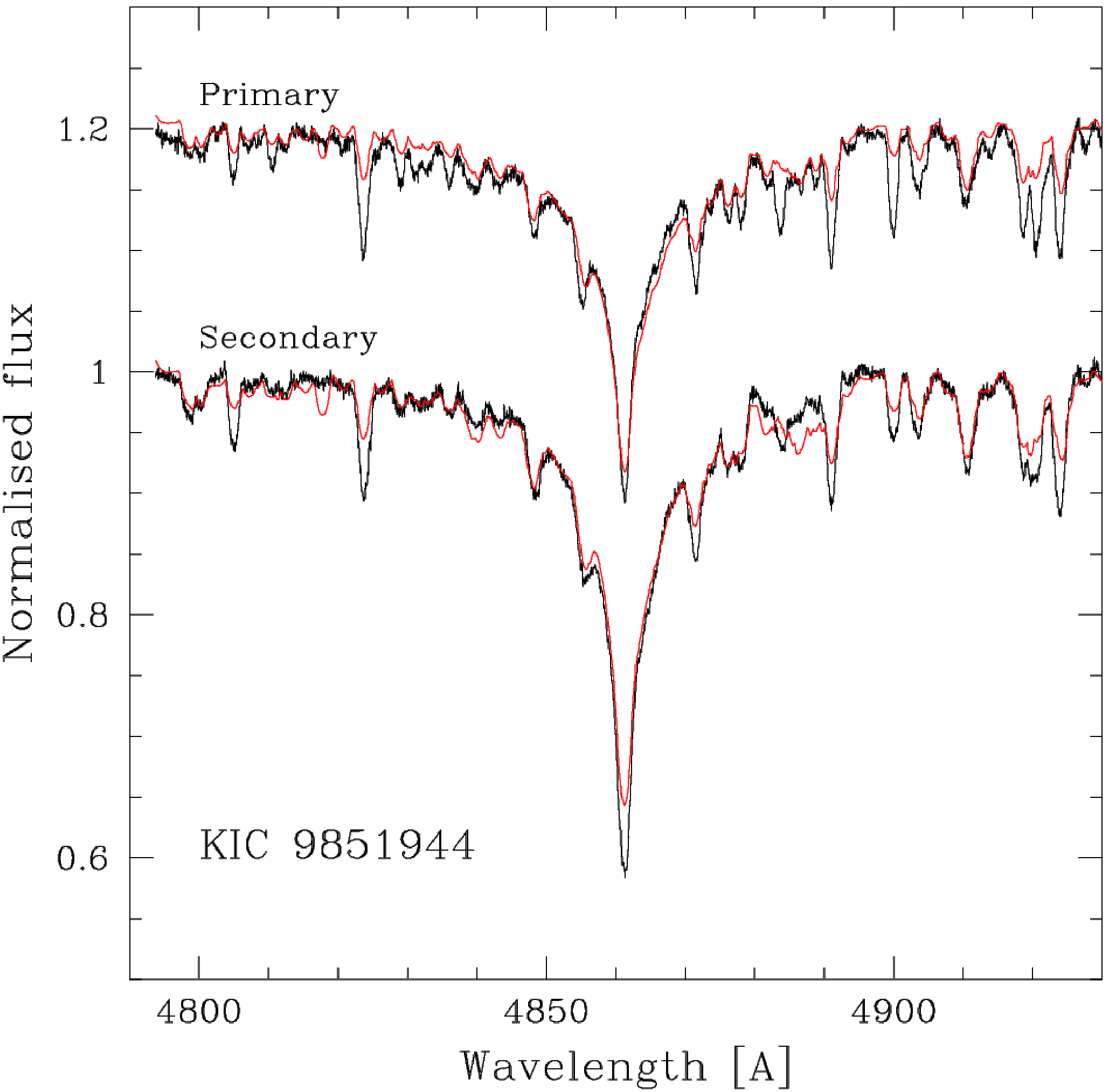}
    \caption{Optimal fitting of the wings for H$\beta$ lines in disentangled spectra of binary system KIC~9851944. The spectral lines of metals were masked during the fitting, so were not included in calculations of the merit function. The disentangled spectra are shown in black, and the optimal fits in red.}
    \label{fig:optimal}
\end{figure}

\begin{table}
\caption{\label{tab:optfits} Determination of the atmospheric parameters from disentangled spectra for the components of KIC 9851944. \reff{The surface gravity for each component was fixed to the values determined from light curve and RVs solution as listed in Table~\ref{tab:absdim}.}}
   \centering
    \begin{tabular}{lccc}
    \hline 
    Parameter & H$\beta$ & H$\alpha$ & Metals  \\
    Wavelength range (\AA)  & 4750--4890 & 6500--6600 & 5120--5220 \\
    \hline
    Primary &  &  &    \\ 
     \Teff [K] & 6980$\pm$55 & 6940$\pm$70 & 7205$\pm90$    \\
    $\log g$ [cgs] & 3.873  & 3.873  & 3.873  \\
    $v \sin i$ [\kms] & 56.2  & 56.22  &  56.5$\pm$1.9  \\
    LDF    & 0.432$\pm$0.005  & 0.434$\pm0.007$ & 0.439$0.009$     \\
    \hline
    Secondary &       &    &      \\
    \Teff [K]   &  6875$\pm$45 & 6770$\pm$65 & 6990$\pm$80  \\
    $\log g$ [cgs] &  3.761  & 3.761  &  3.761   \\
    $v \sin i$ [\kms] & 67.2    & 67.2   & 67.2$\pm$1.5   \\
    LDF        & 0.568$\pm$0.004  & 0.566$\pm0.006$ &  0.561$\pm$0.009      \\
    \hline
\end{tabular} 
\end{table}

Overall, a slight difference in line depth between the two components is seen. The more massive component has deeper lines and slightly faster rotation. The Balmer lines are broadened by Stark broadening, and generally are not sensitive to the rotational broadening. Thus, we first optimised portions of the disentangled spectra free of the Balmer lines, primarily to discern the $v\sin i$ values. We then performed  optimal fitting of disentangled spectra centred on the H$\beta$ and H$\alpha$ lines, with fixed $v \sin i$. 

It is well-known that the hydrogen Balmer lines are excellent diagnostic tools for the determination of the \Teff s for stars with $\Teff < 8000$~K because the degeneracy with the surface gravity vanishes \citep{gray_2005}. Moreover, we can use $\log(g)$ determined for the components since this quantity is determined with high precision from the analysis of DLEBs. In the case of KIC~9851944, the $\log(g)$ values are determined with uncertainties of about 0.01 dex, for both components. Therefore, preference is given to the determination of the \Teff\ for the components in KIC~9851944 from line profile fitting of Balmer lines, with fixed surface gravity. This is advantageous over the calculation of the \Teff\ from metal lines because their strength depends on the metallicity of the stellar atmosphere.

Without any appreciable changes in the light ratio in the course of the orbital cycle, i.e.~no spectra were observed in eclipse, ambiguity exists in the reconstruction of the components' spectra and only separation of the spectra still in the common continuum of the binary system is possible. The components' spectra are correctly reconstructed but with scaling factors, i.e.~the shapes of the spectral lines are correct, but not their strength. In such a case, the light
ratio can be determined from fitting the separated spectra with synthetic (theoretical) spectra, in the course of determination of the atmospheric parameters.
The optimal fitting was performed in constrained mode, as defined in \citet{Tamajo_2011}, with the condition that the sum of the light dilution factors is equal to unity. The {\sc starfit} code \citep{Kolbas_2015} uses a genetic algorithm to search for the best fit within a grid of synthetic spectra pre-calculated using the {\sc uclsyn} code \citep{Smalley_2001}. The uncertainties were calculated using the MCMC approach described in \citet{Pavlovski_2018}. This task was straightforward due to the fixed surface gravities and $v\sin i$ values: only the light ratio and \Teff s were optimised.

The results of the optimal fitting for all three spectral segments are given in Table~\ref{tab:optfits}. We adopt the weighted average of the results from H$\alpha$ and H$\beta$ for \Teff, which is $6964 \pm 43$\,K and $6840 \pm 37$\,K for the primary and secondary, respectively; these values are used to obtain the light curve solution in Section \ref{sec analysis of light curve} and are presented formally in Section \ref{sec: physical properties}. \reff{Our \Teff\ values indicate that the KIC \Teff\ (6204\,K) for this object is underestimated, consistent with prior studies \citep[e.g.,][]{Molenda_Zakowicz, Lehmann_2011, Tkachenko_2013, Niemczura_2015} reporting similar underestimations for stars with $\Teff\sim 7000$\,K.} 

With the \Teff\ determined from the Balmer lines, the depths of the metal lines cannot be matched -- the metal lines for both components are slightly deeper than the synthetic spectra for given parameters. This could be explained with the metallicity effect, which affects metal lines more than broad H$\beta$ and H$\alpha$ lines for which Stark broadening dominates. The effects of metallicity explain the somewhat higher \Teff\ (by about 200~K) found from the metal lines. From a grid of calculated synthetic spectra for given atmospheric parameters derived from Balmer lines, we found that the metallicity for both components is [M/H] $\sim 0.15$~dex. \reff{Since the components of KIC 9851944 are at the cool end of where Am stars are found, this motivated us to disentangle the spectra between 3853\,--\,4067\,\AA\ and compare the Ca K lines to synthetic spectra; the findings did not suggest any Am peculiarity although the Ca K line for the primary appears slightly weakened. Due to the relatively high $v\sin i$ for both components and thus severe line blending, as well as the low S/N of the individual spectra, we did not attempt to determine individual elemental abundances.}

The light ratio from the optimal fitting of the wings of H$\beta$ and H$\gamma$ lines, spectral segments containing only metal lines, and the \ion{Mg}{i} triplet at around 5180~{\AA}, are $1.315 \pm 0.018$, $1.304 \pm 0.025$ and $1.278 \pm 0.033$ (Table~\ref{tab:optfits}). All three values for the light ratio are consistent within their 1$\sigma$ uncertainties, with the light ratio determined from the wings of H$\beta$ line being the most precise of the three.

\citet{Guo_2016} determined atmospheric parameters from tomographically reconstructed spectra of the components. Their spectra cover the wavelength range 3930--4610\,\AA\ at a medium spectral resolution of $R=6000$. The analysis by \citet{Guo_2016} was similar to ours, but with an important difference that they fitted complete separated spectra, whilst we concentrate on the wings of Balmer lines. Our results for the Balmer lines corroborate their findings to within 1$\sigma$. We also find good agreement for the $v \sin i$ values, which are also within 1$\sigma$ for both components. \citet{Guo_2016} obtained a mean light ratio $1.34 \pm 0.03$ from spectra centred at $4275$~{\AA} -- this is slightly larger than but still consistent with our own results.

\subsection{Direct fitting for the light ratio}
\label{sec:lr-det}

We were initially unable to constrain the light ratio of the system from the photometric analysis alone (see Section \ref{sec:ephem}), so estimated it using several independent methods. The first method was the \textsc{todcor} light ratio reported in Table \ref{tab:orbital_params}, the second was the optimal fitting of the disentangled spectra (ODS), and a third method is developed here.

We minimize the sum of the squared residuals between synthetic composite spectra and the observed spectra, where the synthetic composite spectra are built by adding Doppler-shifted synthetic spectra weighted by light fractions according to some value of $\ell_{\rm B}/\ell_{\rm A}$, and with atmospheric parameters from the ODS analysis. We search in a grid of $\ell_{\rm B}/\ell_{\rm A}$ values between 0.8 and 2 with ten uniformly spaced samples; these bounds were determined after an initial trial run with bounds of 0.5 to 3. The minimum of a polynomial fit to the sum of the square residuals against trial values for $\ell_{\rm B}/\ell_{\rm A}$ then yields a best estimate for its value. The only free parameters of the minimization for each trial $\ell_{\rm B}/\ell_{\rm A}$ were the coefficients of a sixth-order polynomial that was used to normalize the observed spectra against the synthetic spectrum. The applied Doppler shifts were fixed according to the corresponding RV values derived in Section~\ref{sec:RV-analysis}. The minimization was carried out using the Nelder-Mead method as implemented in the Scipy python package \textsc{minimize} \reff{\citep{2020SciPy-NMeth}}. 

\begin{figure}%
    \centering
    \includegraphics[width=\columnwidth]{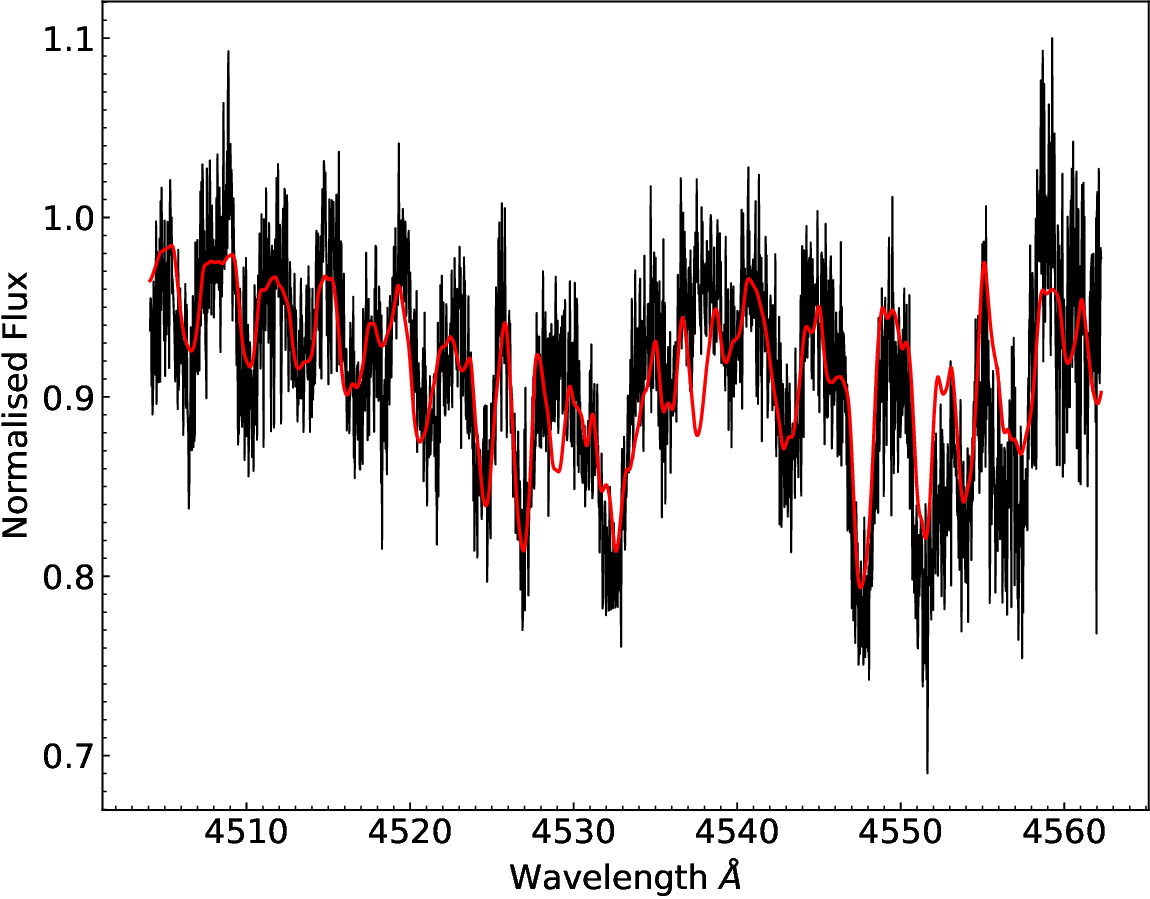}
    \caption{Spectral fit to order 66 using the grid search method with optimal normalization.}%
    \label{fig:LR_GS_fit}
\end{figure}%
\begin{figure}%
    \centering
    \includegraphics[width=\columnwidth]{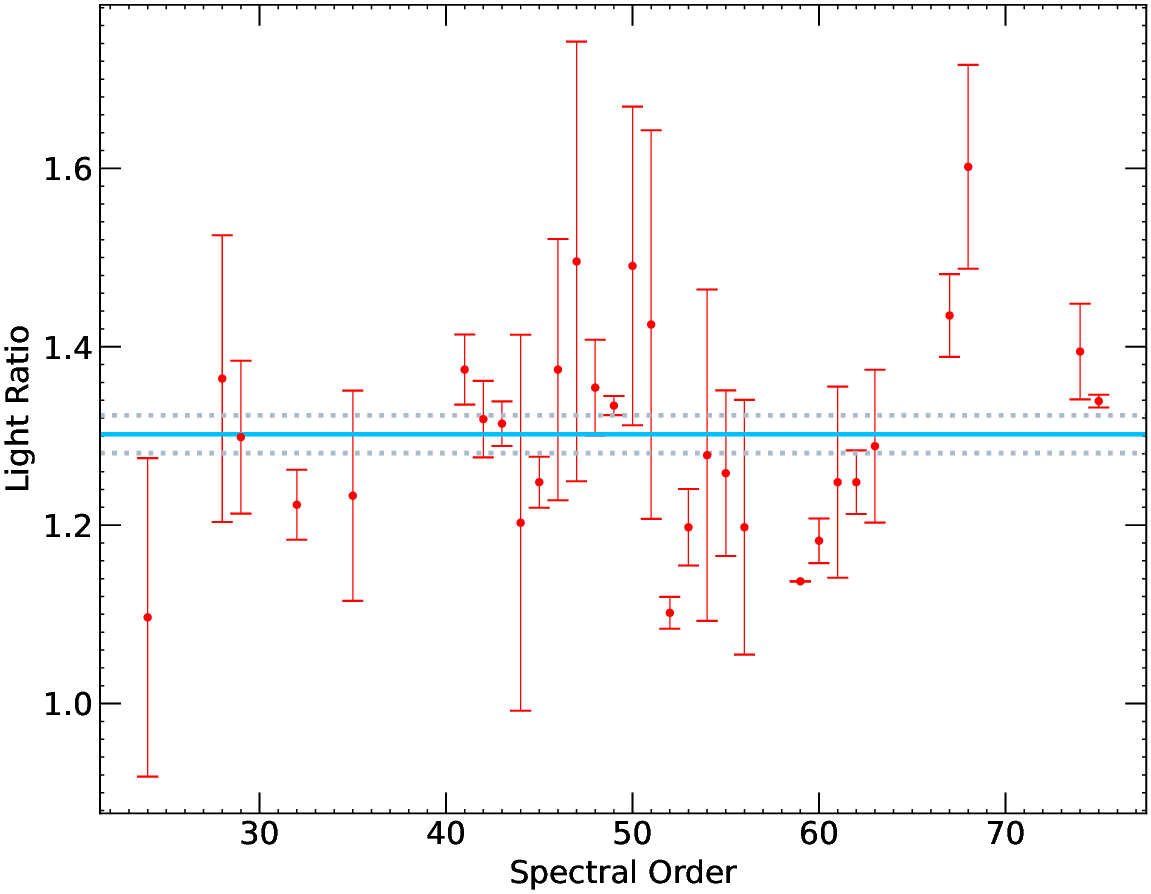}
    \caption{Resulting $\ell_{\rm B}/\ell_{\rm A}$ values from the grid search method averaged over the orders.}%
    \label{fig:LR_GS_results}
\end{figure}%

We repeated the process for all spectral orders showing sufficient well-resolved lines in regions unaffected by tellurics from Earth's atmosphere. To save computing time, only the two observations closest to positions of quadrature were used. This approach resulted in 30 spectral orders showing adequate fits with well defined minima in the sum of the square residuals. Fig.\ref{fig:LR_GS_fit} shows the result of this method applied to order 66, which demonstrates the effectiveness of optimizing the normalization of the observed spectra in the fitting routine for an order where line blending is significant. The resulting values for $\ell_{\rm B}/\ell_{\rm A}$ after applying this method to the selected orders are shown in Fig.\ref{fig:LR_GS_results}, where each result is the average of the result from each of the two observations used. 

Thus, after taking the average of the results in Fig.\ref{fig:LR_GS_results}, which corresponds to the blue line, we have three independent estimations for the light ratio of the system, i.e., \textsc{todcor}, ODS, and direct fitting for the light ratio. These estimations are given in Table \ref{tab:LRs}, where we take the average and its standard deviation from the ODS method. For the rest of this study, we adopt  $\ell_{\rm B}/\ell_{\rm A} = 1.300$ as the \reff{weighted} mean of those values, with an errorbar of $\pm$\,0.036 \reff{which is their standard deviation}. This value is used to ascertain the correct light curve solution is obtained in the next section. 

\begin{table}
\caption{\label{tab:LRs} Light ratios measured and adopted for KIC 9851944.}
   \centering
    \begin{tabular}{l r@{\,$\pm$\,}l}
    \hline 
    Source & \multicolumn{2}{c}{Light Ratio $\ell_{\rm B}/\ell_{\rm A}$} \\
    \hline 
    \textsc{todcor} & 1.222 & 0.136 \\
    ODS & 1.300 & 0.015 \\
    Direct Fit & 1.302 & 0.021 \\[3pt]
    Adopted & 1.300 & 0.036 \\
    \hline
\end{tabular} 
\end{table}


\section{Analysis of the light curve}\label{sec analysis of light curve}

The components of KIC 9841944 are close to each other and thus have a significant tidal deformation. We therefore sought to model the light curve using a code that is based on Roche geometry. We selected the Wilson-Devinney (WD) code \citep{WilsonDevinney71apj,Wilson79apj} for this, and used the 2004 version of the code driven using the {\sc jktwd} wrapper \citep{Me+11mn}. The user guide which accompanies the WD code \citep{WilsonVanhamme04} includes a description of all input and output quantities discussed below.

The WD code is computationally expensive and is not suited to the analysis of the full 500\,000 short-cadence datapoints in one step. We therefore used the orbital ephemeris determined in Section\,\ref{sec:ephem} to convert the datapoints to orbital phase, and then binned them into a much smaller number of points. We chose a bin size of 0.001 orbital phases during the eclipses and 0.005 outside the eclipses, resulting in a total of 456 phase-binned datapoints suitable for analysis with the WD code.

\begin{figure}
\includegraphics[width=\columnwidth]{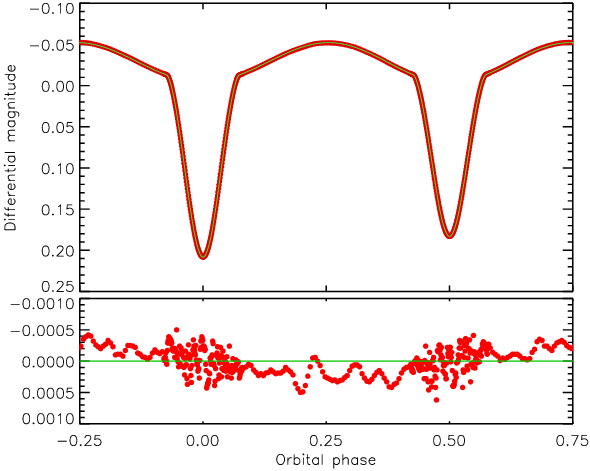} \\
\caption{\label{fig:wd} The best-fitting WD model (green line) to the \textit{Kepler}
phase-binned light curve of KIC 9851944 (red filled circles). The residuals of the fit
are plotted in the lower panel using a greatly enlarged y-axis to bring out the detail.}
\end{figure}

Through a process of trying a large number of different fits with a range of fixed and fitted parameters, we arrived at a good solution to the light curve. We adopt this as the default solution, plot it in Fig.\,\ref{fig:wd}, and give the fitted parameters in Table\,\ref{tab:wd}. It was obtained in Mode $=$ 0 with the following fitted parameters: the light contributions of the two stars; their potentials; their gravity darkening coefficients; the linear coefficients of the logarithmic LD law; and the orbital inclination. We fixed the mass ratio at the spectroscopic value, the orbital eccentricity to zero, the albedos to 1.1, the logarithmic coefficients of LD to values from \citet{Vanhamme93aj}, third light to zero, and the rotation rates to synchronous. We adopted the maximum numerical precision values of \texttt{N1} $=$ \texttt{N2} $= 60$ and \texttt{N1L} $=$ \texttt{N2L} $= 60$, the simple treatment of reflection, and the Cousins $R$-band as a proxy for the \textit{Kepler} passband. This solution gives a light ratio in excellent agreement with the spectroscopic values in Table\,\ref{tab:LRs}. The fractional radii in Table\,\ref{tab:wd} are volume-equivalent values calculated by the {\sc lc} component of the WD code.

The main confounding factors in fitting the light curve were the albedo and gravity darkening. Albedo values around 1.0 give a good fit to the data, but there is a wide local minimum of $\chi^2$ at albedos in the region of 0.0 that gives a worse fit but is frequently found by the steepest-descent minimization method implemented in the WD code. After extensive experimentation, we found that fixing the albedos to values slightly above 1.0 yielded the best fits and did not risk descending into the local minimum at lower abedo values. Fitting for gravity darkening turned out to be a crucial step in obtaining a good model, and the best-fitting values were somewhat varied but generally around 0.8 for the primary star and 1.1 for the secondary star. We interpret this as a systematic issue caused by residual pulsations in the light curve and account for it in the uncertainties.

For the determination of the uncertainties of the fitted parameters we ran a wide range of solutions with a large number of different possible approaches to modelling the light curve (see \citealt{Me20obs}). We calculated the errorbar for each parameter by adding in quadrature the contribution from every model choice, which in turn was taken to be the amount that parameter changed by versus the default solution. The following different approaches were explored.
\begin{enumerate}
\item We varied the amount of binning prior to the WD solution, finding that it had a negligible effect on the fitted parameters.
\item We changed the numerical precision to \texttt{N1} $=$ \texttt{N2} $=$ \texttt{N1L} $=$ \texttt{N2L} $= 50$. This modified the fractional radii by a maximum of 0.5\%.
\item We tried the detailed reflection effect and found almost identical results.
\item Attempts to fit for mass ratio returned a value very close to the spectroscopic one and almost no change in the other parameters.
\item Fitting for the rotation rates of the stars yielded values close to and consistent with synchronous rotation and little change in the other parameters.
\item Fitting for albedo gave a poorer fit but again very little change in the other parameters.
\item Allowing for third light gave very similar parameters and a very small amount of third light consistent with zero.
\item Fitting for one of the temperatures of the stars instead of the two light contributions directly (\reff{namely} Mode $=$ 2) changed the fractional radii by 0.2\%.
\item Fixing the limb darkening coefficients to theoretical values changed the fractional radii by 0.5\%.
\item Using the square-root instead of logarithmic limb darkening law had a negligible effect.
\end{enumerate}
The best-fitting parameters were highly robust against all these experiments. The only significantly discrepant fit (neglecting our original exploratory ones) was when we used the Cousins $I$-band instead of the $R$-band. However, we are able to reject this fit as it is not consistent with the spectroscopic light ratio. The calculated uncertainties are given in Table\,\ref{tab:wd}.

Fig.\,\ref{fig:wd} shows that the best fit is extremely good, with an r.m.s.\ of 0.20\,mmag, but that there are systematics remaining in the residuals. We attribute these to the WD numerical integration for points during the eclipses, and residual pulsations for points outside eclipse. Our approach of phase-binning the data gives a light curve practially without Poisson noise, so makes any imperfections in the fit easily noticable. We note that the systematics in the residuals of the fits we found to the data are too small to show up in plots of unbinned data, so may well be present in previous work on this object.

\begin{table} \label{tab:wd} \centering
\caption{Summary of the parameters for the {\sc wd2004} solutions of the light curves of the systems.
Detailed descriptions of the control parameters can be found in the WD code user guide
\citep{WilsonVanhamme04}. A and B refer to the primary and secondary stars, respectively. Uncertainties are
only quoted when they have been robustly assessed by comparison of a full set of alternative solutions.}
\begin{tabular}{llcccc} \hline
Parameter                            & {\sc wd2004} name    & Value               \\
\hline
{\it Control and fixed parameters:} \\
{\sc wd2004} operation mode          & {\sc mode}           & 0                   \\
Treatment of reflection              & {\sc mref}           & 1                   \\
Number of reflections                & {\sc nref}           & 1                   \\
LD law                               & {\sc ld}             & 2 (logarithmic)     \\
Numerical grid size (normal)         & {\sc n1, n2}         & 60, 60              \\
Numerical grid size (coarse)         & {\sc n1l, n2l}       & 60, 60              \\[3pt]
{\it Fixed parameters:} \\
Mass ratio                           & {\sc rm}             & 1.06                \\
Phase shift                          & {\sc pshift}         & 0.0                 \\
Orbital eccentricity                 & {\sc e}              & 0.0                 \\
$T_{\rm eff}$ star\,A (K)            & {\sc tavh}           & 6964                \\
$T_{\rm eff}$ star\,B (K)            & {\sc tavc}           & 6840                \\
Bolometric albedos                   & {\sc alb1, alb2}     & 1.1, 1.1            \\
Rotation rates                       & {\sc f1, f2}         & 1.0, 1.0            \\
Logarithmic LD coefficients          & {\sc y1a, y2a}       & 0.294, 0.293        \\ [3pt]
{\it Fitted parameters:} \\
Star\,A potential                    & {\sc phsv}           & $5.365  \pm 0.044 $ \\
Star\,B potential                    & {\sc phsc}           & $4.867  \pm 0.093 $ \\
Orbital inclination (\degr)          & {\sc xincl}          & $73.912 \pm 0.044 $ \\
Star\,A gravity darkening            & {\sc gr1}            & $0.75   \pm 0.40  $ \\
Star\,B gravity darkening            & {\sc gr2}            & $1.14   \pm 0.40  $ \\
Star\,A light contribution           & {\sc hlum}           & $5.58   \pm 0.12  $ \\
Star\,B light contribution           & {\sc clum}           & $7.34   \pm 0.12  $ \\
Star\,A linear LD coefficient        & {\sc x1a}            & $0.658  \pm 0.019 $ \\
Star\,B linear LD coefficient        & {\sc x2a}            & $0.662  \pm 0.013 $ \\
Fractional radius of star\,A         &                      & $0.2344 \pm 0.0024$ \\
Fractional radius of star\,B         &                      & $0.2759 \pm 0.0040$ \\[3pt]
\hline
\end{tabular} \end{table}


\section{Physical properties}\label{sec: physical properties}

\begin{table} \label{tab:absdim} \centering
\caption{Physical properties measured for the four systems analysed in this work. The units labelled with a
`N' are given in terms of the nominal solar quantities defined in IAU 2015 Resolution B3 \citep{Prsa+16aj}.}
\begin{tabular}{l r@{\,$\pm$\,}l r@{\,$\pm$\,}l} \hline
Parameter                             & \multicolumn{2}{c}{Star A} & \multicolumn{2}{c}{Star B} \\
\hline
Mass ratio                            & \multicolumn{4}{c}{$1.0667 \pm 0.0038$} \\
Semimajor axis (\Rsunnom)             & \multicolumn{4}{c}{$10.805 \pm 0.019$}  \\
Mass (\Msunnom)                       &    1.749 & 0.010   &    1.866 & 0.010   \\
Radius (\Rsunnom)                     &    2.533 & 0.026   &    2.981 & 0.044   \\
Surface gravity ($\log$[cgs])         &    3.874 & 0.009   &    3.760 & 0.013   \\
Synchronous velocity (\kms)           &    59.2  & 0.6     &    69.7  & 1.0     \\
\Teff\ (K)                            &     6964 & 43      &     6840 & 37      \\
Luminosity $\log(L/\Lsunnom)$         &    1.133 & 0.014   &    1.244 & 0.016   \\
Absolute bolometric magnitude         &    1.907 & 0.035   &    1.631 & 0.039   \\
$E(B-V)$ (mag)                        & \multicolumn{4}{c}{$0.14 \pm 0.02$}     \\
Distance (pc)                         & \multicolumn{4}{c}{$935 \pm 13$}        \\
\hline
\end{tabular} \end{table}

We determined the physical properties of KIC 9851944 from the spectroscopic and photometric results obtained above. For this we used the {\sc jktabsdim} code \citep{Me++05aa}, modified to use the IAU system of nominal solar values \citep{Prsa+16aj} plus the NIST 2018 values for the Newtonian gravitational constant and the Stefan-Boltzmann constant. Errorbars were propagated via a perturbation analysis. The results are given in Table.\,\ref{tab:absdim}.

We determined the distance to the system using optical $BV$ magnitudes from APASS \citep{Henden+12javso}, near-IR $JHK_s$ magnitudes from 2MASS \citep{Cutri+03book} converted to the Johnson system using the transformations from \citet{Carpenter01aj}, and surface brightness relations from \citet{Kervella+04aa}. The interstellar reddening was determined by requiring the optical and near-IR distances to match. We found a final distance of $935 \pm 12$\,pc, which is significantly shorter than the distance of $999 \pm 12$\,pc from the \textit{Gaia} DR3 parallax \citep{Gaia16aa,Gaia21aa}. We have no explanation for this at present, \reff{but note that the \textit{Gaia} DR2 and DR3 parallaxes of this object differ by nearly 2$\sigma$ so might be affected by its binarity.}


\section{Asteroseismic analysis}
\label{sec:asteroseismology}

\subsection{Frequency analysis}
\label{subsec:freq-analysis}
Following the binary modelling, we continue with the asteroseismic analysis of the target. Because the observed pulsations have much smaller amplitudes than the binary signal, the quality of the TESS and WASP data are insufficient for the asteroseismic analysis, and we limit ourselves to using the residual {\em Kepler} light curve. This is the merged light curve of all available {\em Kepler} short-cadence data after subtracting the best-fitting binary model, hereafter referred to as the pulsation light curve. To minimize the impact of outliers and instrumental effects on the asteroseismic analysis of small-amplitude pulsations, we apply additional processing to the data. Firstly, we remove those parts where coronal mass ejections (CMEs) or thermal and pointing changes of the spacecraft, such as at the start of a quarter or after a safe-mode event, have a visible impact on the quality of the light curve. Secondly, we apply preliminary iterative pre-whitening \citep[as described by, e.g.,][]{VanReeth2023} to build a tentative mathematical model of the 20 most dominant pulsations using a sum of sine waves
\begin{equation}
L(t) = \sum_{i=1}^{N=20}a_i\sin\left(2\pi\left[ f_i\left(t - t_0\right) + \phi_i\right]\right),\label{eq:sine_puls_model}
\end{equation}
where $a_i$, $f_i$ and $\phi_i$ are the amplitude, frequency, and phase of the $i^{th}$ sine wave, respectively, and $t_0$ is the average time stamp of all data points in the pulsation light curve. We then identify individual outliers in the residuals using 5$\sigma$~clipping, and remove these data points from the pulsation light curve as well.

Next, we use iterative pre-whitening to measure the pulsation frequencies from the resulting cleaned pulsation light curve. Hereby we iteratively fit additional sine waves to the time series, with frequencies that correspond to the dominant amplitudes in the Lomb-Scargle periodogram \citep{Scargle1982} of the residual pulsation light curve. However, in the particular case of KIC\,9851944, there is a significant amount of red noise present in the data, most likely caused by residual instrumental effects, even after cleaning the pulsation light curve. To ensure that we detect all significant pulsation frequencies (with a signal-to-noise ratio $S/N \geq 5.6$; see e.g.\ \citealt{Baran2015}), we customise our approach. Firstly, we split the frequency range between $0\,\rm d^{-1}$ and the Nyquist frequency ($734.21\,\rm d^{-1}$) in overlapping parts and apply the iterative pre-whitening to these individually: from $0\,\rm d^{-1}$ to $2\,\rm d^{-1}$, from $1\,\rm d^{-1}$ to $6\,\rm d^{-1}$, from $4\,\rm d^{-1}$ to $11\,\rm d^{-1}$, from $9\,\rm d^{-1}$ to $21\,\rm d^{-1}$, from $19\,\rm d^{-1}$ to $51\,\rm d^{-1}$, from $49\,\rm d^{-1}$ to $201\,\rm d^{-1}$, and from $199\,\rm d^{-1}$ to the Nyquist frequency. In this stage, we measure all frequencies with $S/N \geq 4.0$. Secondly, we merge the different frequency lists, keeping only those frequencies that are dominant in a $2.5\,f_{\rm res}$-window, where $f_{\rm res}$ is the frequency resolution of the light curve \citep{Loumos1978} with a value of $0.00208\,\rm d^{-1}$, and that have $S/N \geq 5.6$. Thirdly, we non-linearly optimize this filtered frequency list using the least-squares minimization with the trust region reflective method from the \texttt{lmfit} python package \citep{Newville2019}. After the optimization, we redetermined the $S/N$ that are associated with the different frequencies, again only keeping those with $S/N \geq 5.6$. This leaves us with a final list of 133 measured frequencies in both the p- and g-mode regimes, considerably more than the 89 frequencies reported by \citet{Guo_2016}. Notably, we formally detect one frequency at $391.5095\pm0.0002\,\rm d^{-1}$ with an amplitude of $6.3\pm1.2\,\mu \rm mag$. However, a detailed analysis of this sine wave reveals that it originates from the noise properties of the short-cadence light curve. The signal has a maximal amplitude during the first parts of quarters 13 and 14, but is not detectable in most other parts of the light curve. Hence, we discard this frequency and limit ourselves to the remaining frequencies, with values below $25\,\rm d^{-1}$. These are illustrated in Fig.\,\ref{fig:kepler09851944_fourier} and listed in Table \ref{tab:extracted_freq} in Appendix\,\ref{sec:extracted_freq}.

\begin{figure*}
 \includegraphics[width=\textwidth]{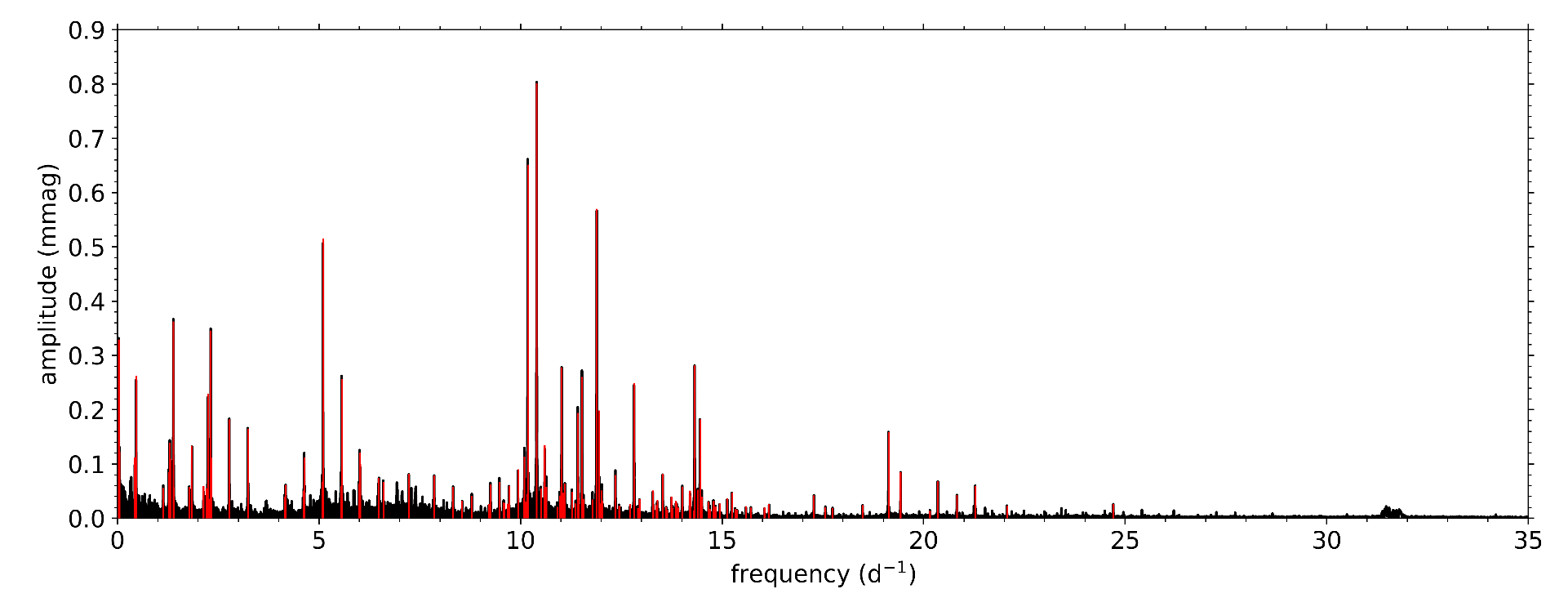}
 \caption{\label{fig:kepler09851944_fourier} Lomb-Scargle periodogram of the short-cadence {\em Kepler} light curve of KIC\,9851944 (black) with the iteratively prewhitened frequencies (full red lines).}
\end{figure*}

\subsection{Tidal perturbation analysis}
\label{subsec:tidal-perturbation}

\begin{figure}
\includegraphics[width=88mm]{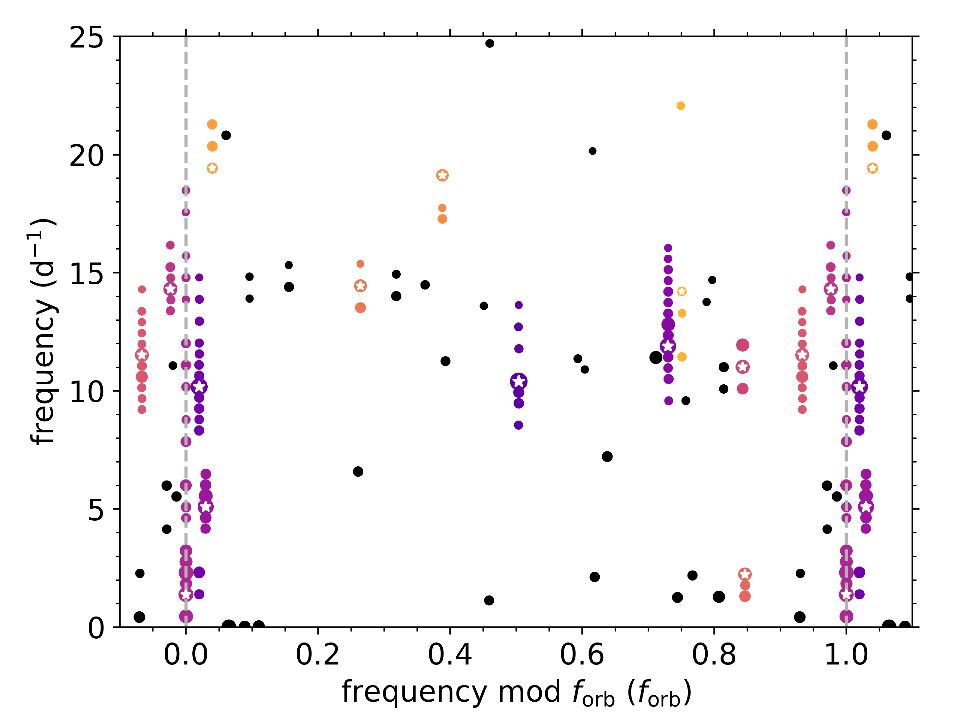}
\caption{\label{fig:kic09851944_echelle} \'Echelle diagram of the prewhitened frequencies of KIC\,9851944, folded with the orbital frequency $f_{\rm orb}$. Identified multiplets are indicated in different colours ranging from purple to dark yellow, with a white star marking the dominant frequency of each multiplet. Single frequencies are shown in black. The marker sizes indicate the associated amplitudes.}
\end{figure}

As already demonstrated by \citet{Guo_2016} and illustrated in Fig.\,\ref{fig:kic09851944_echelle}, many of the detected frequencies form orbital-frequency spaced multiplets, also called ``tidally split multiplets''. In our work, we identify each multiplet by looping over all measured frequencies $f_i$ in order of decreasing amplitude, and consider frequencies $f_j$ to be part of a multiplet around it if 
\begin{equation}
  |f_i - f_j - nf_{\rm orb}| < \Delta f,
\end{equation}
where $n$ is an integer chosen to minimise the left-hand side of the Equation, and $\Delta f$ is a chosen frequency tolerance of $0.002\,\rm d^{-1}$, that is $\approx f_{\rm res}$. In each iteration, the considered frequencies $f_j$ have amplitudes $a_j$ that are smaller than the amplitude $a_i$, associated with $f_i$, and are not yet associated with a different multiplet.

We identify 13 multiplets that consist of three or more components, as listed in Table \ref{tab:extracted_freq} in Appendix\,\ref{sec:extracted_freq}. One of these, with the dominant frequency at $1.386342(4)\,\rm d^{-1}$, consists of 19 frequencies that match integer multiples of the binary orbital frequency $f_{\rm orb}$ within $\Delta f$, and is discussed in detail below in Sect.\,\ref{subsec:tidal-excitation}. In their work, \citet{Guo_2016} explained the other multiplets as consequences of {\em (i)} rotational frequency splitting, and {\em (ii)} partial occultations of the pulsation mode geometries during the eclipses \citep[e.g.,][]{Reed2001,Reed2005,Gamarova2003,Rodriguez2004,Gamarova2005}. However, it is now known that pulsation modes are often tidally perturbed \citep[e.g.,][]{Samadi2018,Steindl2021,VanReeth2023} or tilted \citep[e.g.,][]{Fuller2020,Handler2020,Kurtz2020,VanReeth_2022} in close binaries.

Hence, to determine the true origin of the detected multiplets, we evaluate their dependence on binary orbital phase in detail. Each multiplet with three or more components is studied individually by removing all measured variability that is not associated with that multiplet from the pulsation light curve, that is, we subtract all sine waves with frequencies that are not in the multiplet from the pulsation light curve. We then fold the residual light curve with the orbital period, split the data in 50 bins, and fit the dominant sine wave of the multiplet to the data, optimising the amplitude and phase for each orbital-phase bin. The reason for this approach is twofold: {\em (i)} the frequency spacings within the detected multiplets are not always exact multiples of $f_{\rm orb}$, and {\em (ii)} there can be frequencies missing within the detected multiplets. As a result, the analytical reconstruction from \citet{Jayaraman2022} is not well suited for this star.

The results are illustrated in Fig.\,\ref{fig:kic09851944_multiplet-1039971} for the dominant pulsation, and shown in Appendix\,\ref{appendix:tidal-perturbations} in Figs.\,\ref{fig:kic09851944_multiplet-223972} to \ref{fig:kic09851944_multiplet-19427792} for all pulsations. In each figure, the relevant frequency multiplet is plotted in the left-hand panel, with a white star marking the dominant frequency. On the right-hand side, the middle and bottom panels show the orbit-phase dependence of the pulsation amplitude and phase, respectively. For reference, the top right panel shows the orbital-phase-folded light curve of the binary. From these figures we can draw several conclusions. Firstly, because most pulsation amplitudes and phases vary significantly at all orbital phases, and the scales of the observed pulsation phase modulations are of the order of 0.5 to 1.5\,rad, we can conclude that the observed pulsations are tidally perturbed. While tidally tilted pulsations are expected to have pulsation phase modulations of 0\,rad, $\pi$\,rad or $2\pi$\,rad, which can be smeared out \citep{Fuller2020}, tidal perturbations can result in much smaller pulsation phase modulations \citep[e.g.,][]{VanReeth2023}. Moreover, the observed tidally split multiplets are not perfectly equidistant. This is in contradiction with current theoretical predictions made for tidally perturbed \citep[e.g.,][]{Smeyers_2005} and tidally tilted pulsations \citep{Fuller2020}, indicating that aspects that are currently not included in these theoretical frameworks, such as the Coriolis force, also play a role. Secondly, based on the different morphology of the curves for the different pulsations, we can conclude that the pulsations have different mode geometries \citep{VanReeth2023}. For example, while most observed pulsations are modulated twice per orbit, the amplitude of the pulsation with frequency $f = 5.097165(3)\,\rm d^{-1}$ only reaches one maximum per orbital cycle. Moreover, for some pulsations the pulsation phase decreases as a function of the orbital phase when the observed amplitudes are maximal, while for others the pulsation phase increases. There is no detectable correlation between these effects and the pulsation frequency, but the tidal modulations are most easily observed for pulsations with higher (average) amplitudes. This suggests that most if not all pulsations in this target are tidally perturbed, but that the S/N of the lower-amplitude pulsations is too low for a detection. 

Finally, we can confirm that some pulsations belong to the primary, while others could belong to either the primary or the secondary component, in agreement with the inferences made by \citet{Guo_2016} based on asteroseismic models. As seen in Fig.\,\ref{fig:kic09851944_multiplet-1039971}, the observable amplitude of the p~mode with frequency $10.39971\,\rm d^{-1}$ drops during the primary eclipse, indicating that it belongs to the primary. The higher observed amplitudes just before and after the primary eclipse signifies that the pulsation has a higher amplitude on the side of the primary that is facing the secondary component, similar to what has been observed for $g$-mode pulsations in V456\,Cyg \citep{VanReeth_2022}, and KIC\,3228863 and KIC\,12785282 \citep{VanReeth2023}. By contrast, the observable amplitudes of the p~modes with frequencies $11.52234\,\rm d^{-1}$ and $11.89048\,\rm d^{-1}$ peak during the primary eclipse, as shown in Figs.\,\ref{fig:kic09851944_multiplet-1152234} and \ref{fig:kic09851944_multiplet-1189048}. This can either indicate that these pulsations belong to the secondary component, or it can be caused by reduced geometric mode cancellation during the eclipse, depending on the geometry of these pulsation modes. The origin of the observed pulsations can be investigated further using detailed asteroseismic modelling, which would allow us to calculate the probability that specific pulsations belong to one or the other component. However, such a modelling study lies outside the scope of the current work.

\begin{figure}
\includegraphics[width=88mm]{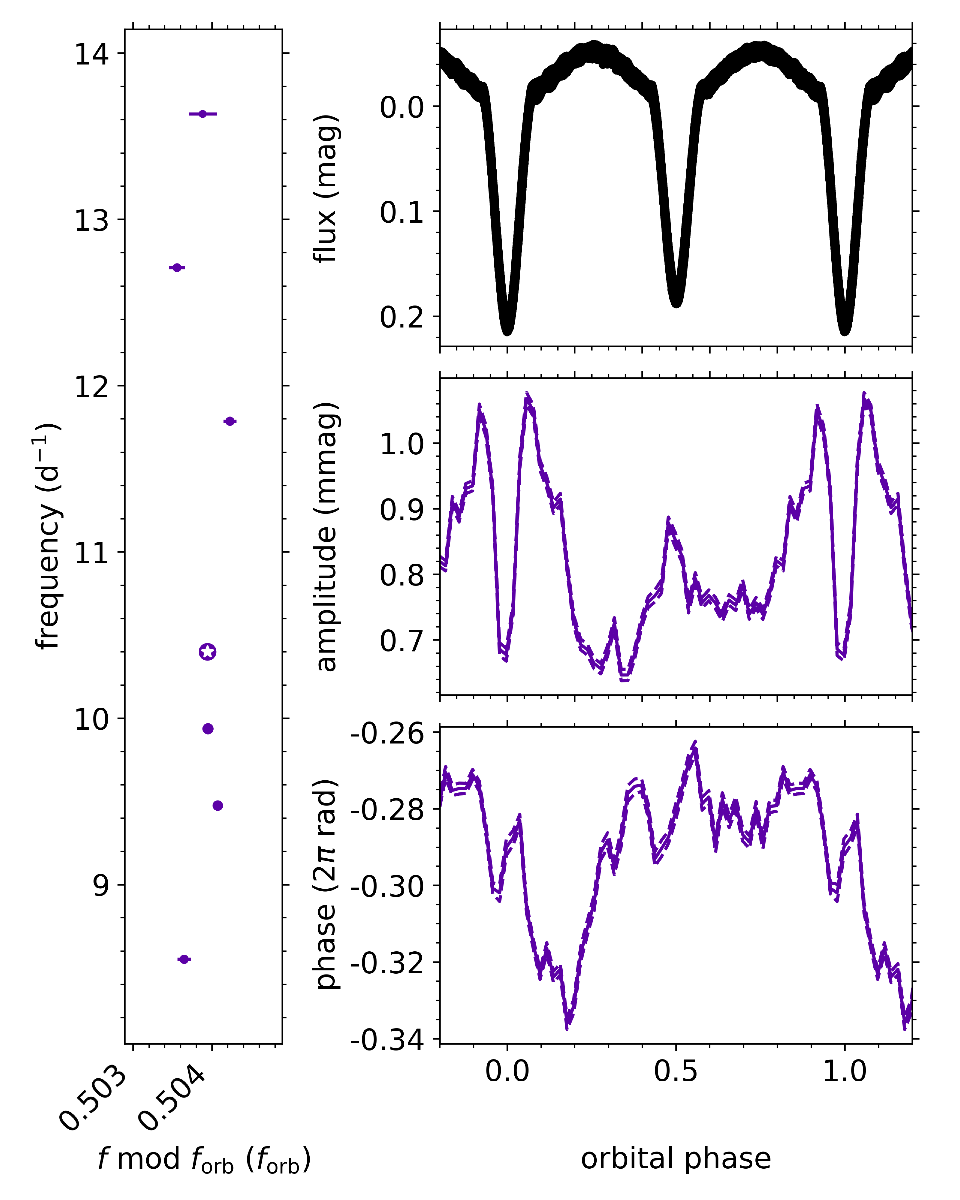}
\caption{\label{fig:kic09851944_multiplet-1039971} Tidally perturbed pulsation with frequency $f = 10.399706(2)\,\rm d^{-1}$. {\em Left:} Associated frequency multiplet, as shown in Fig.\,\ref{fig:kic09851944_echelle}. The white star marks the dominant frequency of the multiplet. {\em Top right:} Orbital-phase folded light curve. {\em Middle right:} Orbital-phase dependent modulations of the pulsation amplitude, calculated in 50 data bins, with the 1$\sigma$ uncertainty range indicated by the dashed lines. {\em Bottom right:} Orbital-phase dependent modulations of the pulsation phase, calculated in 50 data bins, with the 1$\sigma$ uncertainty range indicated by the dashed lines.}
\end{figure}

\subsection{Orbital harmonic frequencies}
\label{subsec:tidal-excitation}
\begin{figure}
\includegraphics[width=88mm]{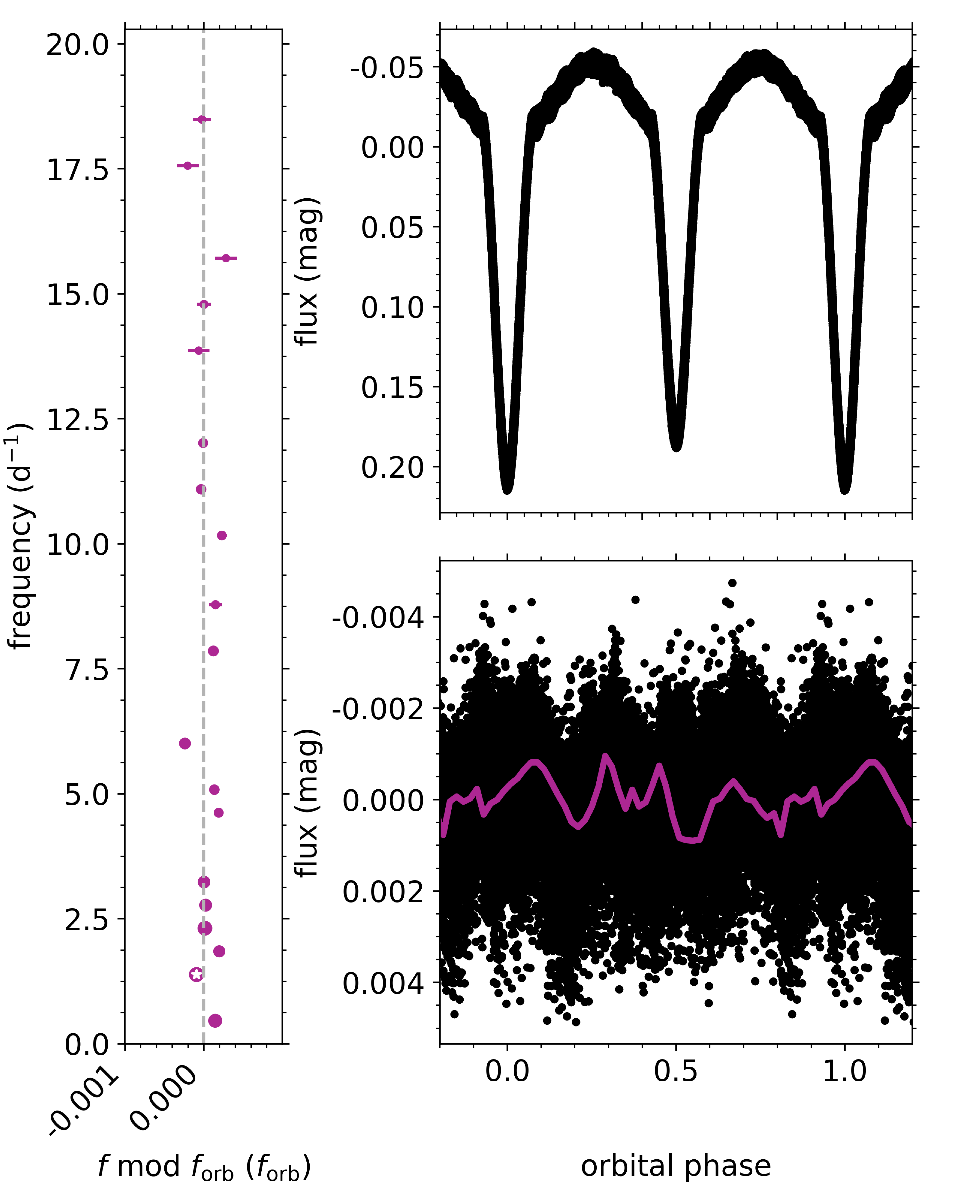}
\caption{\label{fig:kic09851944_tidally-excited} Tidally excited oscillations in KIC\,9851944. {\em Left:} Associated frequency multiplet, as shown in Fig.\,\ref{fig:kic09851944_echelle}. The white star marks the dominant frequency of the multiplet. {\em Top right:} Orbital-phase folded light curve. {\em Bottom right:} Orbital-phase folded residuals of the light curve after subtraction of the binary model. Individual data points are shown in black. The overplotted purple line shows the average variability as a function of orbital phase, evaluated in 50 equal bins.}
\end{figure}

In addition to the tidal perturbation of self-excited pulsations, and in agreement with \citet{Guo_2016}, we also detect an orbital harmonic frequency comb, illustrated in Fig.\,\ref{fig:kic09851944_tidally-excited}, which can indicate tidally excited oscillations. However, we do not recover the same orbital harmonics as \citet{Guo_2016}. They reported the detection of $8f_{\rm orb}$, $22f_{\rm orb}$, $46f_{\rm orb}$, and $50f_{\rm orb}$. While our detected orbital harmonic frequency comb consists of 19 frequencies, we only recover $22f_{\rm orb}$. This discrepancy is likely caused by differences between the reduced short- and long-cadence {\em Kepler} light curves, and between the photometric binary models. Thus, while we confirm the presence of orbital-phase dependent variability in the pulsation light curve, as seen in the bottom right panel of Fig.\,\ref{fig:kic09851944_tidally-excited}, its exact observed characteristics have to be treated with caution.

As pointed out by \citet{Guo_2016}, the detection of an orbital harmonic frequency comb is somewhat unexpected for synchronised binaries with circular orbits, though it has also been detected for other such systems \citep{daSilva2014}. Because the orbital eccentricity of the binary is zero, the equilibrium tides that are responsible for deforming the star and perturbing the pulsations are considerably larger than the dynamical tides that excite oscillations. Hence, this can indicate that this system has a slightly eccentric orbit or that one or both of the components is asynchronously rotating, within the uncertainty margins of our measurements.

\subsection{Gravity-mode period-spacing pattern}
\label{subsec:period-spacings}
\begin{figure}
\includegraphics[width=88mm]{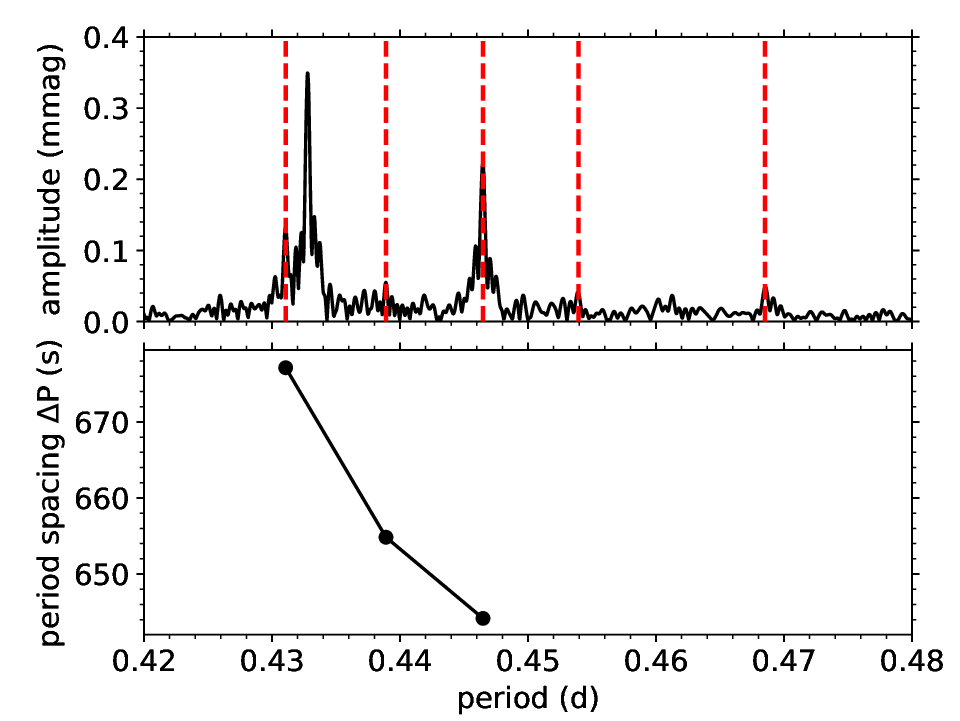}
\caption{\label{fig:kic09851944_period-spacings} Detected period-spacing pattern of g~modes with $(k,m) = (0,2)$ that belong to the primary component of KIC\,9851944. {\em Top:} part of the Lomb-Scargle periodogram of the pulsation light curve (black) with the pulsation periods of the modes that form the pattern (red dashed lines). {\em Bottom:} the period~spacing between consecutive modes in the detected pattern, as a function of the pulsation period. Because there is an undetected mode between the fourth and fifth detected pulsation modes, the fourth period spacing in the pattern is not shown.} The error margins are smaller than the symbol sizes.
\end{figure}

Finally, we analyse the observed g-mode pulsations in detail. \citet{Li_2020b} reported the detection of a short period-spacing pattern of prograde sectoral quadrupole modes, that is with $(k,m) = (0,2)$, and assigned them to the primary component of the system, based on the stellar masses determined by \citet{Guo_2016}. In this work, we confirm the pattern detection by \citet{Li_2020b} using the methodology from \citet{VanReeth2015a}, as shown in Fig.\,\ref{fig:kic09851944_period-spacings}. Only the g~mode with frequency $\sim0.461\,\rm d^{-1}$ was not recovered because of the higher signal-to-noise ratio cutoff that we used, that is, $S/N \geq 5.6$ instead of 4.0. Moreover, the dominant g~mode in the pattern, with a frequency $f$ of $2.239718(6)\,d^{-1}$, was found to be tidally perturbed and exhibit spatial filtering during the primary eclipse, as shown in Fig.\,\ref{fig:kic09851944_multiplet-223972} and with the corresponding multiplet listed in Table\,\ref{tab:extracted_freq}. These observations confirm that the g~modes belong to the primary component of the system: the observed tidal perturbations exhibit a dip in the amplitude and a saw-tooth-like ``glitch'' in the pulsation phase during the primary eclipse, which can only be explained if the pulsation belongs to the primary \citep{VanReeth_2022,VanReeth2023}.

To investigate the potentially asynchronous rotation of the pulsating star, we modelled our detected period-spacing pattern by fitting an asymptotic period-spacing series, following the method described by \citet{VanReeth_2016}. {We confirmed the pulsation mode identification found by \citet{Li_2020b}, $(k,m) = (0,2)$,} and found that the star has a buoyancy radius $\Pi_0$ of $4370^{+690}_{-660}$\,s and a near-core rotation frequency $f_{\rm rot}$ of $0.49^{+0.05}_{-0.06}\,\rm d^{-1}$. These values agree within 1$\sigma$ with the results from \citet{Li_2020b}, who found $\Pi_0$ and $f_{\rm rot}$ values of $3500\pm 500$\,s and $0.41\pm0.05\,\rm d^{-1}$, respectively. Moreover, both sets of values are consistent with synchronous rotation. However, because the detected pattern is so short, the uncertainties on these $f_{\rm rot}$ values are relatively large. Hence, there is still a possibility that the primary is asynchronously rotating, but insufficiently strongly to be detected with the available data.

\section{Discussion}\label{sec:discussion}

We compared the \reff{observed properties} of the components of KIC 9851944 to isochrones calculated by MIST (Mesa Isochrones and Stellar Tracks) using the \textsc{mesa} code \citep{Paxton_2011, Paxton_2013, Paxton_2015, Dotter_2016, Choi_2016, Paxton_2019}. We searched for the best MIST isochrone \reff{using two methods. For method 1, we interpolated radius, \Teff\ and luminosity as functions of mass. We then estimated these parameters using the interpolants at our measured masses. The objective function to minimize is then the sum of the quadrature distances between the interpolated and observed locations of the components with age and metallicity as free parameters. For method 2, we included mass in the calculation of the objective function, i.e., masses are not constrained to the observed values. For the minimization, we used SciPy's implementation of the stochastic differential evolution algorithm \citep{Storn_Price_1997,2020SciPy-NMeth}. Our grid of isochrones spanned from $-$4.0 to 0.5\,dex in [Fe/H] and 0.5 to 10.3\,dex in $\log_{10}(\rm age)$. The grid spacing in [Fe/H] was 0.05\,dex between $-$0.5\ and 0.5\,dex and 0.25\,dex outside that range up to $\pm 2$\,dex, beyond which the grid spacing was 0.5\,dex. A grid spacing of $0.05$\,dex in $\log_{10}(\rm age)$ was used through the full grid.}

\reff{After the best matching coeval isochrone was found, we removed the corresponding metallicity value from our isochrone grid and repeated the procedure to explore the effect of varying [Fe/H] on the predicted age of the system. We did this twice, leading to first, second, and third best matching coeval isochrones, each corresponding to a different metallicity. For method 1, the metallicity values of the first, second and third best matching isochrones were 0.05, 0.1 and 0.0\,dex, respectively. For method 2, these values correspond to 0.05, 0.0 and\,0.1 dex, i.e., the same set, except the second and third best matches are swapped. In all cases, the predicted age of the system is $1.259 \pm 0.073$\,Gyr. The error on this age estimate is taken to be half the average grid spacing either side of the best fitting isochrone. We note that this age estimation is in excellent agreement with the estimation by \citet{Guo_2016} of 1.25\,Gyr.}

\begin{figure*}
    \centering
    \includegraphics[width = \textwidth]{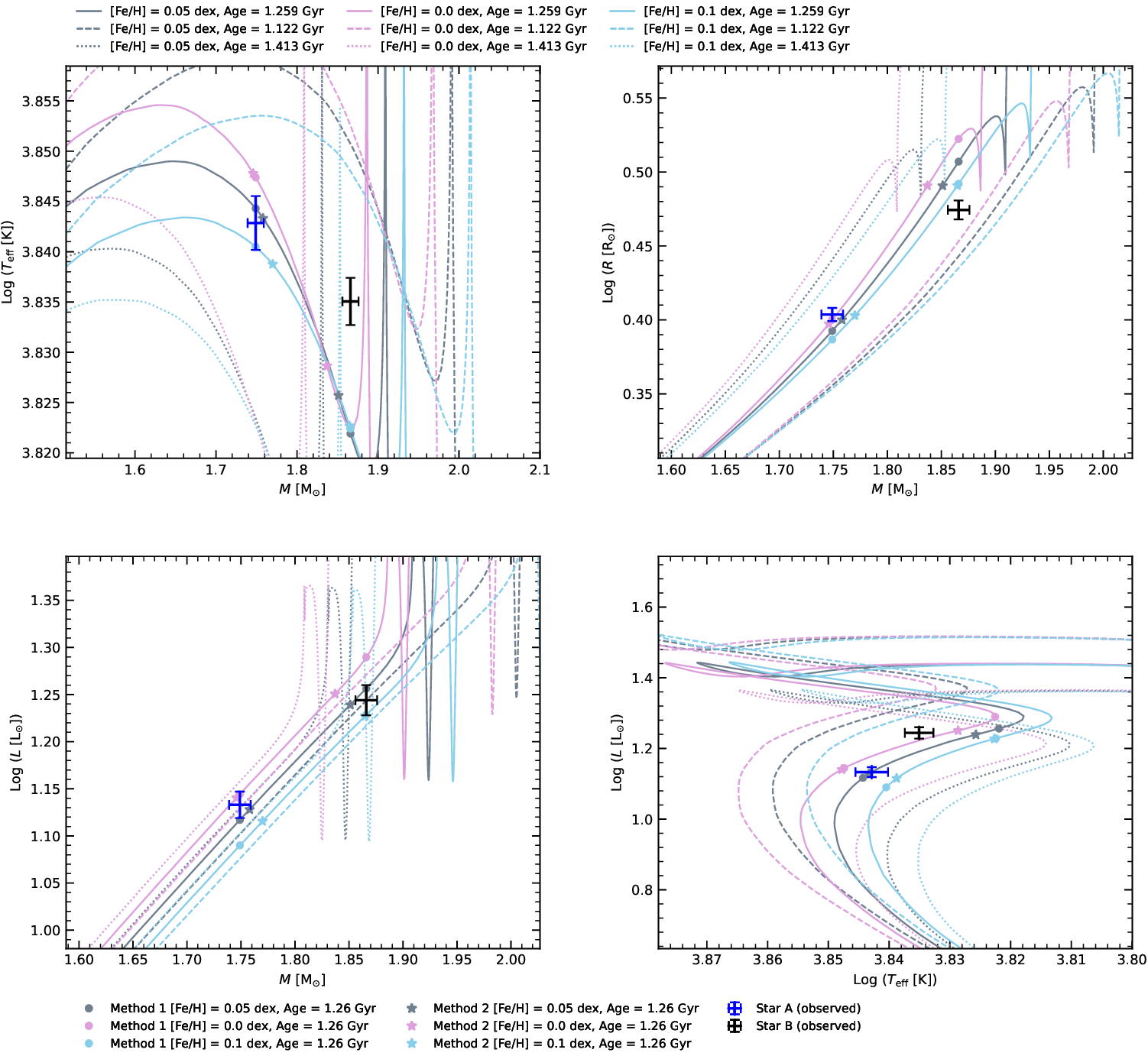}
    \caption{\reff{The three best fitting isochrones with the observed locations of the objects over-plotted in the $M-\log(\Teff)$ (top-left), $M-\log(R)$ (top-right), $M-\log(L)$ (bottom-left) and HR (bottom-right) planes. Also plotted are the best fitting isochrone neighbours, i.e., $\pm$ the grid spacing in age. The locations of the isochrones within those planes corresponding to the model with the smallest quadrature distance to the measured quantities that were included in the objective function are indicated for those from both model comparison methods.}}
    \label{fig: isochrone figure}
\end{figure*}
\reff{The three best-fitting isochrones are shown in Fig.\,\ref{fig: isochrone figure} with the observed locations of the objects over-plotted in the $M-\log(\Teff)$, $M-\log(R)$, $M-\log(L)$ and HR diagrams; also shown are their isochrone neighbours, i.e., $\pm$ the grid spacing in age. For both model comparison methods, the isochrone with [Fe/H] = 0.05\,dex yielded the model with the smallest quadrature distance to the observed quantities that were included in the objective functions so we adopt the corresponding model predictions from now on. Those values and their differences relative to the observed values are given in Table \ref{tab: model_params} for both model comparison methods. The locations of these values in the planes of Fig.\ref{fig: isochrone figure} are also shown; note the relative scales of the axes, particularly for \Teff\ where the scale of the axis is much smaller than the scales of the other axes. }

\reff{Table \ref{tab: model_params} confirms that the luminosity predictions from our best matching isochrone resulting from both comparison methods are accurate for both components and this is also clear in the lower, left panel of Fig.\ref{fig: isochrone figure}. The other parameters are also in good agreement for star A, particularly regarding those resulting from method 2 where the agreement is excellent. However, for star B, the radius is underestimated while the \Teff\ is overestimated, and these discrepancies compensate for each other to yield the accurate luminosity prediction.}

\reff{A possible explanation is that the secondary star is approaching the terminal age main sequence (TAMS), i.e., an evolved stage where the sensitivity of the models increases. Thus, a full evolutionary modelling analysis, such as that carried out by \cite{Guo_2016}, might yield better model predictions because metallicity and other parameters relating to, e.g., overshooting, are included as free parameters so the model is more flexible. This would also provide the means for a more detailed discussion of model comparison with observations but is beyond the scope of this paper. Here, we simply searched in grids of pre-computed models, which are limited by the size of the grid steps in metallicity and age, as well as fixed input physics, to estimate the age of the system. The parameter ranges that the isochrone neighbours span in the planes shown in Fig.\ref{fig: isochrone figure} (i.e., about twice the uncertainty), and the agreement with the value determined by \citet{Guo_2016}, is evidence that this age estimation is accurate.}

\begin{table}
\caption{\label{tab: model_params} \reff{Model parameters of the best fitting isochrone with [Fe/H] = 0.05 dex and an age of $1.259 \pm 0.073$ Gyr. The results obtained from both comparison methods are given. Also given are the differences $\Delta$ between the observations and model for each parameter; we quote this difference in units of the uncertainty associated to the observations in brackets.}}
   \centering
\begin{tabular}{l r r@{\,(}l@{$\sigma$)} c r r@{\,(}l@{$\sigma$)}} \hline
Parameter              & \multicolumn{3}{c}{Method 1}& & \multicolumn{3}{c}{Method 2} \\
                                      & Value    & \multicolumn{2}{c}{$\Delta$[\%]}& 
                                      & Value    & \multicolumn{2}{c}{$\Delta$[\%]}\\
\hline
 Mass ratio                          &   1.0667 &   0.0 & 0.0 & & 1.0530 &  -1.2 & 3.6\\
 $M_{\rm A}$ [\Msun]                 &   1.749  &   0.0 & 0.0 & & 1.758  &  0.5 & 0.9\\
 $M_{\rm B}$  [\Msun]                &   1.866  &   0.0 & 0.0 & & 1.851  &  -0.8& 1.5\\
 $R_{\rm A}$  [\Rsun]                &   2.469  &  -2.5 & 2.5 & & 2.512  &  -0.8& 0.8\\
 $R_{\rm B}$  [\Rsun]                &   3.214  &   7.8 & 5.3 & & 3.096  &   3.8& 2.6\\
 $T_{\rm eff,A}$ [K]                 &   6987   &   0.33& 0.5 & & 6971   &   0.1& 0.2\\
 $T_{\rm eff, B}$  [K]               &   6636   &  -3.0 & 5.5 & & 6694   &  -2.1& 3.9\\
 $\log(L_{\rm A}/\Lsun)$             &   1.117  &  -1.4 & 1.0 & & 1.128  &  -0.4& 0.4\\
 $\log(L_{\rm B}/\Lsun)$             &   1.256  &   1.0 & 0.8 & & 1.239  &  -0.4& 0.3\\
\hline
\end{tabular}
\end{table}

\begin{figure}
    \centering
    \includegraphics[width = \columnwidth]{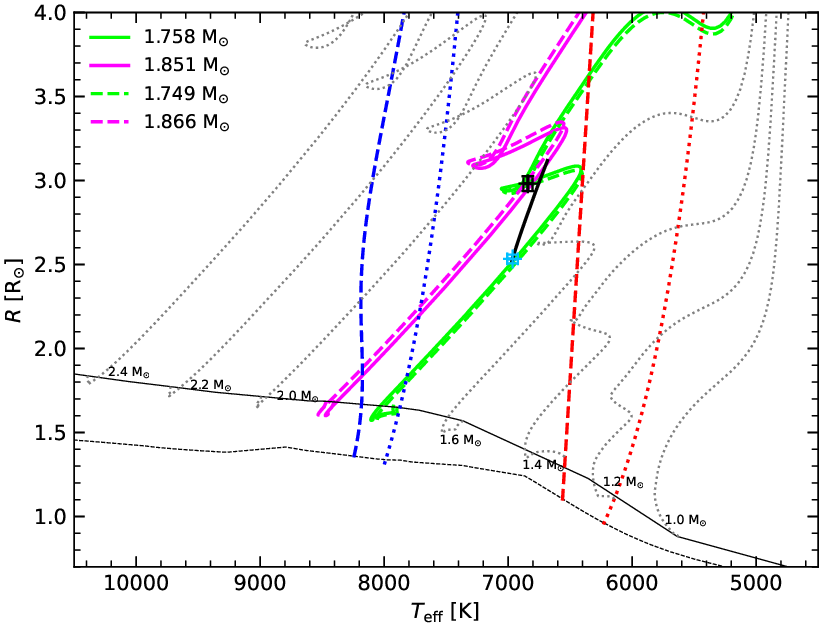}
    \caption{\reff{The \Teff-$R$ plane showing the evolutionary tracks corresponding to the models presented in Table \ref{tab: model_params}. The evolutionary tracks are shown in green and purple for star A and star B, respectively, and their observed locations are indicated by the blue and black markers. The blue (blue lines) and red (red lines) edges of the instability domains for low-order p- and g-modes calculated by \citet{Xiong_2016} for \dsct\ (dashed lines) and \gdor\ (dotted lines) stars are indicated. The thin, black line is the ZAMS corresponding to the models in Table \ref{tab: model_params} ([Fe/H] = 0.05 dex) and the dotted black line is the ZAMS for a metallicity of [Fe/H] $= -0.25$ dex. The thick, black line represents the best fitting isochrone which is shown as a solid grey line in Fig.\ref{fig: isochrone figure}.} Transparent grey dotted lines show the evolutionary tracks of [Fe/H] = 0.05 dex stars with other labelled masses.}
    \label{fig: mass radius evolution}
\end{figure}

\begin{figure}
    \centering
    \includegraphics[width = \columnwidth]{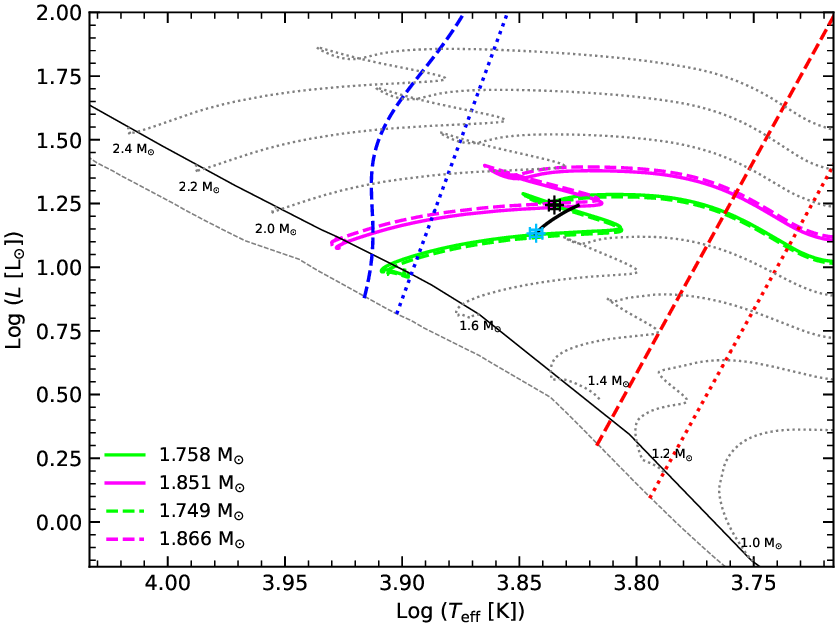}
    \caption{Same as Fig.\ref{fig: mass radius evolution} but in the Hertzsprung-Russell diagram.}
    \label{fig:HR evolution}
\end{figure}

\reff{The evolutionary tracks corresponding to the masses of the models in Table \ref{tab: model_params}} are shown in the \Teff-$R$, and HR diagrams in Figs.\ \ref{fig: mass radius evolution} and \ref{fig:HR evolution}, respectively,\reff{ as well as the observed locations of the components.} While \citet{Guo_2016} find the secondary to be in the hydrogen-shell burning phase for the same age estimate as ours, we find that the secondary has not yet exhausted the hydrogen in its core, and is approaching the TAMS. The scenario found here is more likely to be observed \reff{on a statistical basis} since the evolution up the \reff{HR} diagram after the blue loop occurs on a very short timescale. In any case, the secondary is more evolved and thus larger and more massive than the primary, but cooler. 

Also shown in those figures are is the blue and red edges of the instability strips for low-order p- (dashed lines), and g-modes (dotted lines) in \dsct\ and \gdor\ stars, respectively, from \citet{Xiong_2016}. Both stars are well within both instability domains which is in agreement with the findings in Section \ref{sec:asteroseismology} that both components \reff{might be} pulsating. \citet{Guo_2016} find that the secondary is slightly hotter than the blue edge of the \gdor\ instability domain calculated by \citet{Dupret_2005b}. This reflects the fact that the calculations by \citet{Xiong_2016} (used here) predict a much broader overlap between the \dsct\ and \gdor\ instability domains so stars with a larger range in \Teff\ are expected to pulsate simultaneously in p- and g-modes, i.e., hybrids are expected to be more common. \reff{We also plotted a sub-solar ZAMS ([Fe/H] = -0.25\,dex) as well as the ZAMS corresponding to our best matching isochrone ([Fe/H = 0.05\,dex] in Fig.\ref{fig: mass radius evolution} and Fig.\ref{fig:HR evolution}; this shows how the ZAMS is raised to higher luminosities and radii for higher metallicity.} 

We present the results previously derived by \citet{Guo_2016} in Table \ref{tab: previous results}. The 13 spectroscopic observations \reff{that the previous authors} used to derive their RVs and atmospheric parameters have resolutions of $R=6000$. While it was already discussed in section \ref{sec:atmospheric parameters} that our atmospheric parameters agree with those derived by \citet{Guo_2016}, the higher resolution ($R=60000$) of the spectra, and larger number of observations (33) used in this work yielded a much higher quality RV curve. This is reflected by comparing the reported masses between the two studies; the precision in the mass estimates by \citet{Guo_2016} are $4.0\%$ and $3.9\%$ for the primary and secondary, respectively, while we attain precisions of $0.57\%$ and $0.59\%$. The two sets of mass measurements agree to within $0.2\sigma$ for $M_{\rm A}$ and 1.1$\sigma$ for $M_{\rm B}$. In contrast, our radius measurements differ from those of \citet{Guo_2016} by 6.6$\sigma$ for star~A and 3.5$\sigma$ for star~B. 
\reff{However, \citet{Guo_2016} note the discrepancy between their values for the radius ratio derived from two spectroscopic methods ($k = 1.22 \pm 0.05$ and $k = 1.27 \pm 0.29$) and the value derived from their light curve modelling ($1.41 \pm 0.018$), where they \emph{tentatively} adopt the radii associated to the latter. The radius ratio derived here agrees with the values derived by \citet{Guo_2016} using their spectroscopic methods within 0.8\,$\sigma$ and 0.3\,$\sigma$, but shows the same discrepancy compared to the value derived from their light curve modelling, for which \citet{Guo_2016} note there exists a family of comparable solutions due to the partial nature of the eclipses.}

\reff{We have assumed a circularised and synchronized orbit for KIC\,9851944 because our RV- and light-curve solutions were consistent with a circular orbit, and the values for the component of the synchronous velocity along our line of site are in excellent agreement with the values for \vsini\ derived in section \ref{sec:atmospheric parameters} (see Table \ref{tab:optfits}). Further justification was provided by \citet{Guo_2016} where they examined the eclipse times by \citet{Conroy_2014} and \citet{Giess_2015}, finding that the median deviation of the phase difference between the primary and secondary eclipses from that of a circular orbit suggests $e \geq 0.0001$. Furthermore, the circularization time-scale for a binary system like KIC 9851944 is 600~Myr, and the synchronization timescale is an order of magnitude shorter \citep{Zahn_1977, Khaliullin_2010, Guo_2016}; these timescales are shorter than the age of the system reported by both studies.}

\reff{Our estimations of the near-core and surface rotations reported in section \ref{sec:atmospheric parameters} and \ref{sec:asteroseismology} suggest KIC 9851944 is rotating rigidly and synchronously.} This is similar to the findings by e.g.\ \citet{Guo_2017b, Guo_Li_2019}, that the short-period EBs KIC 9592855 and KIC 7385478 both contain a \gdor\ pulsator that is tidally synchronised at the surface as well as at the near-core regions. Such measurements allow for a calibration of the time-scales for synchronization at the surface compared to the core, and thus the time-scales associated to angular momentum transport, which further aids in the discrimination among different theories. 

\reff{In contrast to the above}, our detection of the orbital harmonic frequency comb for KIC 9851944 either suggests asynchronous rotation or \reff{a non-zero eccentricity that has not been detected due to} observational error because these can be indicative of tidally excited oscillations. It is unclear whether the very small value for the lower limit on the eccentricity, i.e., $e>0.0001$, reported above for this system would be enough to induce tidal excitation of modes at the amplitudes observed here.  

The pulsation analysis here complements the study by \citet{Guo_2016} \reff{who report splittings to some of the pulsation modes. We confirm the detection of tidally split multiplets, and explain their origin; we present evidence to suggest that these are due to perturbations to the pulsation mode cavaties, i.e., tidally perturbed pulsations, by investigating their phase and amplitude dependencies with orbital phase.} \citet{Guo_2016} attempted to interpret the modes by comparing the observations with theoretically computed frequencies, from which they concluded that the observations can be explained by low-order p modes in the primary and the secondary, or the g mode and mixed modes of the secondary \citep{Guo_2016}. \reff{We confirm that some of the p modes belong to the primary, and others could belong to either the primary or the secondary from modulation of the amplitudes during eclipses. Our evidence suggests the primary is the hybrid, and this is because the saw-tooth-like "glitch" in pulsation phase of the tidally perturbed g mode (see Fig.\ref{fig:kic09851944_multiplet-223972}) can only be explained if the pulsation belongs to the primary \citep{VanReeth_2022,VanReeth2023}.} \citet{Guo_2016} report that mode identification was inconclusive, reflecting the difficulty in identifying the p-modes in \dsct\ stars. 
\begin{table} \centering
\caption{\label{tab: previous results} Previously reported results for KIC 9851944 by \citet{Guo_2016}. Also given are the discrepancies $\Delta$ of our results compared to those results, given as a percentage of the previous result as well as units of sigma $\sigma$.}
\setlength{\tabcolsep}{4.5pt}
\begin{tabularx}{\columnwidth}{l r@{\,$\pm$\,}l r@{\,(}l@{$\sigma$)} c r@{\,$\pm$\,}l r@{\,(}l@{$\sigma$)}} \hline
Parameter                             & \multicolumn{4}{c}{Star A} & &\multicolumn{4}{c}{Star B} \\
                                      &  \multicolumn{2}{c}{Value} & \multicolumn{2}{c}{$\Delta$ [\%]}& & \multicolumn{2}{c}{Value}   &\multicolumn{2}{c}{$\Delta$[\%]}\\
\hline
Mass (\Msun)                          &    1.76 & 0.07 & -0.6 & 0.2 & & 1.79 & 0.07 &4.24  & 1.1  \\
Radius (\Rsun)                        &    2.27 & 0.03 & 11.6 & 6.6 & & 3.19 & 0.04 &-6.55 & 3.5  \\
$\log (g\, [\rm cgs])$                &    3.96 & 0.03 & -2.2 & 2.7 & & 3.69 & 0.03 & 1.9  & 2.1  \\
$v_{\rm sync}$ (\kms)                 &    51.4 & 0.7  & 15.2 & 8.4 & & 72.1 & 0.09 & -3.3 & 2.4  \\
\Teff\ (K)                            &    7026 & 100  & -0.9 & 0.6 & & 6902 & 100  & -0.9 & 0.6  \\

$q$                                   & \multicolumn{4}{c}{$1.01 \pm 0.03$} && \multicolumn{4}{c}{5.6\,(1.9$\sigma$)} \\ 
$a$ (\Rsun)                           & \multicolumn{4}{c}{$10.74 \pm 0.014$} && \multicolumn{4}{c}{0.6\,($2.8\sigma$)} \\
\hline
\end{tabularx} \end{table}


\section{Conclusions}\label{sec:conclusion}

We have determined the physical properties of KIC\,9851944, a short-period detached eclipsing binary containing two F-type stars, both of which pulsate. Our analysis is based on 33 \'echelle spectra plus light curves from the \emph{Kepler} and TESS missions. We measure masses and \Teff s to 0.6\%, radii to 1.0\% and 1.5\%, and 133 frequencies due to p- and g-mode pulsations. We find no evidence of a third component, apsidal motion, or eccentricity. We estimate the age of the system to be $\sim$1.26\,Gyr by comparison of the measured properties to the MIST model isochrones.

We investigated the systemic errors associated to using the cross correlation technique to extract RVs, which arise due to blending between the spectral lines of the components in a binary system. We find that the effect is small when using {\sc todcor} but correcting for the issue is still necessary because the resulting shifts are clearly systematic in nature and have a non-negligible impact on the results. We used three independent spectroscopic methods to determine the light ratio for the system. The results are in agreement with each other and consistent with the value obtained from modelling the light curve which supports the reliability of our light curve solution. We compared our results to those reported by \citet{Guo_2016} and find that we have improved the precision of the measured masses significantly, but the precision in the radius estimates are comparable. Both these outcomes are expected since we use much higher-resolution spectra but the same photometry (\emph{Kepler}). 

By analysing the residuals of the light curve model, we confirm the \reff{detection} of tidally perturbed p-mode pulsations, possibly in both components of KIC\,9851944. A short period spacing pattern was detected among the g modes and was assigned to the primary component, where \reff{perturbations to one of the g-modes was detected}. Thus, the primary component is a \dsct/\gdor\ hybrid. If pulsation mode identification can be performed \reff{\citep[as in, e.g.,][]{Bedding_2020}} on the p modes, KIC 9851944 will become a well-equipped laboratory for stellar physics.


\section*{Data Availability}
This paper includes data collected by the \emph{Kepler} and TESS missions which is publicly available at the Mikilski Archive for Space Telescopes (MAST) at the Space Telescope Science institute (STScl) (\href{https://mast.stsci.edu/portal/Mashup/Clients/Mast/Portal.html}{https://mast.stsci.edu}).

\section*{Acknowledgements}
We gratefully acknowledge Barry Smalley for his insightful comments regarding the properties of Am stars, comparing the Ca K lines with synthetics, and sharing his thoughts on the possibility that the components of KIC 9851944 are Am stars. 

We are grateful to Kelsey Clubb for assisting in obtaining and reducing the spectroscopic observations from the Shane telescope.
TVR gratefully acknowledges funding from the KU Leuven Research Council (grant C16/18/005: PARADISE). We gratefully acknowledge financial support from the Science and Technology Facilities Council. This research has made use of the SIMBAD and CDS databases operated by the Centre de Donn\'ees astronomiques de Strasbourg, France. This paper also made use of data from the \emph{Kepler} and TESS missions obtained from the MAST data archive at the Space Telescope Science Institue (STScI). Funding for the \emph{Kepler} and TESS missions is provided by the NASA Science Mission Directorate and NASA Explorer Program, respectively. STScI is operated by the Association of Universities for Research in Astronomy, Inc., under NASA contract NAS 5–26555.

\bibliographystyle{mnras}
\bibliography{main.bib}


\clearpage
\appendix
\section{Ephemeris Determination}
\FloatBarrier
This appendix presents the results from the preliminary \textsc{jktebop} analaysis as well as times of primary minimum and fitted values calculated in section \ref{sec:ephem}.
\begin{table} \caption{\label{tab:pre_lcfit} \reff{Prelimimary light curve results from \textsc{jktebop} fits to the \emph{Kepler} and TESS light curves. We do not present the error bars associated to each parameter in Table \ref{tab:pre_lcfit} because 1; uncertainties from the covariance matrix of a light curve fit are notoriously underestimated and the results from these preliminary fits are not reliable (see below) so we did not attempt to derive better uncertainties, and 2; excluding the error bar from the result makes it clear that these are not our final values for the light curve parameters.}}
\centering
\begin{tabular}{ll }
\hline
Paramter & Value  \\
\hline
$J$           &  0.7956      \\
$r_A+r_A$     &  0.5091      \\
$k$           &  1.591       \\
$i (^\circ)$  &  74.294      \\
$e\cos\omega$ & -0.000120    \\
$e\sin\omega$ & -0.012329    \\
$u_A$         &  0.684       \\
$u_B$         &  0.310       \\
\hline
\end{tabular} \end{table}

\begin{table}\caption{\label{tab: app- O-C data} Measured and fitted times of primary minimum derived in section \ref{sec:ephem}.}
    \centering
    \begin{tabular}{cccc}
    \hline
    Cycle & Observed (BJD)& Calculated (BJD) & O-C (days)\\
    \hline
    -624&  2454958.029393 (85)&  2454958.0293642&   0.0000293\\   
    -168&  2455944.768607 (47)&  2455944.7685736&   0.0000337\\
    -154&  2455975.063169 (69)&  2455975.0631985&  -0.0000287\\  
    -140&  2456005.357832 (82)&  2456005.3578233&   0.0000091\\   
    -128&  2456031.324618 (54)&  2456031.3246446&  -0.0000261\\   
    -113&  2456063.783268 (56)&  2456063.7831712&   0.0000973\\   
    -100&  2456091.913868 (59)&  2456091.9138943&  -0.0000260\\   
    -88&  2456117.880630 (67)&  2456117.8807156&  -0.0000855\\   
    -72&  2456152.503214 (57)&  2456152.5031440&   0.0000708\\   
    -56&  2456187.125554 (54)&  2456187.1255724&  -0.0000181\\   
      0&  2456308.30403 (14)&  2456308.3040718&  -0.0000409\\   
     14&  2456338.598621 (52)&  2456338.5986967&  -0.0000748\\  
     31&  2456375.385071 (58)&  2456375.3850268&   0.0000442\\   
     44&  2456403.515741 (61)&  2456403.5157499&  -0.0000083\\   
     53&  2456422.99070 (15)&  2456422.9908659&  -0.0001687\\   
   1106&  2458701.57955 (17)&  2458701.5794350&   0.0001219\\   
   1119&  2458729.71047 (17)&  2458729.7101581&   0.0003167\\   
   1439&  2459422.15886 (22)&  2459422.1587261&   0.0001436\\   
   1600&  2459770.54729 (25)&  2459770.5469118&   0.0003871\\   
   1613&  2459798.67785 (22)&  2459798.6776349&   0.0002222\\   
   1626&  2459826.80849 (21)&  2459826.8083580&   0.0001361\\  
   \hline
    \end{tabular}
\end{table}
\clearpage
\FloatBarrier
\section{Iteratively prewhitened frequencies}
\label{sec:extracted_freq}
\begin{table*}\label{tab:extracted_freq} \centering 
\caption{Parameter values associated with the iteratively prewhitened frequencies, which were obtained as described in Sec.\,\ref{subsec:freq-analysis}. The frequencies are grouped, with the first set consisting of those labelled as independent frequencies, and the following groups consisting of combination frequencies, sorted according to the dominant parent frequency. In the last column, we list the identified combinations. Combination frequencies that are found within the frequency resolution $f_{\rm res}$ but not within $3\sigma$, are indicated with ${}^*$ in the first column and with $\approx$ in the final column.}
\begin{tabular}{rrrrrl}
\hline
& frequency $f$ & amplitude $a$ & phase $\phi$ & signal-to-noise $S/N$ & comments\\
& ($\rm d^{-1}$) & (mmag) & ($2\pi$\,rad) & & \\
\hline
$f_{1}$ & 0.029864(4) & 0.3292(12) & 0.4796(6) & 20.42 & \\
$f_{2}$ & 0.04104(12) & 0.1145(12) & 0.246(2) & 7.15 & \\
$f_{3}$ & 0.050798(11) & 0.1231(12) & -0.482(2) & 7.72 & \\
$f_{4}$ & 0.429539(12) & 0.1126(12) & -0.412(2) & 8.25 & \\
$f_{6}$ & 1.13642(2) & 0.0559(12) & 0.196(3) & 5.68 & \\
$f_{7}$ & 1.26821(2) & 0.0855(12) & 0.151(2) & 9.41 & \\
$f_{8}$ & 1.297029(9) & 0.1389(12) & 0.044(14) & 15.59 & \\
$f_{14}$ & 2.1344(2) & 0.0598(12) & 0.345(3) & 10.12 & \\
$f_{15}$ & 2.20293(2) & 0.0554(12) & -0.028(3) & 9.51 & \\
$f_{16}$ & 2.239718(6) & 0.23(12) & -0.0492(8) & 39.16 & \\
$f_{17}$ & 2.27839(4) & 0.0363(12) & -0.258(5) & 6.12 & \\
$f_{22}$ & 4.14574(3) & 0.04(12) & -0.092(5) & 5.63 & \\
$f_{27}$ & 5.097165(3) & 0.5158(12) & -0.301(4) & 55.8 & \\
$f_{28}$ & 5.53879(2) & 0.0597(13) & -0.271(3) & 5.91 & \\
$f_{30}$ & 5.99419(2) & 0.0644(12) & 0.442(3) & 5.87 & \\
$f_{34}$ & 6.59017(2) & 0.0672(12) & 0.469(3) & 6.12 & \\
$f_{35}$ & 7.22671(2) & 0.0824(12) & -0.288(2) & 8.08 & \\
$f_{45}$ & 9.59223(5) & 0.0284(12) & -0.486(7) & 7.4 & \\
$f_{53}$ & 10.176017(2) & 0.6515(12) & 0.429(3) & 172.1 & \\
$f_{54}$ & 10.399706(2) & 0.8015(12) & -0.2904(2) & 221.19 & \\
$f_{58}$ & 10.90789(6) & 0.0211(12) & 0.045(9) & 5.79 & \\
$f_{60}$ & 11.00533(2) & 0.0579(12) & 0.059(3) & 16.02 & \\
$f_{61}$ & 11.018536(5) & 0.2785(12) & 0.3112(7) & 76.68 & \\
$f_{63}$ & 11.08192(6) & 0.0237(12) & 0.408(8) & 6.61 & \\
$f_{66}$ & 11.27246(3) & 0.0487(12) & 0.117(4) & 13.83 & \\
$f_{67}$ & 11.3651(5) & 0.0275(12) & -0.247(7) & 7.76 & \\
$f_{68}$ & 11.41982(7) & 0.1941(12) & -0.3938(10) & 54.46 & \\
$f_{70}$ & 11.43819(3) & 0.0376(12) & -0.15(5) & 10.57 & \\
$f_{71}$ & 11.52234(5) & 0.2606(12) & -0.4854(7) & 73.33 & \\
$f_{74}$ & 11.890477(2) & 0.5704(12) & 0.1807(3) & 164.93 & \\
$f_{90}$ & 13.61013(6) & 0.0213(12) & 0.222(9) & 7.7 & \\
$f_{93}$ & 13.76588(7) & 0.019(12) & 0.161(10) & 7.05 & \\
$f_{98}$ & 14.01097(2) & 0.0589(12) & -0.244(3) & 23.17 & \\
$f_{100}$ & 14.21089(3) & 0.0417(12) & -0.354(5) & 16.79 & \\
$f_{102}$ & 14.315077(5) & 0.2819(12) & 0.4401(7) & 115.05 & \\
$f_{103}$ & 14.39805(2) & 0.0527(12) & -0.117(4) & 21.96 & \\
$f_{104}$ & 14.448108(7) & 0.1824(12) & -0.3085(11) & 77.2 & \\
$f_{105}$ & 14.4932(3) & 0.0396(12) & 0.333(5) & 16.83 & \\
$f_{107}$ & 14.69426(7) & 0.0175(12) & 0.023(11) & 7.6 & \\
$f_{111}$ & 14.83253(6) & 0.022(12) & 0.083(9) & 9.79 & \\
$f_{125}$ & 19.126701(8) & 0.1584(12) & -0.4735(12) & 91.81 & \\
$f_{126}$ & 19.42779(2) & 0.0847(12) & -0.419(2) & 47.64 & \\
$f_{127}$ & 20.15601(9) & 0.0139(12) & -0.124(14) & 7.51 & \\
$f_{129}$ & 20.82366(3) & 0.0432(12) & 0.429(4) & 21.19 & \\
          &             &            &          &       & \\
$f_{5}^*$ & 0.462195(5) & 0.2625(12) & -0.3711(7) & 19.39 & $\approx 1f_{\rm orb}$\\
$f_{10}^*$ & 1.386342(4) & 0.3628(12) & -0.3605(5) & 44.54 & $\approx 3f_{\rm orb}$\\
$f_{13}^*$ & 1.848604(10) & 0.1336(12) & 0.2485(14) & 20.41 & $\approx 4f_{\rm orb}$\\
$f_{18}$ & 2.310647(4) & 0.3462(12) & 0.3839(6) & 58.23 & 5$f_{\rm orb}$\\
$f_{20}$ & 2.77278(7) & 0.1827(12) & 0.0065(11) & 32.88 & 6$f_{\rm orb}$\\
$f_{21}$ & 3.234899(8) & 0.1651(12) & -0.3581(12) & 29.22 & 7$f_{\rm orb}$\\
$f_{24}^*$ & 4.62137(3) & 0.051(12) & 0.015(4) & 6.14 & $\approx 10f_{\rm orb}$\\
$f_{26}^*$ & 5.08347(2) & 0.0676(12) & -0.349(3) & 7.34 & $\approx 11f_{\rm orb}$\\
$f_{31}^*$ & 6.007556(11) & 0.1215(12) & -0.13(2) & 11.09 & $\approx 13f_{\rm orb}$\\
$f_{36}^*$ & 7.85624(2) & 0.0794(12) & 0.441(2) & 9.71 & $\approx 17f_{\rm orb}$\\
$f_{39}$ & 8.7805(4) & 0.0331(12) & -0.314(6) & 7.19 & 19$f_{\rm orb}$\\
$f_{52}^*$ & 10.16693(3) & 0.0453(12) & -0.448(4) & 11.96 & $\approx 22f_{\rm orb}$\\
$f_{64}$ & 11.09106(2) & 0.0642(12) & -0.195(3) & 17.87 & 24$f_{\rm orb}$\\
$f_{77}$ & 12.01533(3) & 0.0517(12) & 0.058(4) & 15.06 & 26$f_{\rm orb}$\\
$f_{95}$ & 13.86382(6) & 0.021(12) & 0.064(9) & 7.98 & 30$f_{\rm orb}$\\
$f_{109}$ & 14.7881(4) & 0.0326(12) & 0.314(6) & 14.44 & 32$f_{\rm orb}$\\
\hline\end{tabular}\end{table*}
\begin{table*} \centering\contcaption{}
\begin{tabular}{rrrrrl}
\hline
& frequency $f$ & amplitude $a$ & phase $\phi$ & signal-to-noise $S/N$ & comments\\
& ($\rm d^{-1}$) & (mmag) & ($2\pi$\,rad) & & \\
\hline
$f_{118}$ & 15.71249(6) & 0.0201(12) & -0.435(10) & 9.87 & 34$f_{\rm orb}$\\
$f_{122}$ & 17.56078(6) & 0.0202(12) & -0.407(10) & 10.72 & 38$f_{\rm orb}$\\
$f_{124}$ & 18.48512(5) & 0.0242(12) & -0.172(8) & 13.89 & 40$f_{\rm orb}$\\
  &        &             &            &          &       \\
$f_{9}$ & 1.315491(12) & 0.1099(12) & 0.252(2) & 12.54 & $ f_{16}$ - 2$f_{\rm orb}$\\
$f_{12}^*$ & 1.77778(2) & 0.0577(12) & -0.476(3) & 8.7 & $\approx f_{16}$ - $f_{\rm orb}$\\
  &        &             &            &          &       \\
$f_{23}^*$ & 4.17278(2) & 0.062(12) & -0.128(3) & 8.61 & $\approx f_{27}$ - 2$f_{\rm orb}$\\
$f_{25}$ & 4.635033(12) & 0.1126(12) & 0.394(2) & 13.54 & $ f_{27}$ - $f_{\rm orb}$\\
$f_{29}$ & 5.559303(5) & 0.2567(12) & -0.1643(8) & 25.3 & $ f_{27}$ + $f_{\rm orb}$\\
$f_{32}^*$ & 6.021343(13) & 0.1002(12) & -0.298(2) & 9.15 & $\approx f_{27}$ + 2$f_{\rm orb}$\\
$f_{33}^*$ & 6.48364(2) & 0.0743(12) & 0.31(3) & 6.72 & $\approx f_{27}$ + 3$f_{\rm orb}$\\
 &         &             &            &          &       \\
$f_{131}$ & 22.06635(5) & 0.0237(12) & -0.077(8) & 12.1 & $f_{52}$ + $f_{73}$\\
$f_{132}$ & 24.70539(5) & 0.0264(12) & 0.45(7) & 14.83 & $f_{73}$ + $f_{81}$\\
 &         &             &            &          &       \\
$f_{11}$ & 1.39562(2) & 0.0572(12) & 0.312(3) & 7.06 & $ f_{53}$ - 19$f_{\rm orb}$\\
$f_{19}$ & 2.319817(12) & 0.1116(12) & -0.005(2) & 18.75 & $ f_{53}$ - 17$f_{\rm orb}$\\
$f_{37}$ & 8.32746(2) & 0.0582(12) & -0.077(3) & 9.42 & $ f_{53}$ - 4$f_{\rm orb}$\\
$f_{40}$ & 8.78961(3) & 0.0418(12) & 0.348(5) & 9.09 & $ f_{53}$ - 3$f_{\rm orb}$\\
$f_{42}$ & 9.25174(2) & 0.0625(12) & 0.162(3) & 15.84 & $ f_{53}$ - 2$f_{\rm orb}$\\
$f_{47}$ & 9.71391(2) & 0.0599(12) & -0.347(3) & 15.77 & $ f_{53}$ - $f_{\rm orb}$\\
$f_{57}$ & 10.63816(2) & 0.058(12) & -0.107(3) & 15.93 & $ f_{53}$ + $f_{\rm orb}$\\
$f_{65}^*$ & 11.10018(3) & 0.0449(12) & -0.321(4) & 12.47 & $\approx f_{53}$ + 2$f_{\rm orb}$\\
$f_{72}^*$ & 11.56252(4) & 0.0362(12) & 0.098(5) & 10.34 & $\approx f_{53}$ + 3$f_{\rm orb}$\\
$f_{78}$ & 12.02459(4) & 0.0327(12) & -0.062(6) & 9.52 & $ f_{53}$ + 4$f_{\rm orb}$\\
$f_{84}$ & 12.94893(4) & 0.0356(12) & 0.191(5) & 11.25 & $ f_{53}$ + 6$f_{\rm orb}$\\
$f_{96}$ & 13.87319(5) & 0.0283(12) & 0.428(7) & 10.81 & $ f_{53}$ + 8$f_{\rm orb}$\\
$f_{110}^*$ & 14.79735(8) & 0.0163(12) & -0.299(12) & 7.22 & $\approx f_{53}$ + 10$f_{\rm orb}$\\
 &         &             &            &          &       \\
$f_{38}^*$ & 8.55106(4) & 0.0327(12) & -0.308(6) & 6.3 & $\approx f_{54}$ - 4$f_{\rm orb}$\\
$f_{43}^*$ & 9.47551(2) & 0.067(12) & 0.496(3) & 17.51 & $\approx f_{54}$ - 2$f_{\rm orb}$\\
$f_{48}$ & 9.93758(14) & 0.0905(12) & 0.007(2) & 23.92 & $ f_{54}$ - $f_{\rm orb}$\\
$f_{73}^*$ & 11.78622(4) & 0.0348(12) & 0.096(6) & 10.11 & $\approx f_{54}$ + 3$f_{\rm orb}$\\
$f_{81}^*$ & 12.71017(5) & 0.0264(12) & 0.386(7) & 8.21 & $\approx f_{54}$ + 5$f_{\rm orb}$\\
$f_{91}^*$ & 13.63457(8) & 0.0158(12) & -0.362(12) & 5.72 & $\approx f_{54}$ + 7$f_{\rm orb}$\\
 &         &             &            &          &       \\
$f_{49}$ & 10.08101(3) & 0.0377(12) & -0.22(5) & 9.93 & $ f_{60}$ - 2$f_{\rm orb}$\\
 &         &             &            &          &       \\
$f_{50}$ & 10.094274(11) & 0.1133(12) & 0.134(2) & 29.92 & $ f_{61}$ - 2$f_{\rm orb}$\\
$f_{75}$ & 11.942814(7) & 0.1987(12) & 0.13(10) & 57.43 & $ f_{61}$ + 2$f_{\rm orb}$\\
 &         &             &            &          &       \\
$f_{41}$ & 9.21174(5) & 0.0262(12) & 0.358(7) & 6.61 & $ f_{71}$ - 5$f_{\rm orb}$\\
$f_{46}$ & 9.67396(5) & 0.0273(12) & -0.086(7) & 7.21 & $ f_{71}$ - 4$f_{\rm orb}$\\
$f_{51}$ & 10.13587(4) & 0.0313(12) & -0.405(6) & 8.28 & $ f_{71}$ - 3$f_{\rm orb}$\\
$f_{56}$ & 10.598064(10) & 0.1353(12) & 0.2578(14) & 37.3 & $ f_{71}$ - 2$f_{\rm orb}$\\
$f_{62}$ & 11.06027(3) & 0.0476(12) & -0.186(4) & 13.31 & $ f_{71}$ - $f_{\rm orb}$\\
$f_{76}$ & 11.9845(5) & 0.0274(12) & 0.103(7) & 8.01 & $ f_{71}$ + $f_{\rm orb}$\\
$f_{80}$ & 12.44658(5) & 0.0237(12) & -0.383(8) & 7.04 & $ f_{71}$ + 2$f_{\rm orb}$\\
$f_{83}$ & 12.90873(7) & 0.0189(12) & 0.327(10) & 5.96 & $ f_{71}$ + 3$f_{\rm orb}$\\
$f_{87}$ & 13.37075(5) & 0.0246(12) & -0.004(8) & 8.46 & $ f_{71}$ + 4$f_{\rm orb}$\\
$f_{101}$ & 14.29517(9) & 0.0146(12) & 0.283(13) & 5.95 & $ f_{71}$ + 6$f_{\rm orb}$\\
 &         &             &            &          &       \\
$f_{44}$ & 9.58038(4) & 0.0297(12) & -0.181(6) & 7.74 & $ f_{74}$ - 5$f_{\rm orb}$\\
$f_{55}^*$ & 10.50457(2) & 0.0551(12) & 0.106(3) & 15.18 & $\approx f_{74}$ - 3$f_{\rm orb}$\\
$f_{59}$ & 10.96637(3) & 0.042(12) & -0.292(5) & 11.57 & $ f_{74}$ - 2$f_{\rm orb}$\\
$f_{69}$ & 11.42854(2) & 0.0649(12) & 0.357(3) & 18.21 & $ f_{74}$ - $f_{\rm orb}$\\
$f_{79}$ & 12.35285(2) & 0.08(12) & -0.375(2) & 23.84 & $ f_{74}$ + $f_{\rm orb}$\\
$f_{82}^*$ & 12.814957(5) & 0.2491(12) & 0.0752(8) & 77.37 & $\approx f_{74}$ + 2$f_{\rm orb}$\\
$f_{85}$ & 13.27707(3) & 0.0514(12) & -0.101(4) & 17.07 & $ f_{74}$ + 3$f_{\rm orb}$\\
$f_{92}$ & 13.73916(3) & 0.0399(12) & -0.488(5) & 14.76 & $ f_{74}$ + 4$f_{\rm orb}$\\
\hline\end{tabular}\end{table*}
\begin{table*}\centering\contcaption{}
\begin{tabular}{rrrrrl} 
\hline
& frequency $f$ & amplitude $a$ & phase $\phi$ & signal-to-noise $S/N$ & comments\\
& ($\rm d^{-1}$) & (mmag) & ($2\pi$\,rad) &    & \\
\hline
$f_{99}$ & 14.2014(3) & 0.0509(12) & 0.167(4) & 20.4 & $ f_{74}$ + 5$f_{\rm orb}$\\
$f_{106}$ & 14.66346(4) & 0.031(12) & -0.274(6) & 13.37 & $ f_{74}$ + 6$f_{\rm orb}$\\
$f_{113}$ & 15.12563(4) & 0.0365(12) & 0.415(5) & 17.04 & $ f_{74}$ + 7$f_{\rm orb}$\\
$f_{117}$ & 15.5876(6) & 0.0227(12) & -0.035(8) & 11.38 & $ f_{74}$ + 8$f_{\rm orb}$\\
$f_{119}$ & 16.04978(7) & 0.0198(12) & -0.329(10) & 10.23 & $ f_{74}$ + 9$f_{\rm orb}$\\
 &         &             &            &          &       \\
$f_{112}$ & 14.93524(5) & 0.0278(12) & 0.498(7) & 12.66 & $ f_{98}$ + 2$f_{\rm orb}$\\
 &         &             &            &          &       \\
$f_{86}$ & 13.28676(5) & 0.0259(12) & 0.304(7) & 8.65 & $ f_{100}$ - 2$f_{\rm orb}$\\
 &         &             &            &          &       \\
$f_{88}$ & 13.39085(4) & 0.0341(12) & 0.204(6) & 11.76 & $ f_{102}$ - 2$f_{\rm orb}$\\
$f_{94}^*$ & 13.85328(4) & 0.0326(12) & -0.282(6) & 12.38 & $\approx f_{102}$ - $f_{\rm orb}$\\
$f_{108}$ & 14.77747(4) & 0.031(12) & -0.005(6) & 13.62 & $ f_{102}$ + $f_{\rm orb}$\\
$f_{114}$ & 15.2393(3) & 0.0473(12) & -0.302(4) & 22.42 & $ f_{102}$ + 2$f_{\rm orb}$\\
$f_{120}$ & 16.16355(5) & 0.0262(12) & -0.061(7) & 13.58 & $ f_{102}$ + 4$f_{\rm orb}$\\
 &         &             &            &          &       \\
$f_{115}$ & 15.32214(8) & 0.0173(12) & -0.351(11) & 8.23 & $ f_{103}$ + 2$f_{\rm orb}$\\
 &         &             &            &          &       \\
$f_{89}$ & 13.52388(2) & 0.0822(12) & 0.432(2) & 29.63 & $ f_{104}$ - 2$f_{\rm orb}$\\
$f_{116}$ & 15.37227(8) & 0.0156(12) & -0.108(12) & 7.54 & $ f_{104}$ + 2$f_{\rm orb}$\\
 &         &             &            &          &       \\
$f_{97}$ & 13.90834(6) & 0.02(12) & 0.279(10) & 7.68 & $ f_{111}$ - 2$f_{\rm orb}$\\
 &         &             &            &          &       \\
$f_{121}$ & 17.27819(3) & 0.0423(12) & -0.464(5) & 22.4 & $ f_{125}$ - 4$f_{\rm orb}$\\
$f_{123}$ & 17.74024(7) & 0.019(12) & 0.113(10) & 10.38 & $ f_{125}$ - 3$f_{\rm orb}$\\
 &         &             &            &          &       \\
$f_{128}$ & 20.3521(2) & 0.0686(12) & -0.152(3) & 35.16 & $ f_{126}$ + 2$f_{\rm orb}$\\
$f_{130}$ & 21.27623(2) & 0.0601(12) & -0.429(3) & 29.62 & $ f_{126}$ + 4$f_{\rm orb}$\\
\hline
\end{tabular}
\end{table*}

\clearpage
\section{Detected tidally perturbed pulsations}
\label{appendix:tidal-perturbations}
In this Appendix we illustrate all detected tidally perturbed pulsations of KIC\,9851944. In each figure, the left panel shows a zoom-in of the associated frequency multiplet, as plotted in Fig.\,\ref{fig:kic09851944_echelle}. In the top right panel, we see the orbital-phase folded light curve of the binary. In the middle and bottom right panels, we see the orbital-phase dependent modulations of the pulsation amplitudes and phases, respectively. These are calculated in 50 data bins, with the 1$\sigma$ uncertainty range indicated by the dashed lines.
\FloatBarrier
\begin{figure}
\centering
\includegraphics[width=84.5mm]{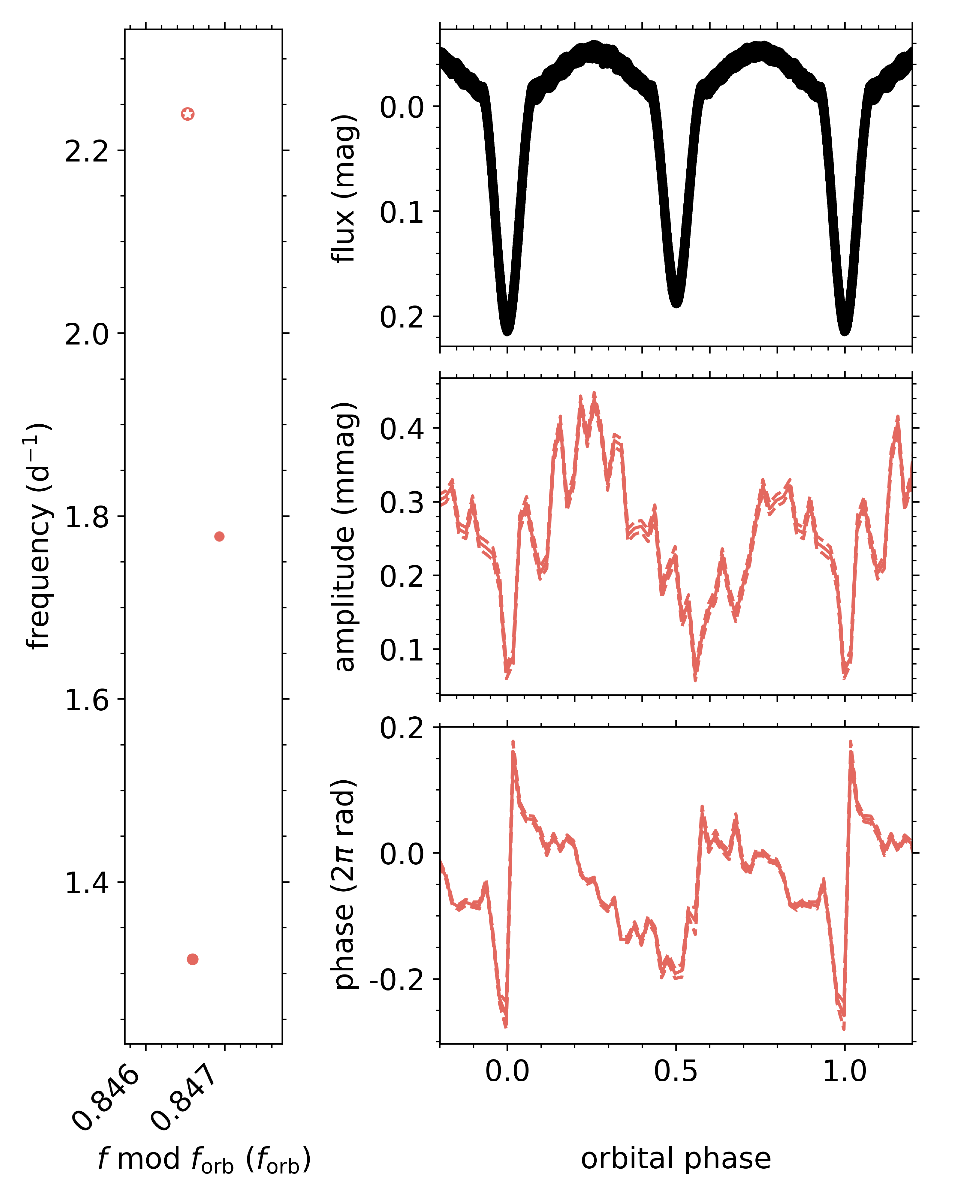}
\caption{\label{fig:kic09851944_multiplet-223972} Tidally perturbed pulsation with frequency $f = 2.239718(6)\,\rm d^{-1}$.}
\end{figure}

\begin{figure}
\includegraphics[width=84.5mm]{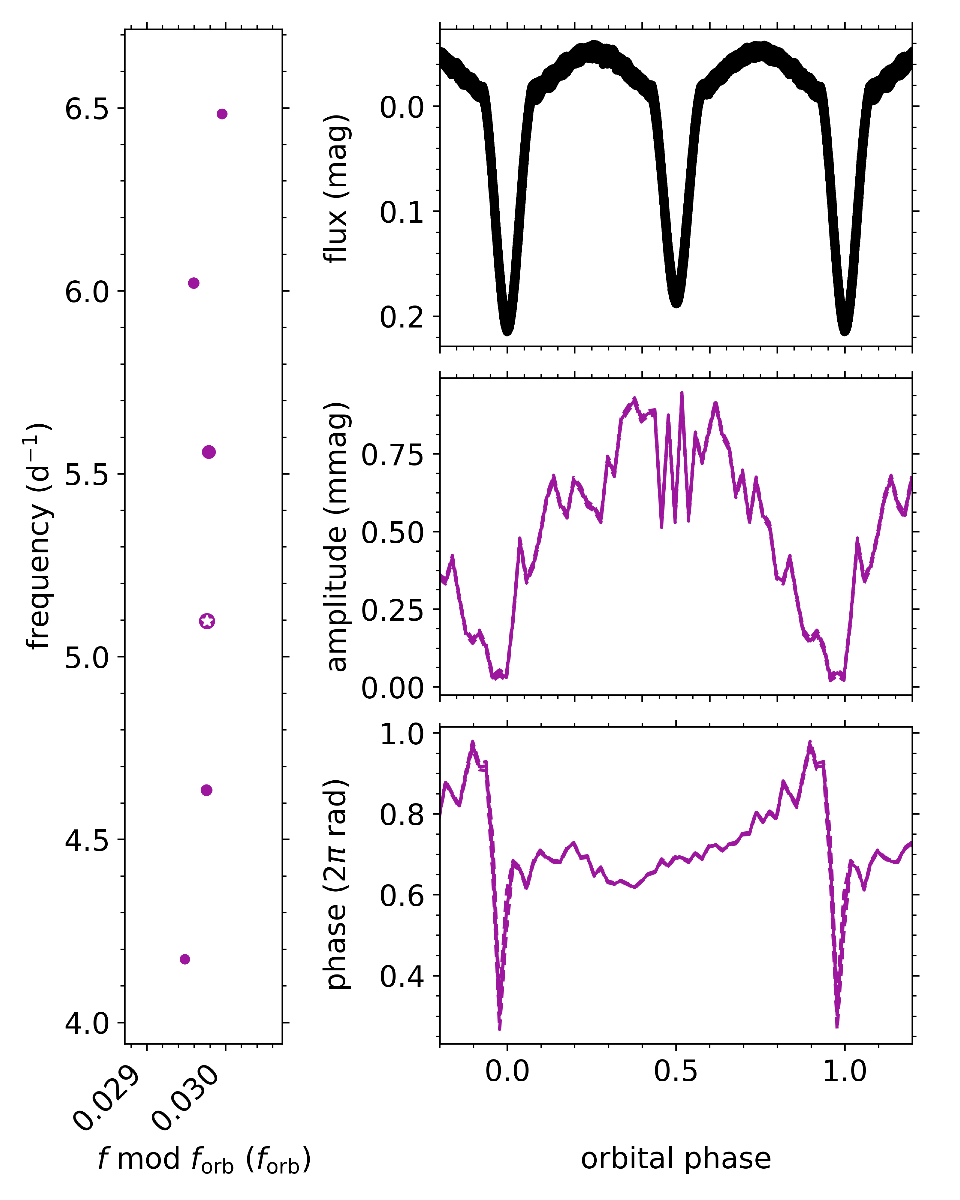}
\caption{\label{fig:kic09851944_multiplet-509716} Tidally perturbed pulsation with frequency $f = 5.097165(3)\,\rm d^{-1}$.}
\end{figure}

\begin{figure}
\includegraphics[width=84.5mm]{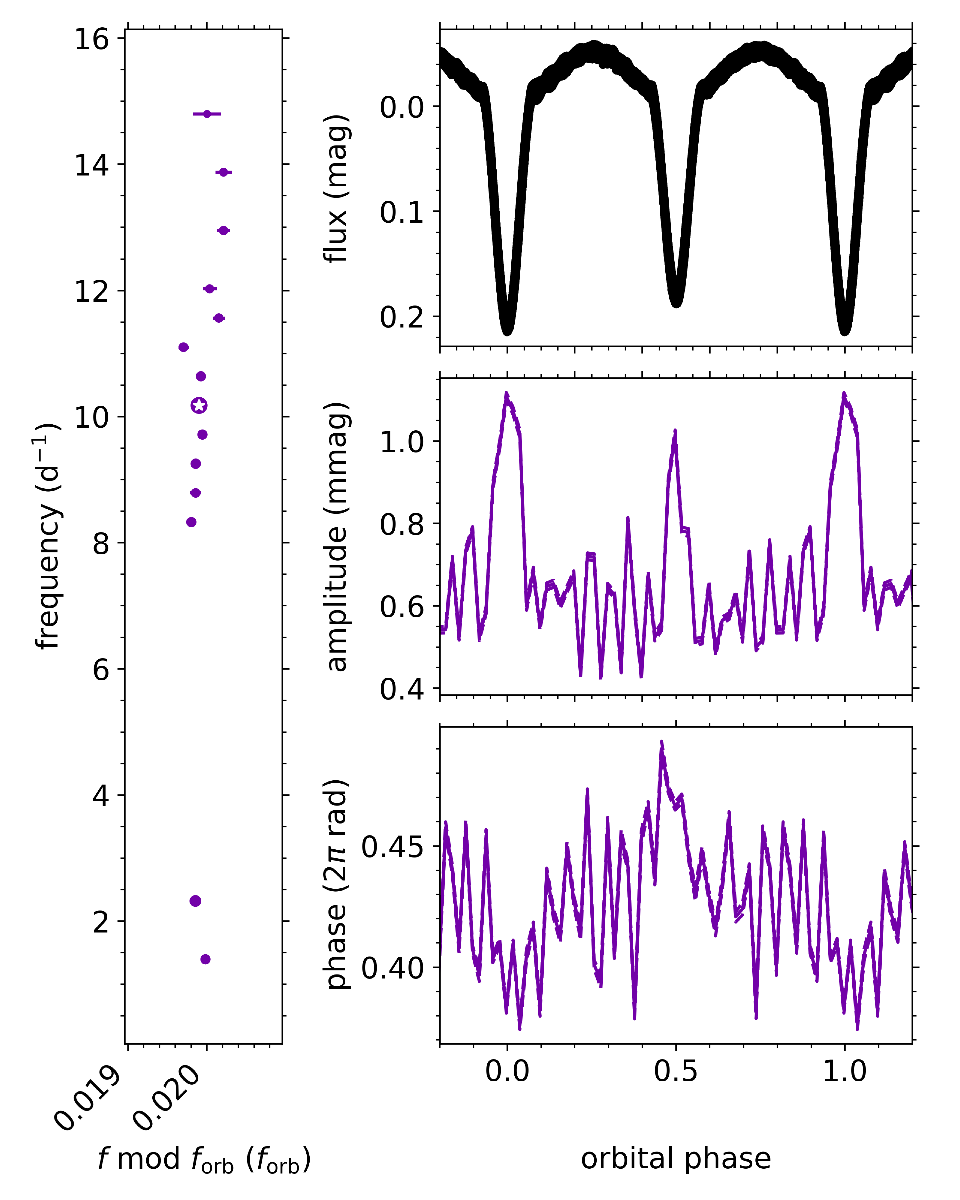}
\caption{\label{fig:kic09851944_multiplet-1017602} Tidally perturbed pulsation with frequency $f = 10.176017(2)\,\rm d^{-1}$.}
\end{figure}

\begin{figure}
\includegraphics[width=82mm]{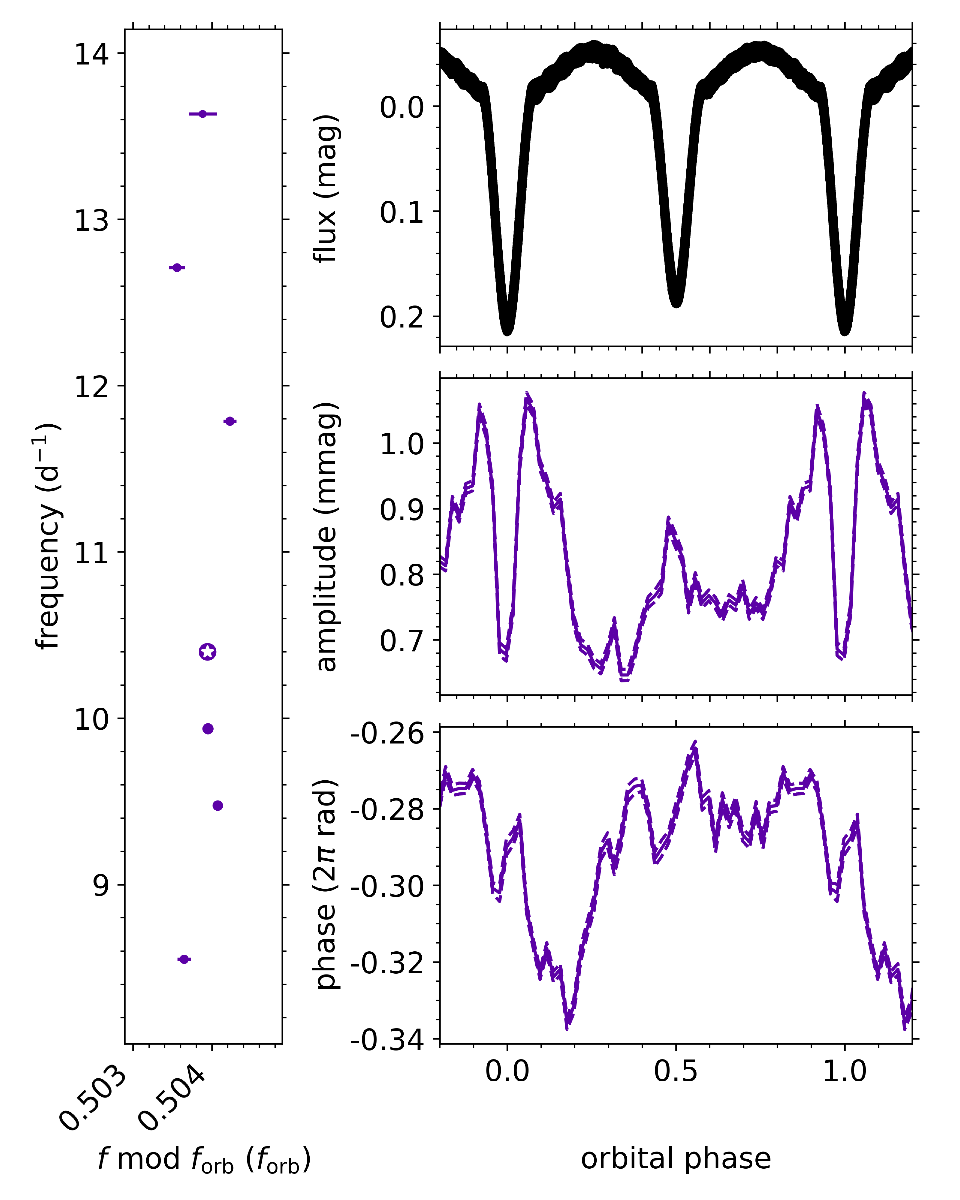}
\caption{\label{fig:kic09851944_multiplet-1039971-b} Tidally perturbed pulsation with frequency $f = 10.399706(2)\,\rm d^{-1}$.}
\end{figure}

\begin{figure}
\includegraphics[width=82mm]{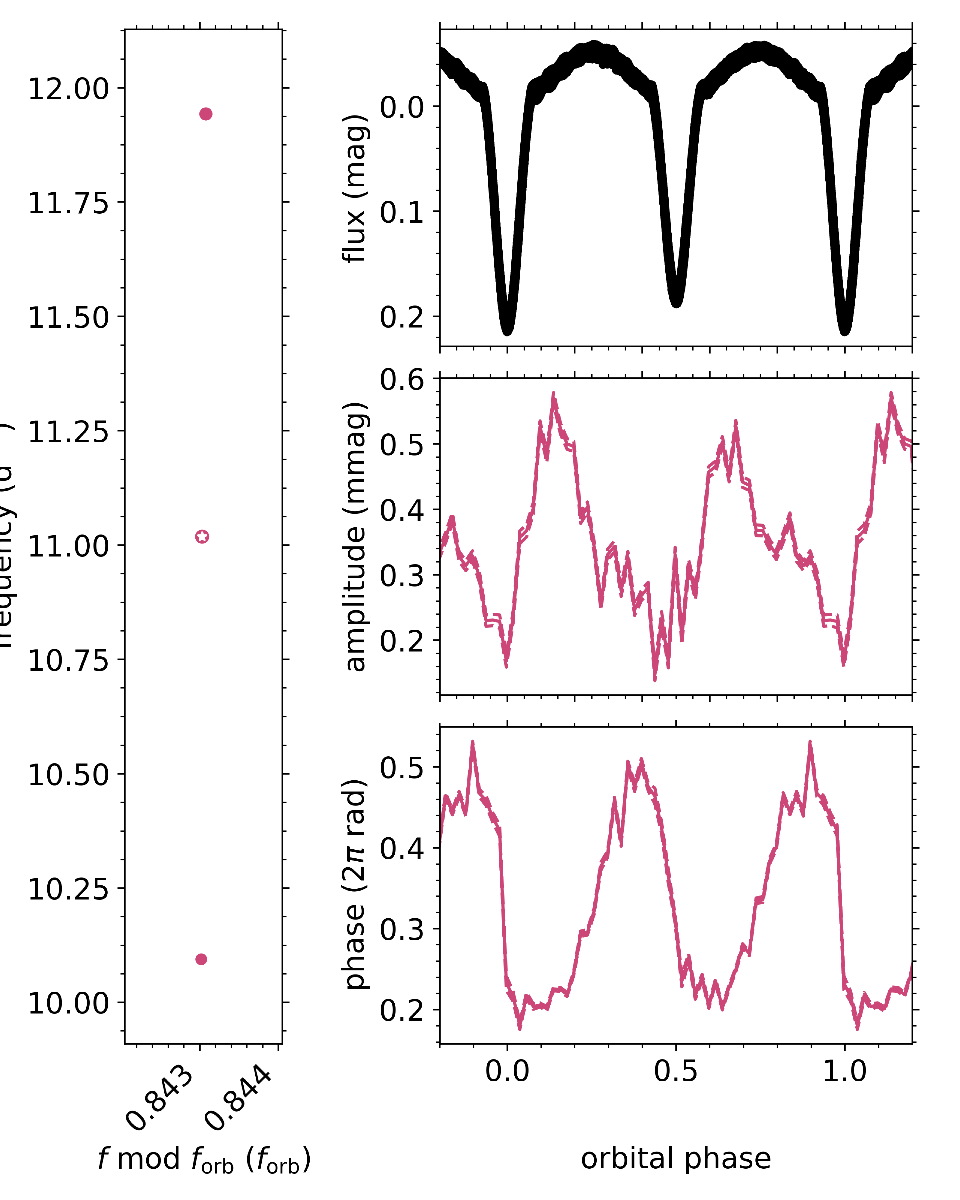}
\caption{\label{fig:kic09851944_multiplet-1101854} Tidally perturbed pulsation with frequency $f = 11.018536(5)\,\rm d^{-1}$.}
\end{figure}

\begin{figure}
\includegraphics[width=82mm]{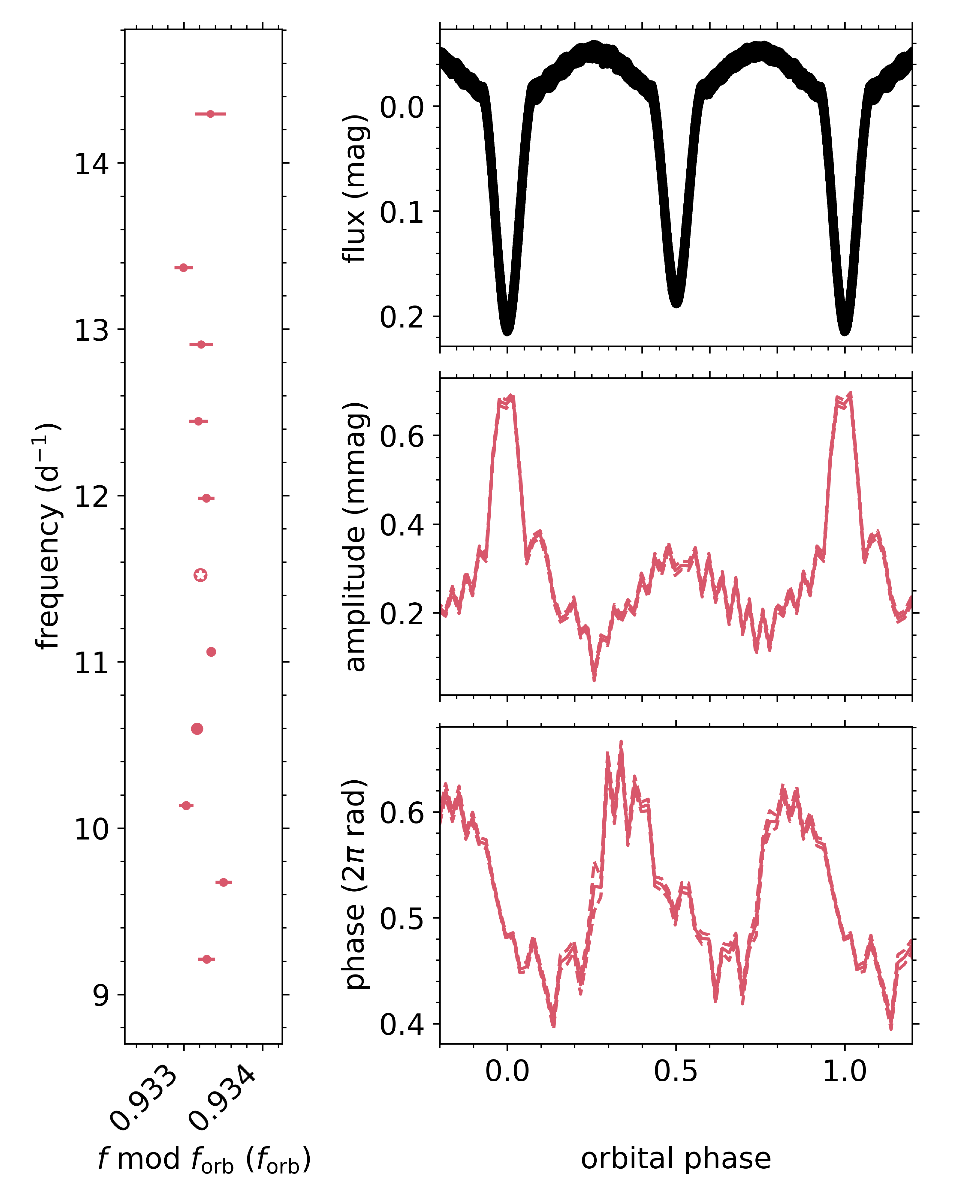}
\caption{\label{fig:kic09851944_multiplet-1152234} Tidally perturbed pulsation with frequency $f = 11.522340(5)\,\rm d^{-1}$.}
\end{figure}

\begin{figure}
\includegraphics[width=82mm]{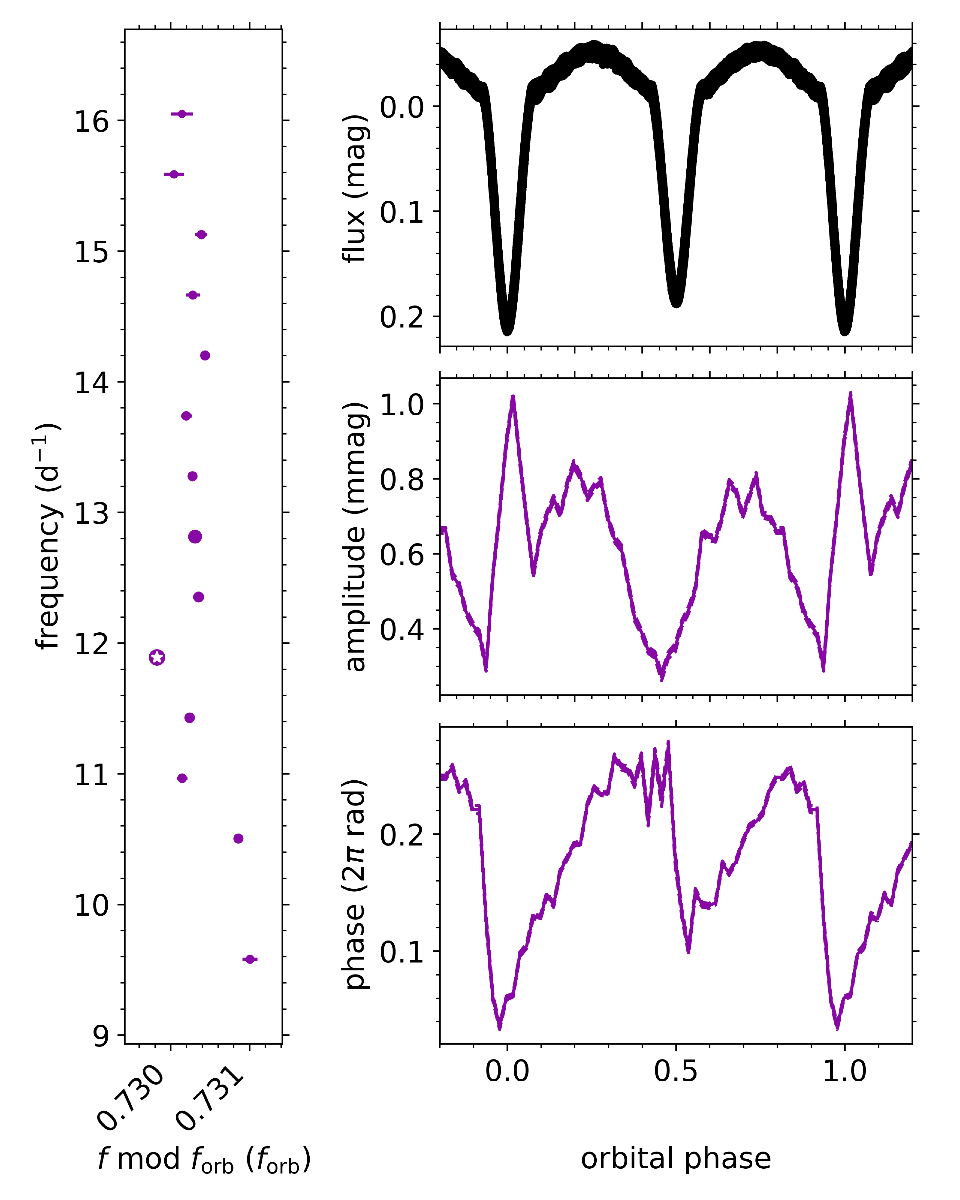}
\caption{\label{fig:kic09851944_multiplet-1189048} Tidally perturbed pulsation with frequency $f = 11.890477(2)\,\rm d^{-1}$.}
\end{figure}

\begin{figure}
\includegraphics[width=82mm]{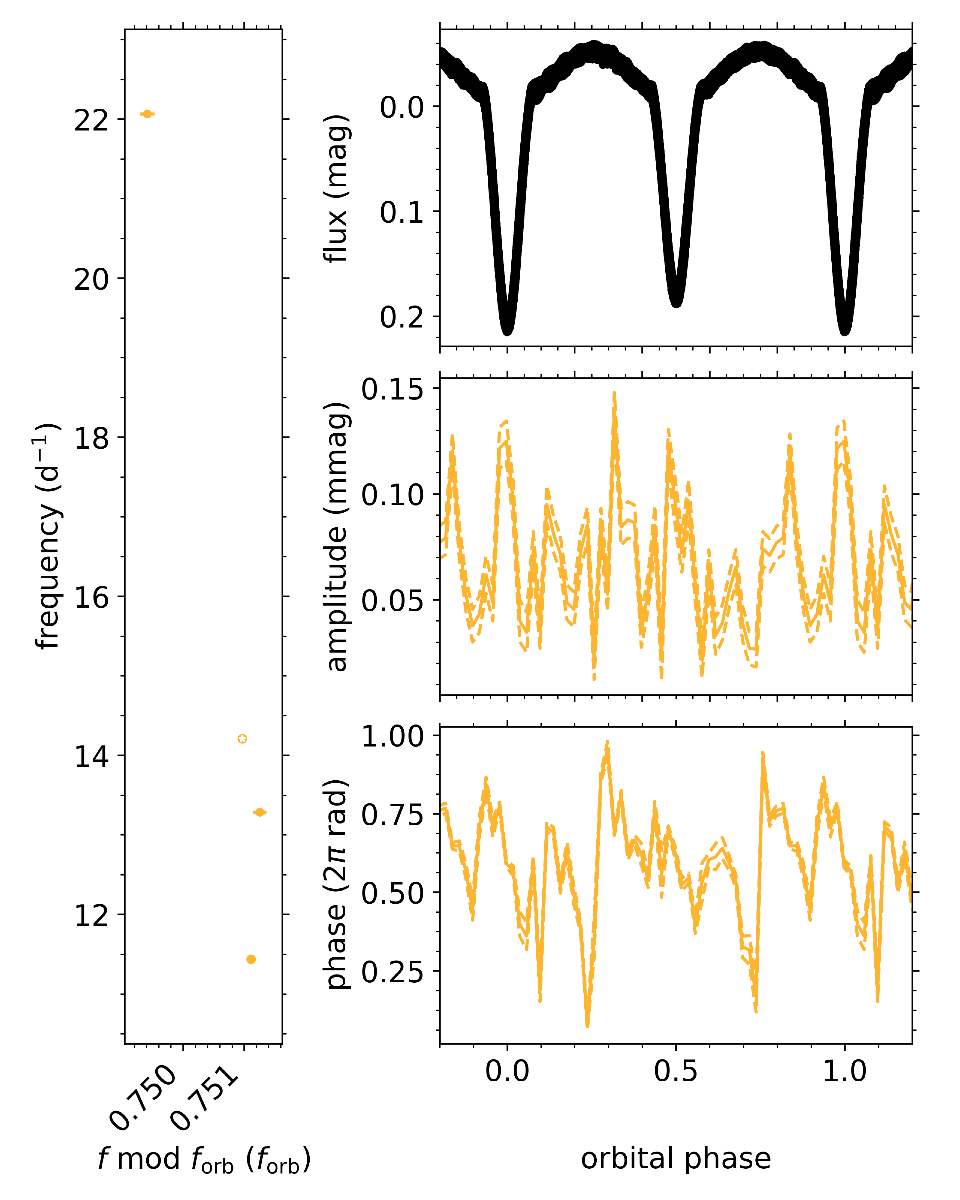}
\caption{\label{fig:kic09851944_multiplet-1421089} Tidally perturbed pulsation with frequency $f = 14.210888(3)\,\rm d^{-1}$.}
\end{figure}

\begin{figure}
\includegraphics[width=82mm]{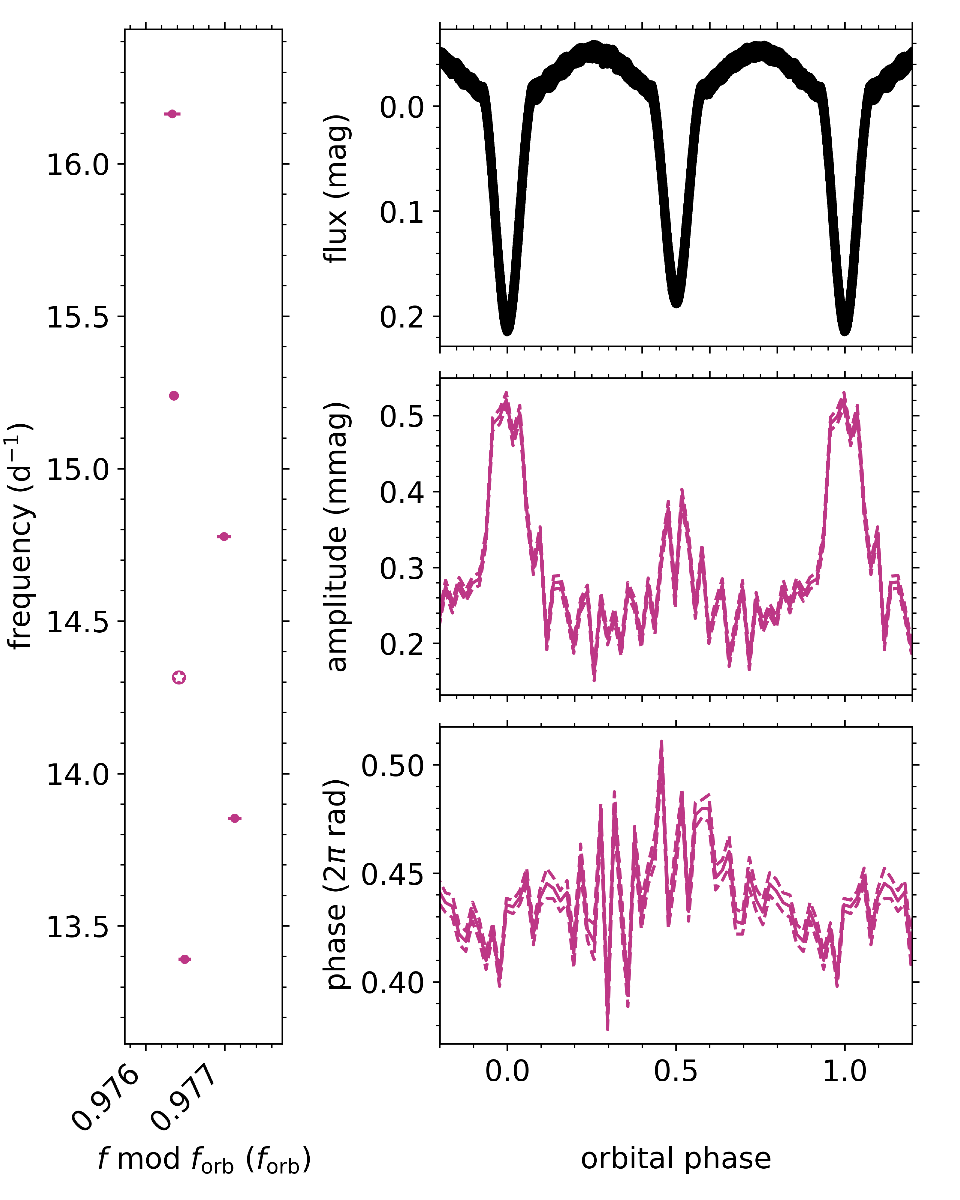}
\caption{\label{fig:kic09851944_multiplet-1431508} Tidally perturbed pulsation with frequency $f = 14.315077(5)\,\rm d^{-1}$.}
\end{figure}

\begin{figure}
\includegraphics[width=82mm]{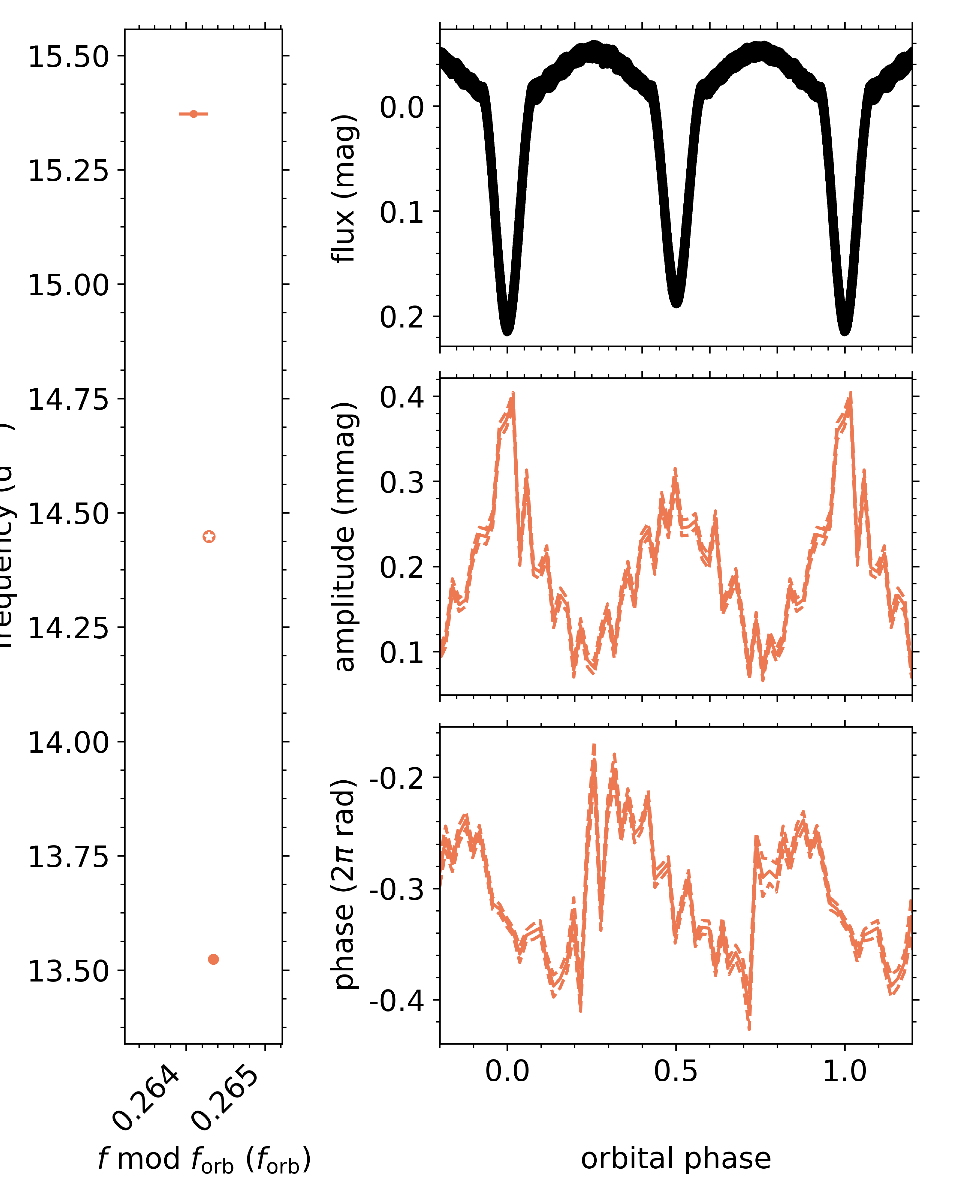}
\caption{\label{fig:kic09851944_multiplet-1444811} Tidally perturbed
pulsation with frequency $f = 14.448108(7)\,\rm d^{-1}$.}
\end{figure}

\begin{figure}
\includegraphics[width=82mm]{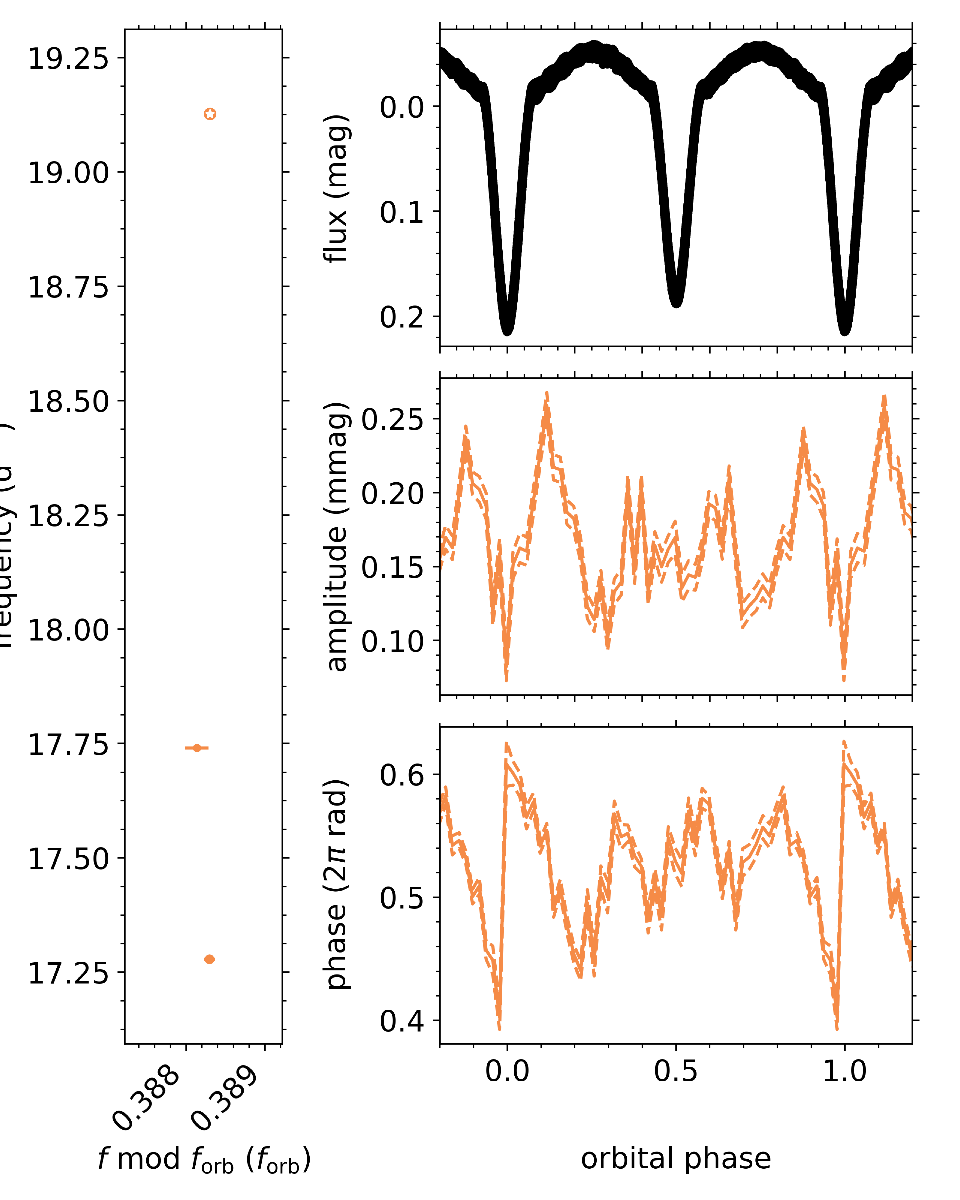}
\caption{\label{fig:kic09851944_multiplet-1912670} Tidally perturbed pulsation with frequency $f = 19.126701(8)\,\rm d^{-1}$.}
\end{figure}

\begin{figure}
\includegraphics[width=82mm]{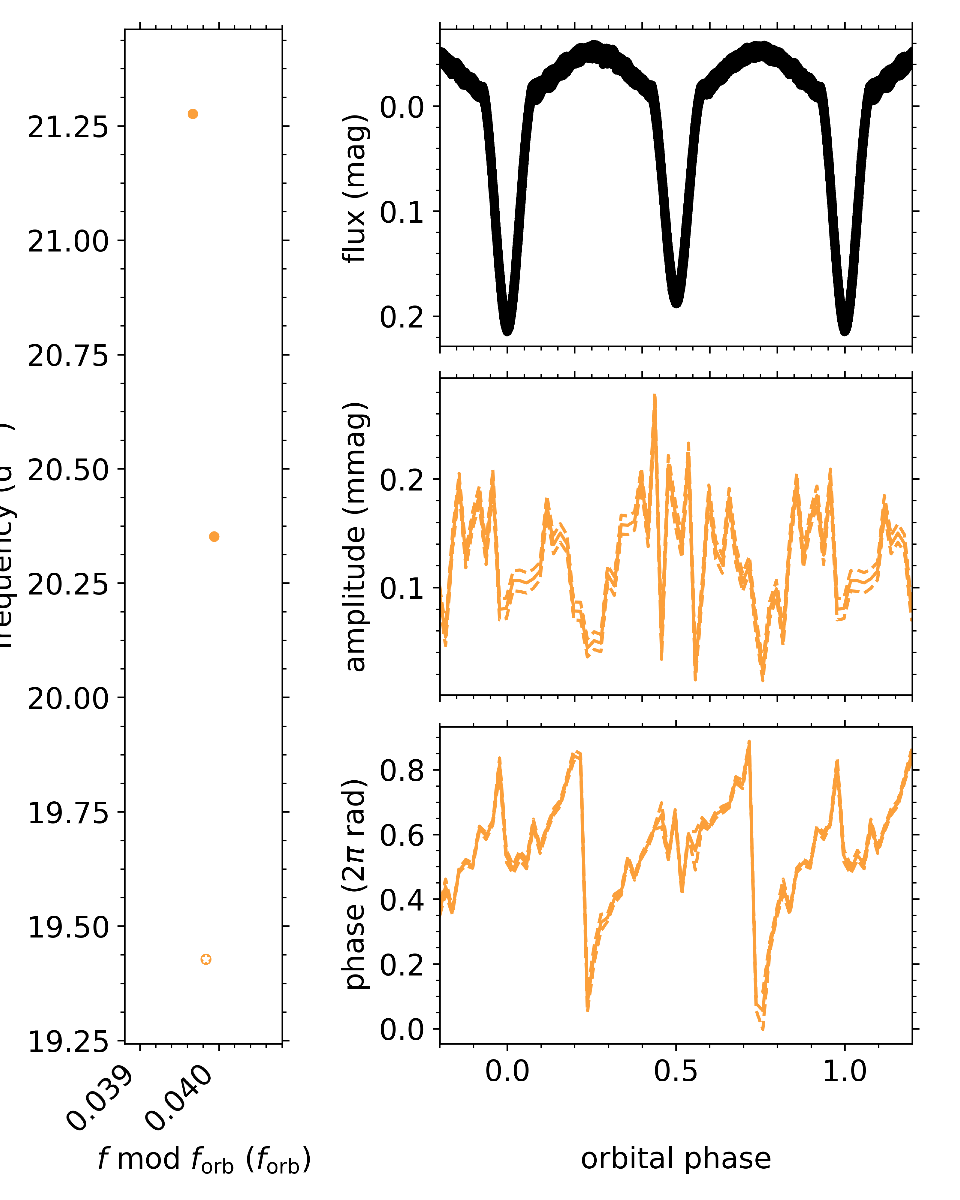}
\caption{\label{fig:kic09851944_multiplet-19427792} Tidally perturbed pulsation with frequency $f = 19.427792(15)\,\rm d^{-1}$.}
\end{figure}


\bsp	
\label{lastpage}
\end{document}